\documentclass[lineno]{jfm}

\usepackage{graphicx}
\usepackage{newtxtext}
\usepackage{newtxmath}
\usepackage{natbib}
\usepackage{hyperref}
\usepackage[dvipsnames]{xcolor}
\hypersetup{
    colorlinks = true,
    urlcolor   = blue,
    citecolor  = MidnightBlue,
}

\newcommand{\RomanNumeralCaps}[1]
\linenumbers

\usepackage{hvfloat}
\usepackage{newfloat}
\DeclareFloatingEnvironment[name=Example,placement=htbp,fileext=lop]{example}

\newsavebox{\myimage}

\usepackage{svg}

\usepackage{comment}

\usepackage{csquotes}

\shorttitle{Structures driving trailing-edge noise. Part II - Numerical investigation}
\shortauthor{Z. Yuan and others}

\title{Identification of structures driving trailing-edge noise. Part II - Numerical investigation}

\author{
Zhenyang Yuan\aff{1}\corresp{\email{zhenyang@kth.se}},
Simon Demange\aff{2}, 
Kilian Oberleithner\aff{2},
Andr\'e V. G. Cavalieri\aff{3}
\and
Ardeshir Hanifi\aff{1}
}

\affiliation{
    \aff{1}FLOW, Department of Engineering Mechanics, KTH Royal Institute of Technology, Stockholm, Sweden
    \aff{2}Laboratory for Flow Instability and Dynamics, Technische Universität Berlin, 10623 Berlin, Germany
    \aff{3}Divisao de Engenharia Aeronáutica, Instituto Tecnológico de Aeronáutica, São José dos Campos, Brazil
}

\begin{document}

\newcommand{\ZY}[1]{\textcolor{blue}{#1}}

\maketitle

\begin{abstract}

The aim of the present work is to investigate the mechanisms of broadband trailing-edge noise generation to improve prediction tools and control strategies. We focus on a NACA 0012 airfoil at 3 degrees angle of attack and chord Reynolds number $Re = 200,000$. A high-fidelity wall-resolved compressible implicit large eddy simulation (LES) is performed to collect data for our analysis.
The simulation is designed in close alignment with the experiment described in detail in the companion paper (Demange \textit{et al}. 2024\textit{b}). Zig-zag geometrical tripping elements, added to generate a turbulent boundary layer, are meshed to closely follow the experimental setup. A large spanwise domain is used in the simulation to include propagative acoustic waves with low wavenumbers.
An in-depth comparison with experiments is conducted showing good agreement in terms of mean flow statistics, acoustic and hydrodynamic spectra, and coherence lengths.
Furthermore, a strong correlation is found between the radiated acoustics and spanwise-coherent structures. To investigate the correlation for higher wavenumbers, spectral proper orthogonal decomposition (SPOD) is applied to the spanwise Fourier-transformed LES dataset. The analysis of all SPOD modes for the leading spanwise wavenumbers reveals streamwise-travelling wavepackets as the source of the radiated acoustics. This finding, confirming observations from experiments in the companion paper, leads  to a new understanding of the turbulent structures driving the trailing-edge noise. By performing extended SPOD based on the acoustic region, we confirm the low rank nature of the acoustics, and a reduced-order model based on acoustic extended SPOD is proposed for the far-field acoustic reconstruction.

\end{abstract}
\section{Introduction}

Trailing-edge noise, also known as airfoil self-noise, is one of the main sources of noise pollution from airfoils. It is a significant concern in aviation and onshore wind energy studies due to its substantial contribution to  noise pollution and its potential impact on communities located near airports or wind farms. Efforts are made to understand, model and mitigate trailing-edge noise in order to develop quieter and more environmentally friendly technologies, but this remains a challenge today.

Trailing-edge noise is caused by  the interaction between the airfoil blade and the turbulence produced in its own boundary layer and near wake \citep{BrooksThomasF.1989Asap}. It 
can be divided into two main categories: tonal noise and broadband noise.  For many low-to-moderate Reynolds number flow conditions, trailing-edge tonal noise is due to an acoustic feedback loop \citep{LonghouseR.E.1977Vsno, ArbeyH.1983Ngba, PröbstingS.2015Rotn, GolubevVladimir2021RAiA, RicciardiTulioR.2022Tiap}. Acoustic waves generated at the trailing edge propagate upstream, where they trigger instabilities in the boundary layer within its receptive region. These perturbations are then amplified, leading to the formation of two-dimensional vortices above a laminar separation bubble near the trailing edge. These spanwise coherent structures then scatter acoustic waves, closing the feedback loop. Alternatively, when the transition of the boundary layer over an airfoil is induced by freestream turbulence at high Reynolds numbers or by surface roughness elements positioned well upstream of the trailing edge, the resulting noise follows a broadband pattern. Unlike tonal noise, broadband noise manifests itself as a bump across the spectrum rather than a distinct tonal peak. A comprehensive review by \citet{LeeSeongkyu2021Tblt} summarises the large amount of work on modelling broadband noise over the past decades, including theoretical, numerical, and experimental approaches. However, the underlying noise generation mechanism is still unclear. Therefore, in this study, we focus on the simulation of broadband noise and target the noise generation mechanism.

The classic theoretical model for trailing-edge noise prediction is implied by Ffowcs Williams and Hall \citep{williams_hall_1970} using the classic Lighthill analogy \citep{LighthillMichaelJames1952Osga} with a tailored Green's function. This theoretical model employs the Green's function as the transfer function, and the measurement of the turbulence in the region close to the trailing edge is of significant importance \citep{LeeSeongkyu2021Tblt}. In most acoustic models, turbulence is represented by the  surface pressure coherence length. Moreover, the Amiet approach \citep{AMIET1975407, AMIET1976387} considers the surface pressure spectrum in the spanwise direction and has been successfully applied to the problems with more complex geometries, such as trailing edge serrations \citep{LyuB.2016Ponf}.
Nevertheless, the experimental measurement of the surface pressure coherence length or spectrum is typically challenging, given that the coherence length scale is comparable to the boundary layer thickness \citep{HerrigAndreas2013BATN}.

In a recent investigation, \citet{Nogueira2017Ampf} studied the phenomenon of installed jets and discovered that the sound scattered by the trailing edge is only weakly dependent on the two-point coherence of the source. Moreover, the authors found that the scattering condition depends on the spanwise wavenumber of the source. Only the source with a spanwise wavenumber ($k_z$) lower than the acoustic wavenumber ($k_0$) can radiate sound. This finding offers the potential for a novel approach for acoustic modelling: a surface pressure field can be understood as the superposition of spanwise Fourier modes, and only those with wavenumbers $k_z < k_0$  are responsible for acoustic radiation. Hence, according to this theory, the spanwise coherent pressure fluctuations are radiating sound. This has been validated by recent works \citep{Sano2019, Abreu2021}, which focus on the spanwise averaged fluctuations only. The findings indicate that streamwise non-compact coherent wavepacket-like structures have strong correlations with the radiating acoustics.  However, the relatively small width of the numerical domain in that study did not allow for the investigation of  higher wavenumbers nor did it come with experimental validation. Both of these shortcomings are addressed in the present work.

Although the identification of wavepackets and their relationship to acoustic radiation is a relatively new topic in trailing-edge noise research, it is a well-established area within the field of jet noise research, as evidenced by numerous studies and references \citep{CavalieriV.G2019WMfJ, schmidt_guide_2020, Pickering_Rigas_Schmidt_Sipp_Colonius_2021}. The mean field of a turbulent round jet is homogenous in azimuthal direction, which naturally allows for a  Fourier decomposition of the flow snapshots in azimuthal modes and  frequencies. This technique in combination with proper orthogonal decomposition established that the source of radiated acoustics is the non-compact wavepackets in the frequency range of the broadband noise. 

However, considering finite-span airfoils, it is challenging to extract flow structures at low spanwise wavenumbers due to the spanwise boundary conditions. The companion paper \citep{Demange2024arxiv} addresses this challenge by investigating the cross-spectrum density (CSD) of a spanwise surface pressure sensor array. By investigating the eigenvalues of the CSD matrix, it is shown that the dominant structures are very similar to Fourier modes. 
This provides a strong justification that spanwise Fourier decomposition leads to optimal bases in the spanwise direction.

Furthermore, the companion paper identifies spanwise coherent  structures (with spanwise wavenumber $k_z = 0$) that correlate strongly with the far-field acoustics at the same wavenumber. However, in these experiments, the limited number of surface sensors did not allow for correlations at higher spanwise wavenumbers nor could identify the shape of the coherent structures causing the correlation. Therefore, the main objective of this paper is to investigate such correlations and find the structures responsible for the sound radiation. 
%To the best of the authors' knowledge, this work is the first to investigate the higher wavenumber contents and their relationship with acoustics in the trailing edge noise research.

With increasing computational power, high fidelity simulations, such as direct numerical simulations and large-eddy simulations, are now capable of resolving turbulent structures in detail and may reveal their correlation with acoustics \citep{OberaiAssadA2002CoTN, SandbergR.D.2009Dnso}. 
In addition, data-driven methods such as proper orthogonal decomposition (POD) \citep{lumey_1967, lumey_stochastic_1970} and its spectral version SPOD \citep{PICARD2000359,SieberMoritz2016Spod} have been used to identify the coherent structures in the various flow conditions \citep{schmidt_guide_2020}. SPOD has been used extensively in previous work, such as \citet{SchmidtOliverT.2018Saoj, CavalieriV.G2019WMfJ, AbreuLeandraI.2020Spod, demange2024} numerically and \citet{HOLMES1997337, Arndt1997, SUZUKI_COLONIUS_2006, gudmundsson_colonius_2011} experimentally.  In particular, the SPOD method is successfully applied to the trailing-edge broadband noise problem in the work of \citet{Sano2019, Abreu2021} mentioned above.

In the present study, a highly resolved compressible LES is performed to resolve the turbulent flow around a NACA 0012 airfoil. The configurations include a freestream Mach number of $M = 0.3$, a chord-based Reynolds number of $Re = 200,000$ and three degrees angle of attack. A geometric tripping element is used to assure turbulent transition in the boundary layer and broadband noise. 
Unlike the previous simulations by \citet{Sano2019, Abreu2021}, we designed our computational domain with respect to the scattering condition. Here, the span width is larger than 40\% of the chord length to accommodate more propagative wavenumbers. This allows the exploration of the structures associated with noise generation also for non-zero spanwise wavenumbers. 

The structure of this paper is as follows: \S \ref{numerical_setup} describes the numerical setup and its relation to that of the experiments. \S \ref{results} provides the in-depth comparison between the numerical and experimental datasets and the flow visualization over the airfoil. The comparison includes statistics, spectral analysis of the acoustic and hydrodynamic datasets and addressing the scattering condition. The SPOD analysis is presented in \S \ref{wavepacket} including  acoustic and hydrodynamic based extended SPOD (ESPOD) \citep{BOREEJ2003Epod}. The analysis in \S \ref{HSPOD} focuses on the identification of structures that generate acoustics, and in \S \ref{ASPOD} focuses on  the reduced-order modelling for acoustic prediction. Finally, \S \ref{conclusion} presents the conclusions and perspectives of this work.

\section{Numerical simulation}
\label{numerical_setup}
The simulation is designed to operate under conditions similar to those of the experiment, as described in the companion paper \citep{Demange2024arxiv}, to allow for full validation of the LES results by the experimental data. However, due to the fact that the simulation requires relatively a higher Mach number to speed up the convergence rate and a lower Reynolds number to reduce the number of grid points, some compromises are made on both the simulation and experimental sides to ensure consistency of designs. In this section, we briefly introduce the experimental setups, with detailed descriptions of the facilities and measurements available in the companion paper \cite{Demange2024arxiv}, and focus on the LES configurations and the comparison between experimental and simulation setups.

\subsection{Experimental condition and airfoil model}

The airfoil investigated in the experiment is a modified NACA 0012 airfoil with a rounded trailing edge of radius 0.4 mm. The chord length of the model is $c = 100$ mm. The airfoil is mounted between two vertical side plates in a large anechoic chamber. The freestream is provided by a rectangular open jet nozzle, the distance from the nozzle to the leading edge of the airfoil is 1.5$c$.  The width of the nozzle is 4$c$, which is equal to the span of the airfoil. For the following discussions, we define a Cartesian coordinate system ($x$, $y$, $z$) which defines the streamwise (horizontal), vertical and spanwise directions, respectively.
In order to investigate the broadband trailing-edge noise, a tripping device (a zig-zag shaped tape) with a height of 0.42 mm, an angle of 60 degrees and a streamwise extend of 8.3 mm is placed at 5\% chord length on both the pressure and suction sides. The experimental campaigns are operated at freestream velocities of 30 m/s and 45.4 m/s, giving freestream Mach numbers of 0.088 and 0.133 and chord-based Reynolds numbers of $2 \times 10^5$ and $3 \times 10^5$.  

\subsection{Solver, mesh and setups}
The open-source numerical framework for high-order flux reconstruction PyFR  \citep{WITHERDEN20143028} is used for wall-resolved implicit large eddy simulations (iLES).
The compressible Navier-Stokes equations are solved in their conservative format for flow quantities $q$ = [$\rho, \rho u, \rho v, \rho w, E$], where $\rho$ is density, $u,\ v,\ w$ are Cartesian velocity components and $E = \frac{p}{\gamma-1}+0.5\rho|\mathbf{u}|_2^2$. The ideal gas law is also used to complete the set of equations and Sutherland's law is used for the viscosity correction.
The discrete set of equations is obtained from the spatial-spectral ($hp$) discretisation, which first decomposes the computational domain into  a set of mesh elements ($h$ refinement) and uses high-order polynomials ($p$ refinement) within each element to represent the flow fields. The supported element types are hexahedra, prisms, tetrahedra and pyramids. The equations are then integrated in time using an adaptive six-stage Runge-Kutta method.

We use the same airfoil geometry as in the experiments, including the zig-zag tripping elements. The advantage of using these physical surface tripping elements is that the position of transition to turbulence is fixed, making the results less sensitive to the low level of inflow turbulence in the experiments (turbulence intensity = $0.2\%$ in the experiments at freestream velocities of 30 m/s \citep{schneehagen_design_2021}), which is ignored in the LES. Further, compared to conventional tripping technique using volume forcing \citep{SchlatterPhilipp2012Tbla}, the noise generated by physical tripping elements could be more closely compared to the experiments. 

Due to computational cost and practicality, the numerical simulation targets the case with Reynolds number $Re = 2 \times 10^5$, which requires a smaller numerical mesh, and Mach number $M = 0.3$, which improves the convergence of the simulation. As mentioned in the introduction, the far-field acoustics can be affected by the scattering condition \citep{Nogueira2017Ampf}. Therefore, we use a  spanwise length of $L_z = 43.75$ mm, which fits seven periods of the zig-zag tripping elements in the LES. This span width  ensures that a sufficient number of propagative acoustic waves can appear in the simulation. Further results on the scattering of propagating waves are detailed in \S \ref{scattering_condition}. 

A significant difference between LES and experiments is the effect of installation of the model in the wind tunnel. To keep the computational costs reasonable, in the simulation, the airfoil is placed in a uniform clean flow where the installation effect is not considered. We are aware of possible differences this may cause, however, as it will be demonstrated in \S \ref{results}, a good agreement between experimental and numerical results is achieved.  

PyFR supports structured and unstructured mesh topology. In order to exploit this, the following meshing strategy is implemented: a wall resolved O-grid with the maximum resolution at the mid-chord is given in terms of wall units: $\Delta x^+ < 10, \Delta z^+ < 5$ and $\Delta y^+ < 0.9$. At the trailing edge, resolution in the streamwise direction is kept as $\Delta x^+ < 6$ and $\Delta y^+ < 0.6$.
The structured mesh is applied in the near wall field and wake resolved region until 1.2 chords after the trailing edge. 
The first layer of elements to the airfoil surface has an initial height of $\Delta s = 0.000245c$ and the growth of the structured layers has a rate of 1.08. 
Close to the tripping region, an unstructured mesh (prisms layers) is employed instead to avoid stretching the grid and preserve the resolution. The prism layer has much smaller elements with $\Delta s = 0.0031c$, which can reduce the mesh-to-mesh discontinuity due to the discontinuous nature of the flux reconstruction method. 

An unstructured mesh is also used in the far field to avoid large aspect ratios and reduce computational cost.  At a streamwise position $x/c = 1$ (corresponding to the trailing edge), the mesh element spacing in the acoustic field is set to $\Delta s$ = [$0.02c$, $0.73c$, $0.10c$] when $y$ = [$0.25c$, $1.0c$, $5.0c$] respectively. The aspect ratio of the elements is kept close to one. Analysis of the simulation data shows that the acoustic waves are well-resolved  for frequencies up to $St = 25$ within $y/c < 5$. Note that the Strouhal number is defined based on the freestream velocity $U_{\infty}$ and chord length $St = fc/U_{\infty}$. Acoustic data obtained for $y/c > 10$ are not considered in the following analysis. The computational domain is extended to 20 chords away from the airfoil surface to reduce the influence of the outflow boundary conditions. 

\begin{figure}
    \centering
    \includegraphics[width = 0.8\linewidth]{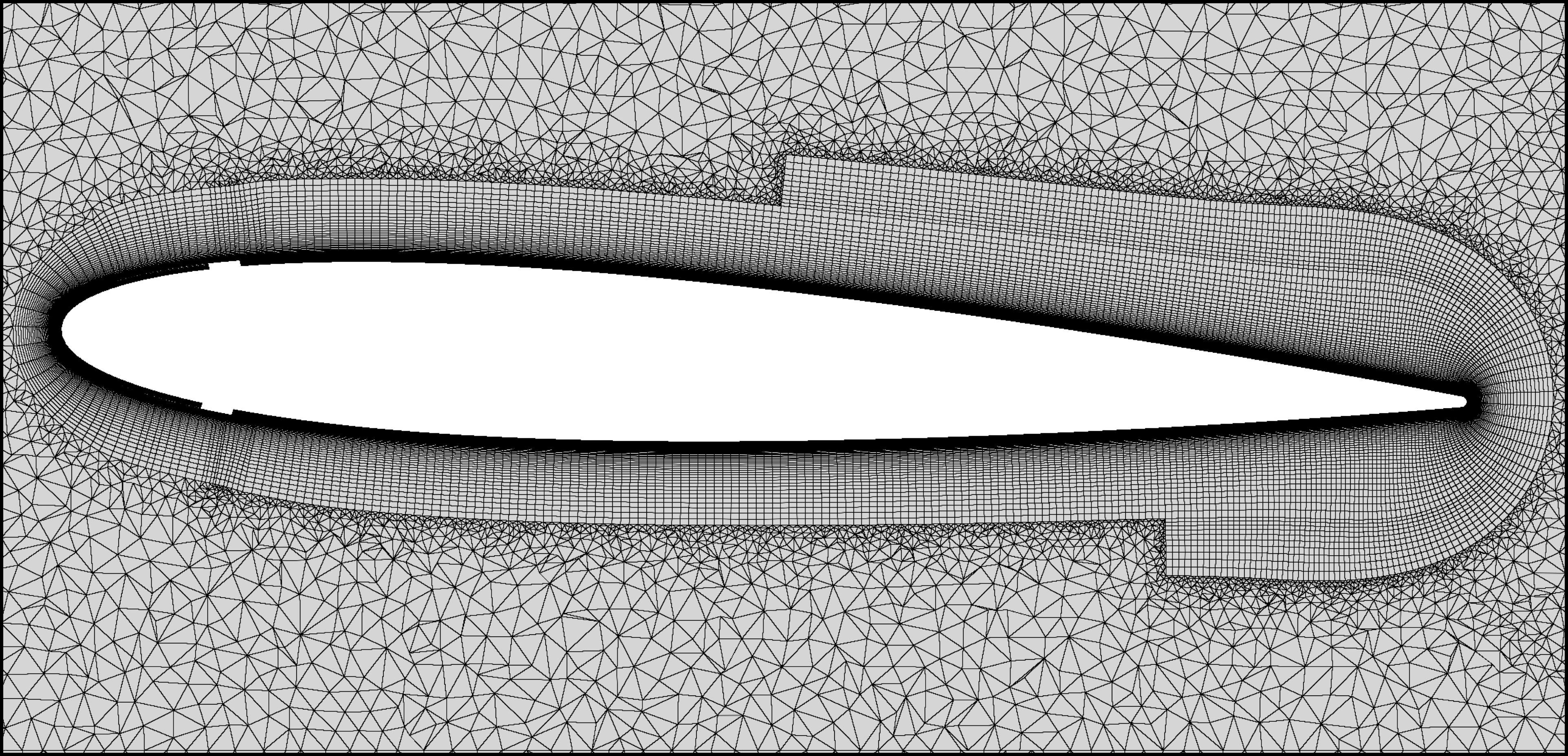}
    \includegraphics[width = 0.8\linewidth]{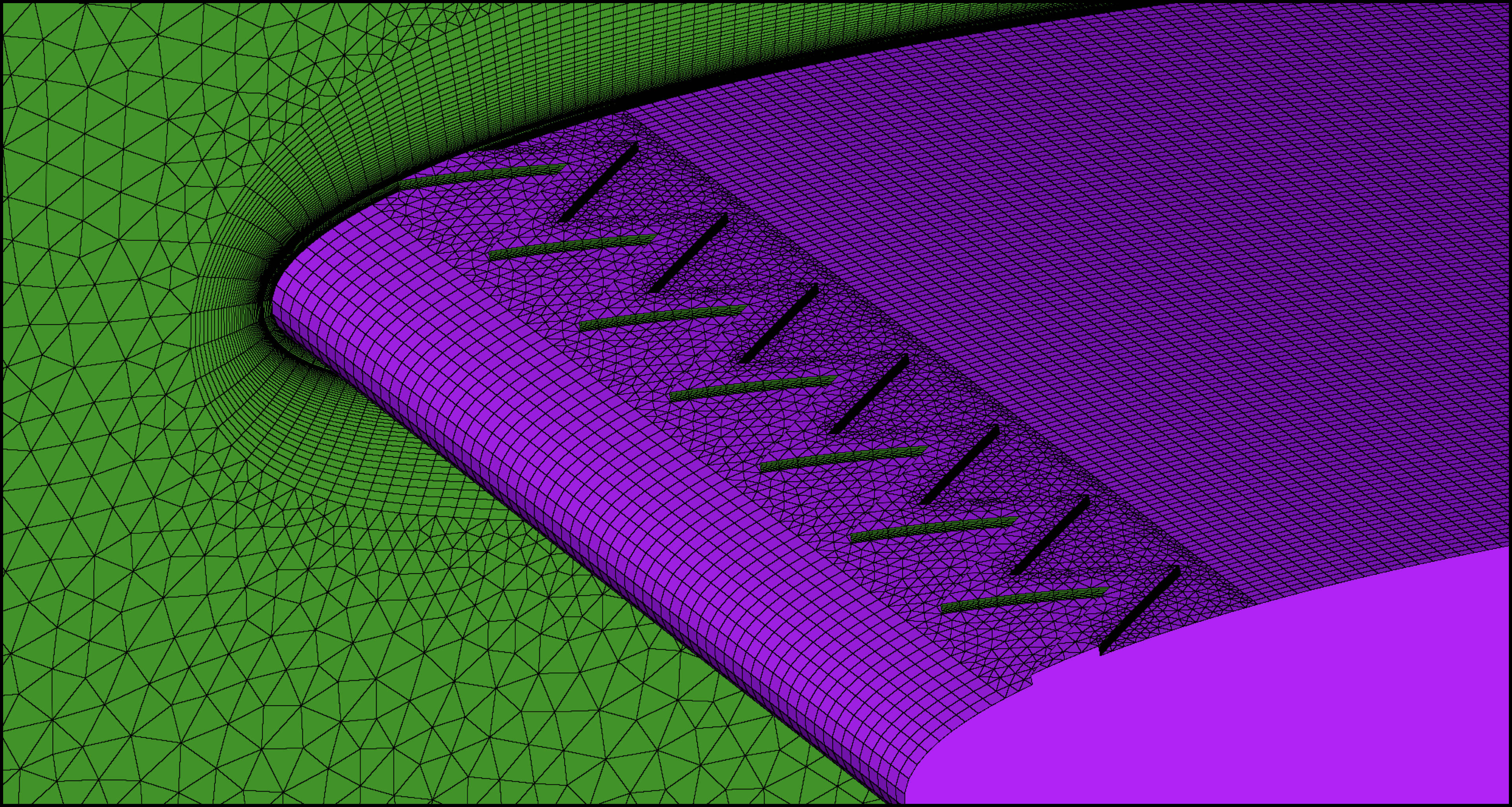}
    \caption{High order computational mesh elements near the airfoil, top: the wall resolved boundary layer mesh and the near field acoustic mesh region; bottom: a close look of surface mesh around tripping elements which has been lifted to the second order to fit geometry curvatures. Grid points inside the elements are skipped for a better visualization.}
    \label{fig:mesh}
\end{figure}

A designed sponge zone of seven chords length is added around the outer boundaries to absorb hydrodynamic wake and acoustic waves \citep{FreundJ.B1997PIBC,BodonyDanielJ.2006Aosz}. The sponge is added to the system of equations as a forcing term in the form $-\sigma(q - q_0)$. The sponge strength parameter $\sigma$ is carefully chosen so that no significant wave reflection is observed. The subscript 0 here indicates the freestream flow quantities. The outflow boundaries use a non-reflecting Riemann invariant boundary condition to work with sponges, which ensures that a non-reflecting condition is provided. The inflow boundary also uses a Riemann invariant condition to provide a homogeneous potential flow with $M = 0.3$, ambient density and pressure. 
A periodic boundary condition is employed in the spanwise direction. Additionally, a non-slip, adiabatic wall condition is applied to the airfoil surface. The original mesh is elevated to the second order in order to align with the geometry curvatures. The detailed mesh is illustrated in figure \ref{fig:mesh}. 

The simulations are conducted with the polynomial order four on LUMI-G AMD MI250X 128Gb GPU nodes. The quadrature order employed for anti-aliasing is one degree of freedom higher than that of the solver. This yields a total grid number of approximately 500 million. 

%-------------------------------------------------------------------------
\subsection{Data sampling}

A total of 7813 snapshots with a sampling rate of $St = 266.7$ are stored and employed for spectral analysis. The high sampling rate ensures that aliasing from high frequencies will have a negligible energy contribution to the spectrum. Ultimately, the resulting numerical dataset is approximately 200 TB in size.

Additionally, a number of probe points are defined in the mesh in order to emulate the experimental surface pressure sensors and the acoustic microphone line array configuration presented in the companion paper. Two lines of surface sensors are positioned directly on the surface of the airfoil, at $x/c = 0.88$ and $0.92$ with a spacing between sensors of $0.24$ mm, to record the hydrodynamic pressure fluctuations. Furthermore, acoustic microphones situated at a distance of $\delta y$ = [-300, -100, +100, +200, +300] mm above (+) or below (-) the trailing edge are employed to record acoustic pressure fluctuations. The arrays comprise 187 and 60 sensors for the surface and acoustic lines, respectively. All sensors utilise the identical sampling frequency as that employed for the snapshots, but with a considerably longer time signal.

\section{Validation of the simulation results against experimental data}
\label{results}

This section presents a comprehensive comparison between the simulation and experiment, with the objective of fully validating the simulation results. Despite the differences in flow configurations, as previously discussed, the simulation results have been processed in a manner that allows for a quantitative comparison with the experimental results, where possible. In addition to the experimental results, the simulation data are used to gain a deeper insight into the physical phenomenon by means of the spectral analysis. 

\subsection{Instantaneous flow field}
\begin{figure}
    \centering
    \includegraphics[width = 1\linewidth]{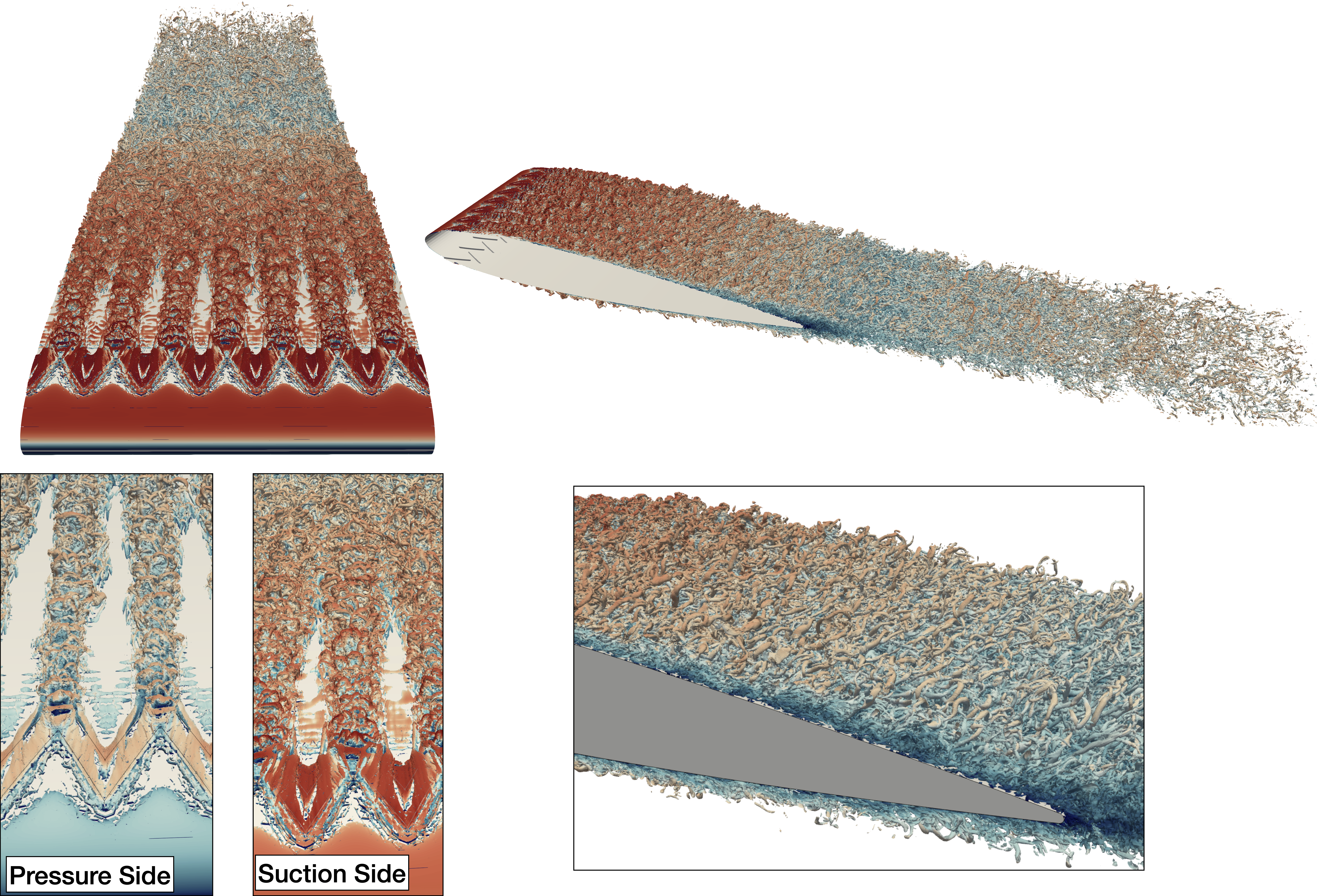}
    \caption{Instantaneous iso-surfaces of the Q-criterion colored by streamwise $u$-velocity. Left: Leading edge tripping elements triggering transition; right: fully developed turbulence close to the trailing edge and wake region. The value Q = 0.001 is chosen for visualization.}
    \label{fig:Q}
\end{figure}

In order to visualize the turbulent structures around the airfoil, figure \ref{fig:Q} illustrates Q-criterion coloured by the streamwise velocity $u$. The leading-edge tripping elements trigger immediate transition to turbulence. However, the tripping also creates streamwise streaky structures. These structures continue to propagate downstream and eventually merge. On the pressure side, due to the favourable pressure gradients, it takes a longer distance to reach spanwise homogeneous turbulent field. Between the tripping elements, small spanwise coherent structures can be identified. These structures have a small streamwise wave length. Close to the trailing edge, the Q-criterion indicates that the flow is fully developed to spanwise homogeneous turbulence, indicating low coherence of the flow structures. Due to this low coherence close to the trailing edge (discussed in the following sections), a broadband noise pattern is expected. 

%-------------------------------------------------------------------
\subsection{Mean flow profiles and turbulence statistics}
\label{statistics}

\begin{figure}
    \centering
    \includegraphics[width = \linewidth]{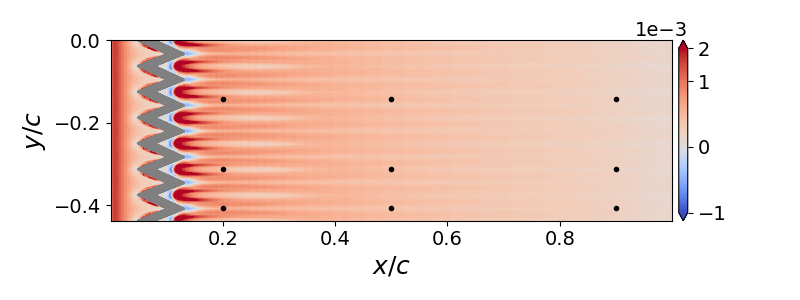}
    \caption{Skin friction coefficient ($C_f$) map of time averaged field. Colour map is saturated such that blue and red show negative and positive $C_f$, respectively.}
    \label{fig:cf}
\end{figure}

\begin{figure}
    \centering
    \begin{subfigure}{0.32\textwidth}
        \includegraphics[width = \linewidth]{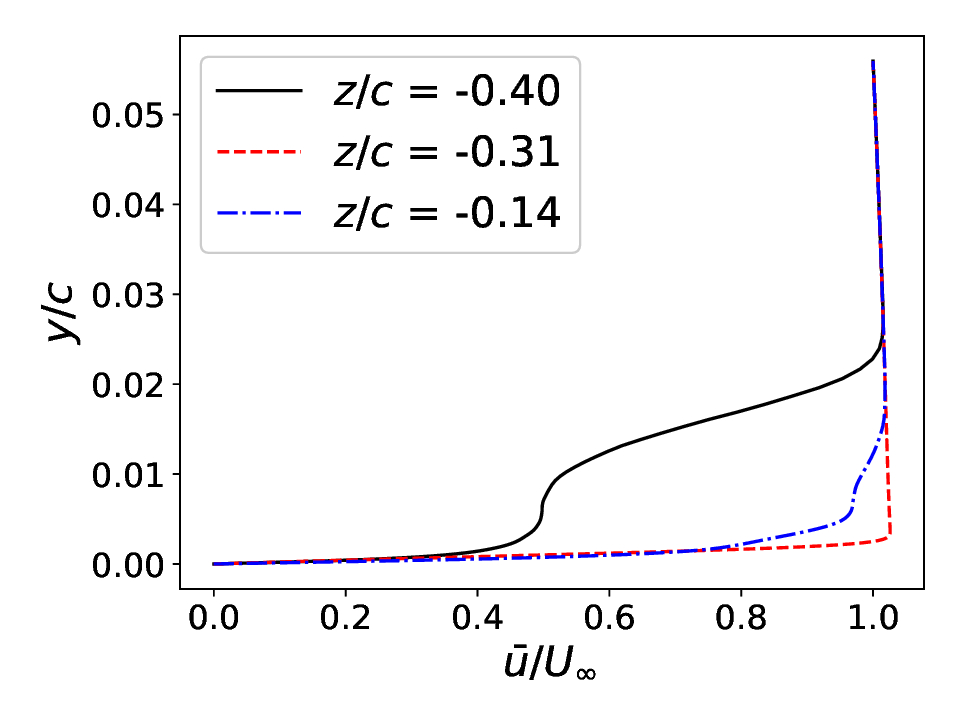}
        \caption{$x/c = 0.2$}
    \end{subfigure}
    \begin{subfigure}{0.32\textwidth}
        \includegraphics[width = \linewidth]{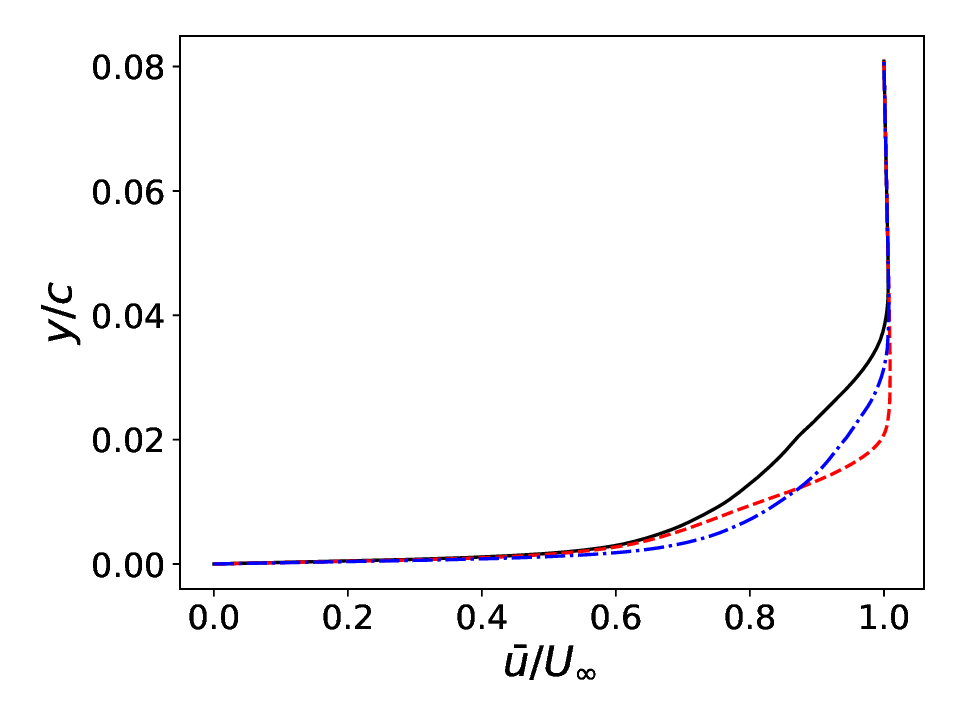}
        \caption{$x/c = 0.5$}
    \end{subfigure}
    \begin{subfigure}{0.32\textwidth}
        \includegraphics[width = \linewidth]{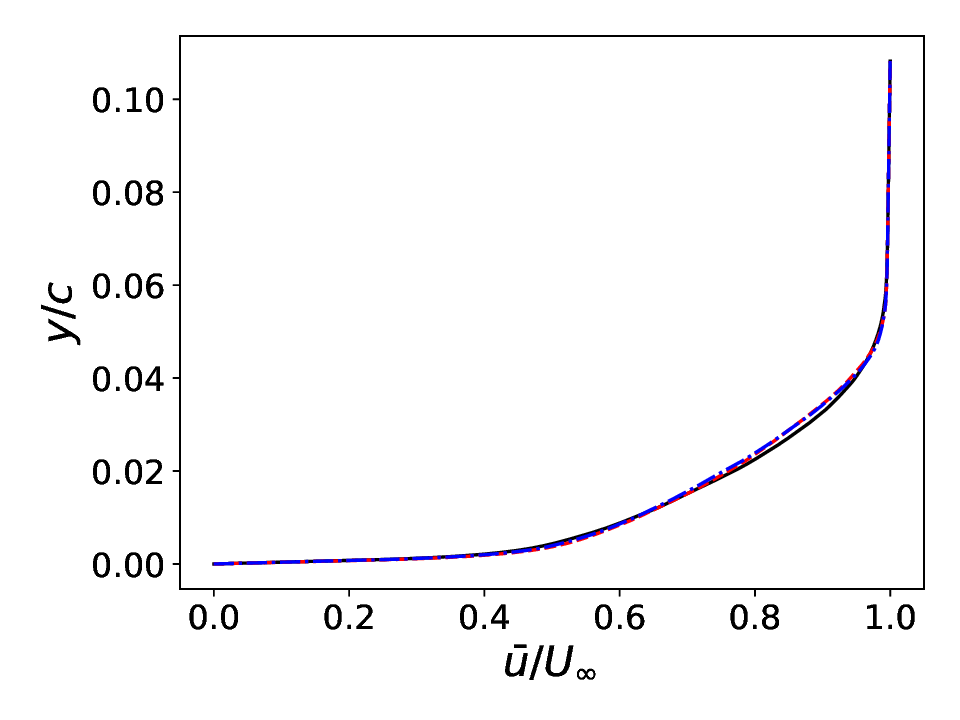}
        \caption{$x/c = 0.9$}
    \end{subfigure}
    \caption{Mean velocity profiles obtained at the streamwise/spanwise locations marked as black dots in figure \ref{fig:cf}.}
    \label{fig:streamwise_mean_profile}
\end{figure}

Figure \ref{fig:cf} illustrates the time averaged skin friction coefficient ($C_f$) on the suction side, defined as:
\begin{equation}
    C_f = \frac{\tau_w}{0.5\rho_{\infty}U_{\infty}}
\end{equation}
where $\tau_w $ represents the wall shear stress. In the vicinity of the tripping elements, a minor recirculation region forms between the tripping elements. It is notable that streamwise elongated structures are evident immediately after these elements. This observation is consistent with the Q-criterion visualizations. These streaky structures generate spanwise inhomogeneity. In order to check the spanwise homogeneity evolution along the different streamwise locations,  mean velocity profiles are shown in figure \ref{fig:streamwise_mean_profile}. The sampling locations are marked with black dots in figure \ref{fig:cf}. At $x/c = 0.2$ the velocity profiles show a distinct behaviour as the streaks are very strong. This difference is maintained at $x/c = 0.5$. Further downstream, at $x/c = 0.9$, the velocity profiles converge with insignificant difference. This indicates the development of fully turbulent flow and spanwise homogeneity in the vicinity of the trailing edge. 

\begin{figure}
    \centering
    \begin{subfigure}{0.45\textwidth}
        \centering
        \includegraphics[width = \linewidth]{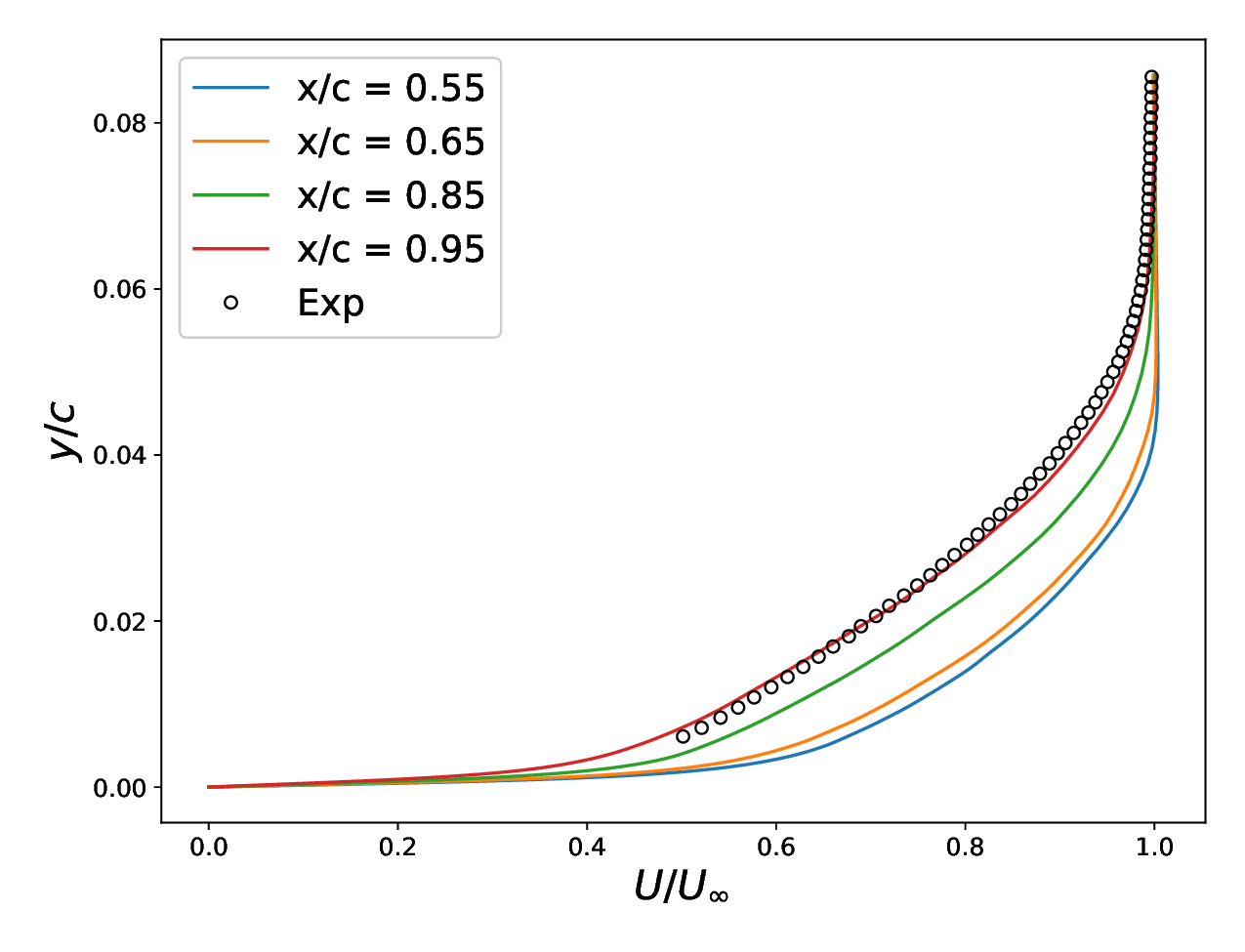}
        \caption{Mean velocity measured in the outer units.}
        \label{fig:mean_u_outer_scale}
    \end{subfigure}
    \begin{subfigure}{0.45\textwidth}
        \centering
        \includegraphics[width = \linewidth]{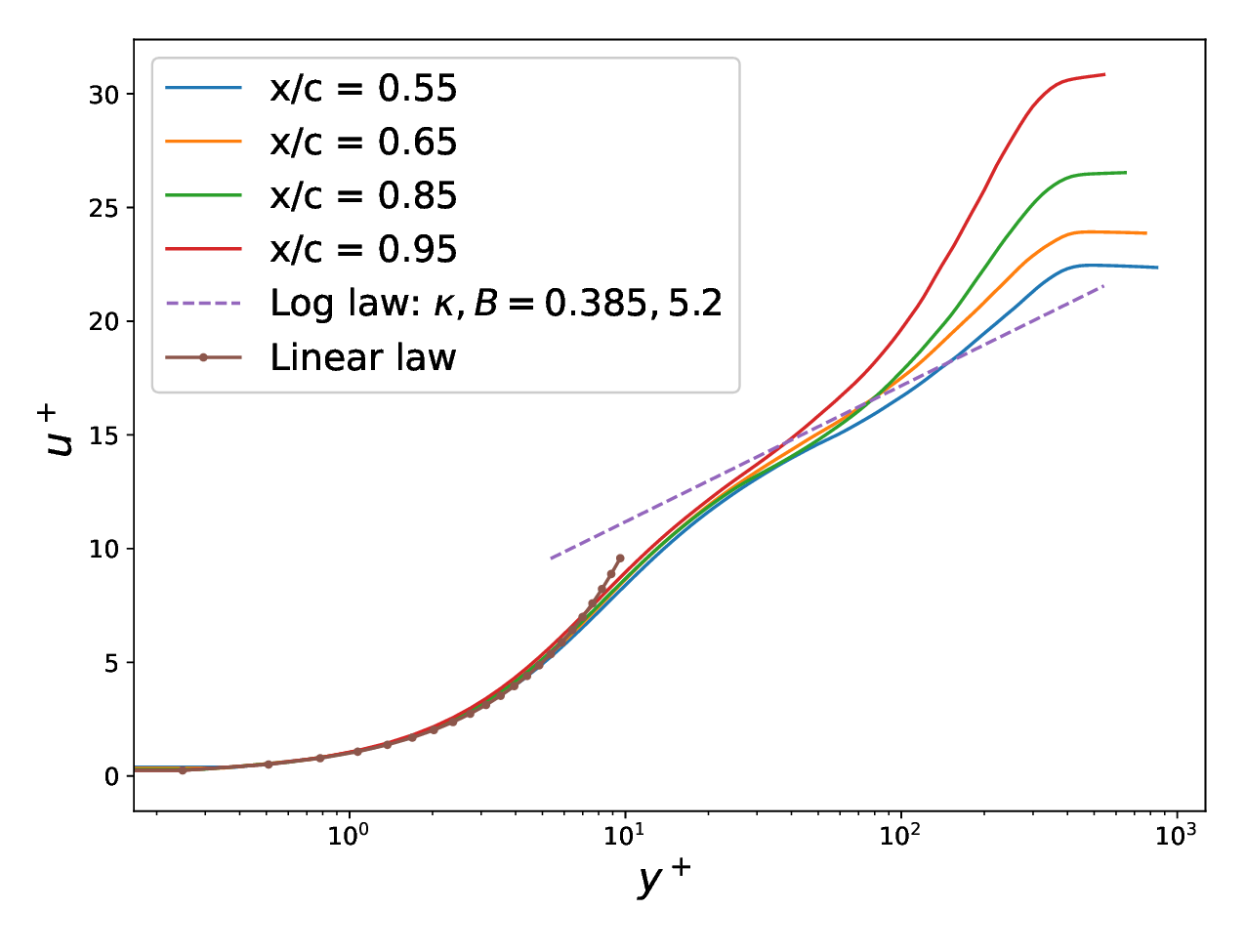}
        \caption{Mean velocity measured in the inner units.}
        \label{fig:mean_u_inner_scale}
    \end{subfigure}
    \caption{Mean velocity profile at the different 
 streamwise stations: $x/c = 0.55$, $x/c = 0.65$, $x/c = 0.85$ and $x/c = 0.95$. Log law is defined as $u^{+}=1/\kappa \ln y^{+} + B$ where $\kappa = 0.385$ and $B = 5.2$.}
    \label{fig:mean_u}
\end{figure}

In order to get an insight into the flow state on the airfoil surface, mean total velocity ($U$) profiles are computed from the LES data at streamwise locations $x/c$ = [$0.55$, $0.65$, $0.85$, $0.95$] at the mid-span location. In the experiments, the velocity profile was measured, using hot-wire anemometry, at $x/c \approx 0.95$. 
The velocity profiles are shown in the figure \ref{fig:mean_u_outer_scale}. Here, the streamwise velocity is normalised by its local freestream value $U_{\infty}$ and the wall-normal distance is normalised to the chord length. The shape of the experimental velocity profile is found to be in close agreement with the numerical data in the region close to the trailing edge.

Figure \ref{fig:mean_u_inner_scale}  shows the   velocity profiles from the LES  plotted in the inner units, denoted by $()^+$. This scaling is made using the wall friction velocity, $u_{\tau}= \sqrt{\tau_w/\rho}$, as the reference velocity and the viscous length scale, $\nu/u_{\tau}$, as the reference length. The mean flow profiles show a commendable fit with a linear increment within the viscous sub-layer and a logarithmic growth region. In this context, the logarithmic law is defined as $u^{+}=1/\kappa \ln y^{+} + B$, with $\kappa = 0.385$ and $B = 5.2$. The choice of $\kappa$ comes from the  work of \citet{HultmarkM2012Tpfa} and \citet{MarusicIvan2013Otlr}.
The experimental data is not shown due to  the lack of access to $u_{\tau}$.

\begin{figure}
    \centering
    \begin{subfigure}{\textwidth}
        \centering
        \includegraphics[width = 0.9\linewidth]{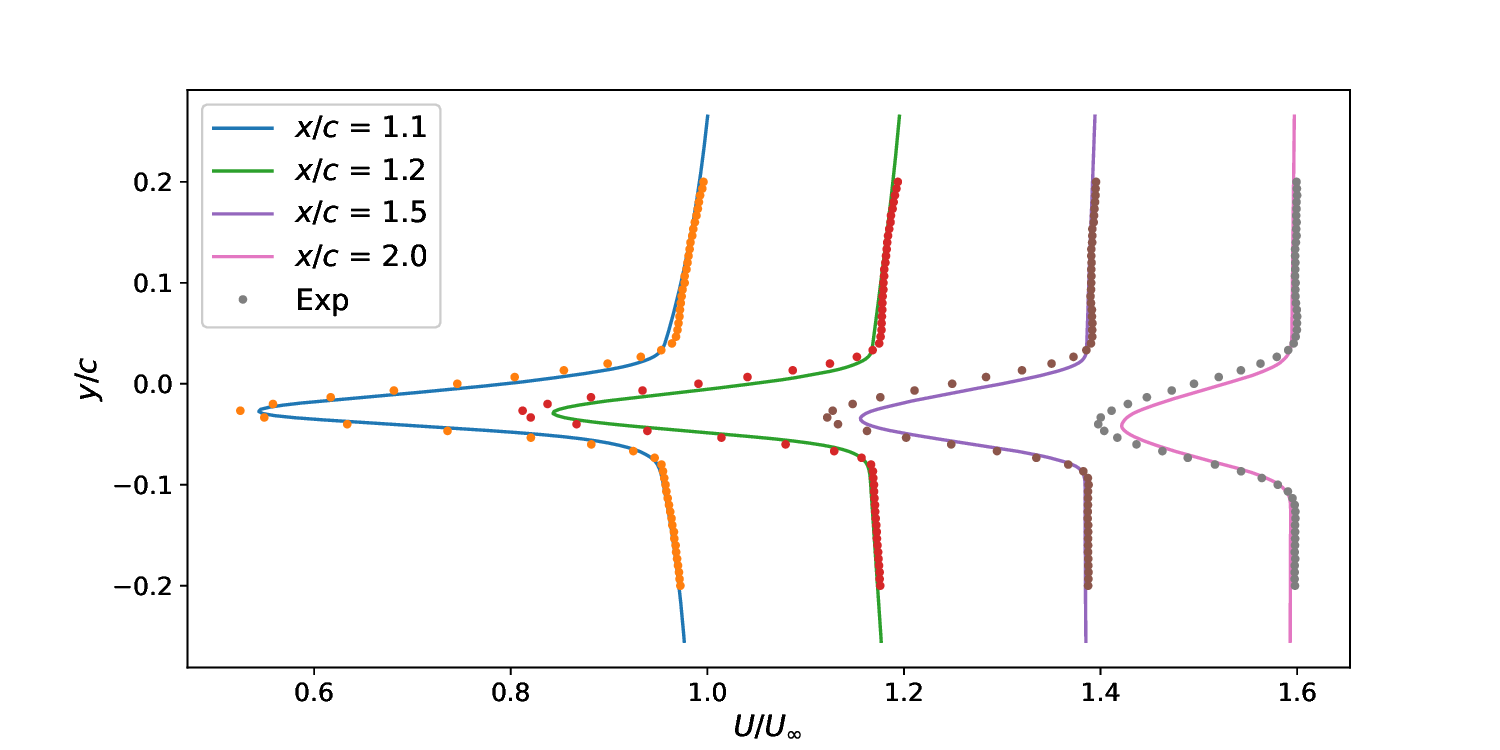}
        \caption{Mean velocity profile.}
        \label{fig:mean_u_wake}
    \end{subfigure}
    \begin{subfigure}{\textwidth}
        \centering
        \includegraphics[width = 0.9\linewidth]{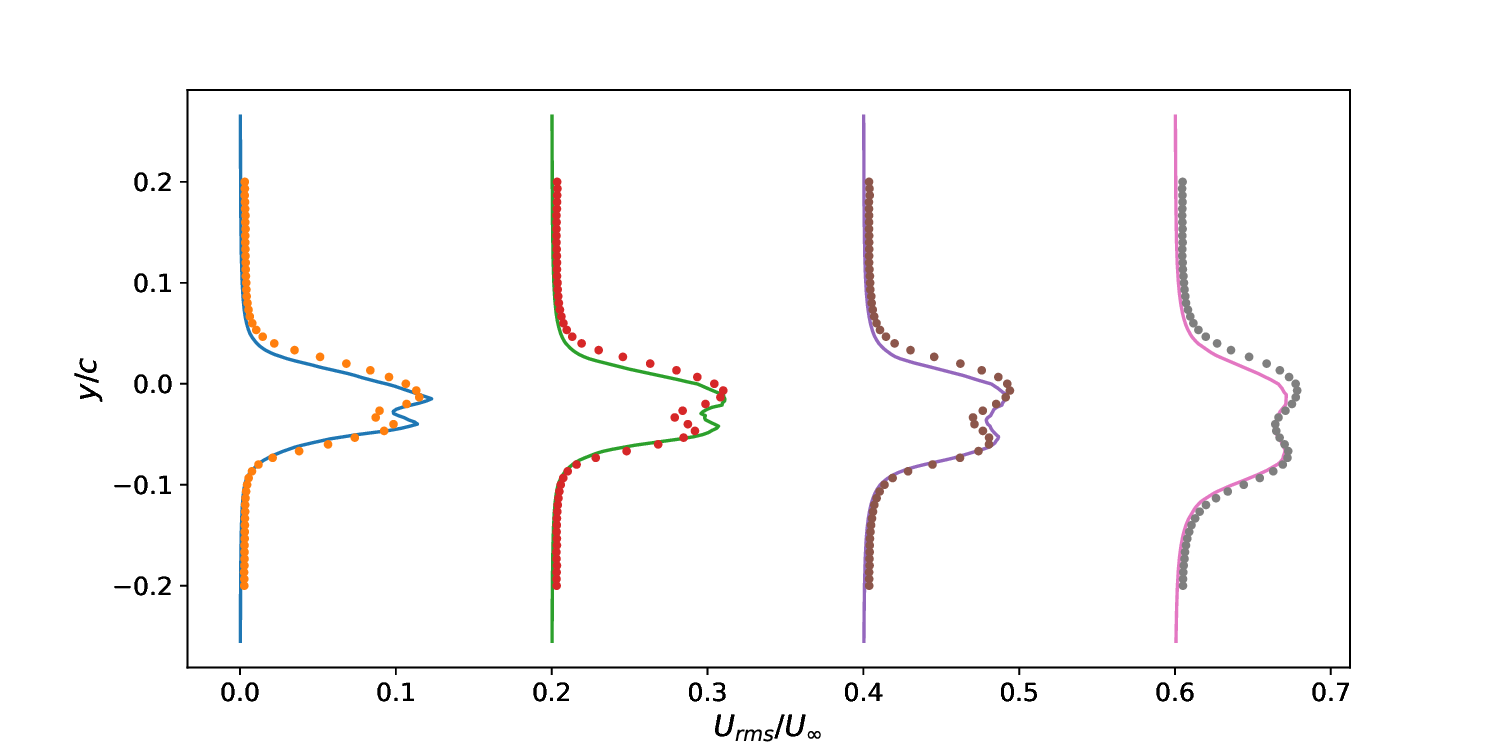}
        \caption{RMS velocity profile.}
        \label{fig:rms_u_wake}
    \end{subfigure}
    \caption{Mean and RMS velocity profiles measure at $x/c = 1.1$, $x/c = 1.2$, $x/c = 1.5$ and $x/c = 2.0$. Velocity profiles are shifted by 0.2$U_{\infty}$ for downstream measured locations.}
    \label{fig:mean_rms_wake}
\end{figure}

The simulated mean velocity and the turbulence intensity profiles in the near-wake region are compared to the experimental data in figures \ref{fig:mean_u_wake} and \ref{fig:rms_u_wake}, respectively. The experimental measurements were conducted using a single hot-wire probe, and thus the simulation data are presented as the mean total velocity. For better visualisation, the profiles are shifted by 0.2 $U_{\infty}$, respectively. Overall, the LES data demonstrate close agreement with the experiments, albeit a discrepancy is observed in capturing the wake center velocity.  This difference is attributed to the diffusion effect inherent in the LES due to its coarser resolution in this particular region. However, the wake region contributes much less to the acoustic radiation compared to the trailing edge region. Therefore, an over refined mesh is not necessary here. The RMS profiles shown in figure \ref{fig:rms_u_wake} also show a good agreement with the experimental dataset, indicating the similarity of the turbulent boundary layer state over the airfoil in both cases.

%-----------------------------------------------------------------------
\subsection{Spectral analysis of surface and farfield line array data}
\label{spec}

\begin{figure}
    \centering
    \begin{subfigure}{0.45\textwidth}
        \centering
        \includegraphics[width = \linewidth]{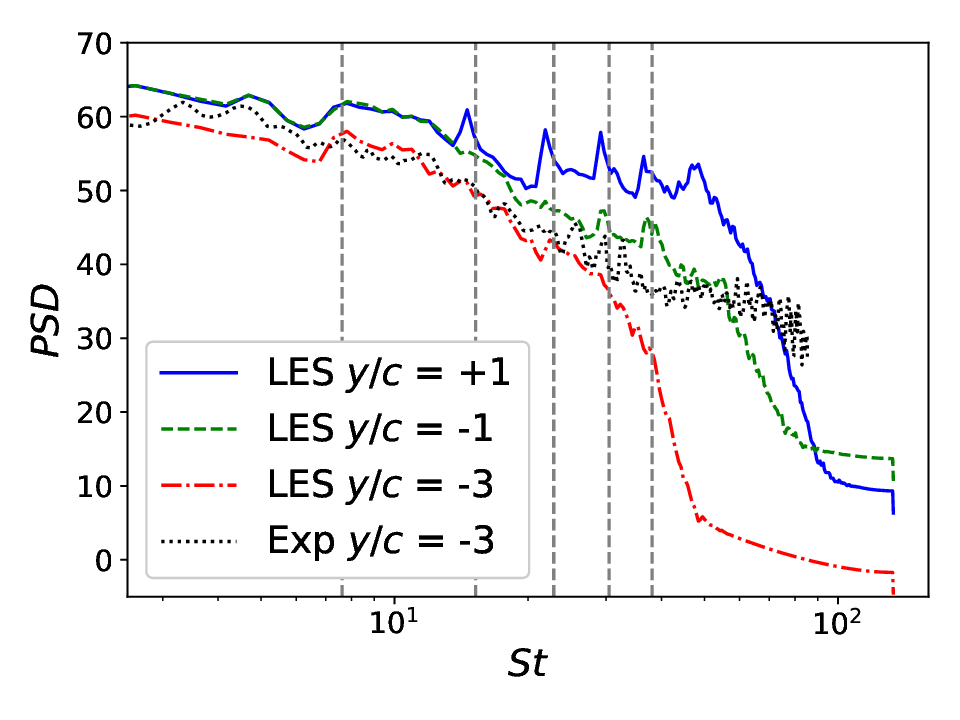}
        \caption{Acoustic spectrum}
        \label{fig:spectra_acous}
    \end{subfigure}
    \begin{subfigure}{0.45\textwidth}
        \centering
        \includegraphics[width = \linewidth]{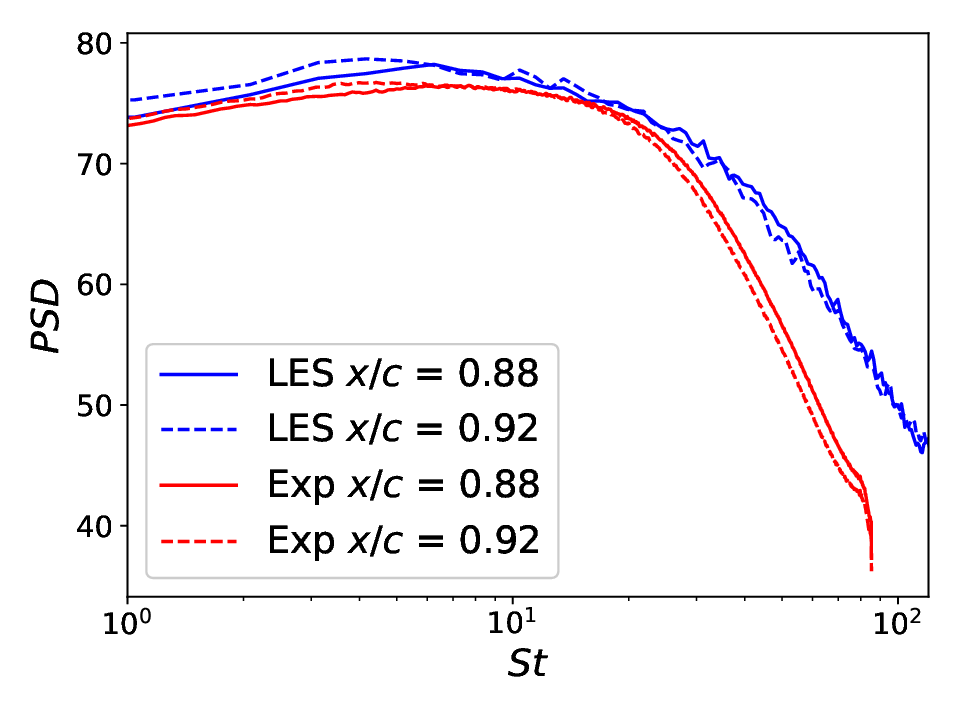}
        \caption{Hydrodynamic spectrum}
        \label{fig:spectra_hydro}
    \end{subfigure}
    \caption{Power spectra density (PSD) calculated at: (a) the acoustic line array at $x/c = 1, y/c = [-1, 1, -3]$ (LES) and $x/c = 1, y/c = -3$ (Exp). The pressure fluctuation is normalised by $M^5$; (b) the surface line array at $x/c = 0.88, 0.92$ for both LES and Exp. Hydrodynamic pressure $\rho U_{\infty}^2$ is used for normalization. The vertical dashed grey line indicate scattering condition limited by airfoil spanwise length in LES.}
    \label{fig:spectra}
\end{figure}

The power spectral density (PSD) of a single-point pressure recording at the location of the farfield line array is illustrated in figure \ref{fig:spectra_acous}. Note that, due to different values of the Mach numbers in the simulation and experiment, an appropriate scaling of the data is essential. The most commonly used acoustic scaling is the Mach number scaling \citep{williams_hall_1970, BrooksThomasF1985Soas}, i.e. $M^5$. The companion paper \citep{Demange2024arxiv} also confirms the good collapse of all cases using the $M^5$ scaling for the frequency range of $3 \le St \le 13$. The figure shows a good agreement between the numerical and experimental results at $x/c = 1$ and $y/c = -3$ up to a frequency $St = 30$, which is close to the mesh cut-off frequency.

The vertical dashed lines in figure \ref{fig:spectra_acous} indicate the cut-off radiating frequency of the successive spanwise wavenumbers $k_z$ according to the scattering condition \citep{Nogueira2017Ampf}. It expresses that propagative sound waves exist only if the spanwise wavenumber satisfies the  condition $k_z < k_0$. In this context, $k_0 = \omega/a_0$ represents the acoustic wavenumber, with [$\omega$, $a_0$] denoting the angular frequency and acoustic phase speed, respectively. As shown in the PSD, we see peaks around the scattering threshold frequency, especially at the suction side location $y/c = 1$. Peaks can also be observed at the location $y/c = -1$ at high frequencies $St = 29.75$ and $37.58$, but much weaker. 
Note that the larger span width of the experimental setup will  allow for a lager number of propagative spanwise wavenumbers to be present. However, according to the data for $y/c = -3$, this has a small impact on the spectrum. Furthermore, it can be observed that the acoustic radiation on the suction side is stronger than on the suction side.

In addition, the PSD at the  surface line array sensors is shown in  figure \ref{fig:spectra_hydro}. Hydrodynamic scaling, i.e. $0.5\rho U_{\infty}^2$, is applied to the surface pressure fluctuations. As seen there, the simulation data shows a good agreement with the experimental results at both streamwise locations on the suction side of the airfoil. The agreement shows a good collapse up to $St \approx 30$, corresponding to the broadband noise frequency regime.

\subsection{Scattering condition}
\label{scattering_condition}
The scattering condition for trailing-edge noise was shown in the work of \citet{Nogueira2017Ampf} through an investigation of the model for sound generation by a jet in the vicinity of a flat plate. This study shows that the  sound scattered by the trailing edge is only weakly dependent on the details of the jet wavepacket envelope and on the two-point coherence of the source. In fact, the salient feature for the noise generation mechanism is the wavepacket phase speed. The work further shows that only wavenumbers corresponding to  supersonic  spanwise phase velocity are propagative, while others are evanescent. Thus, the following cut-on relation can be derived:
\begin{equation}
    He_{n_z} = 2\pi M St_{n_z} = k_{n_z} = \frac{2\pi n_zc}{L_z} \text{~~,  $n_z = 0,1,2...$}
\label{eq:scatter}
\end{equation}
where $He$ is the Helmholtz number and $k_{n_z}$ represents the $n$th discrete spanwise wavenumber and $He_{n_z}$ is the corresponding sonic frequency. For a given $k_{n_z}$, if $He > He_{n_z}$ the wave will be propagative. Note that the Helmholtz number is used here because it takes into account the difference in Mach number between the experiment and the simulation to give a better alignment.
In the companion paper \citep{Demange2024arxiv}, this scattering condition is validated experimentally.

To investigate the scattering condition with numerical and experimental results, a frequency-wavenumber decomposition is conducted. The analysis is based on the cross spectral density (CSD) matrix, which is calculated from the  pressure fluctuation signals $p$:
\begin{equation}
    C_{ij} = E\{p_i(\omega),p_j(\omega)^H\}
\end{equation}
where the superscript $H$ stands for the complex conjugate operation and pressure fluctuations are presented in the frequency domain.
The CSD matrix is approximated using the Welch method \citep{WelchP.1967Tuof}, in which the signal is divided into overlapping segments and a Fourier transform is performed. For each discrete frequency, the complex-valued sound pressure values $p$ are then averaged over all N segments:
\begin{equation}
    C_{ij} = \frac{1}{N}{\sum_{k=1}^{N}p_{i,k}(\omega)p_{j,k}(\omega)^H}
\end{equation}

Next, for the frequency-wavenumber decomposition, the CSD must be  Fourier transformed  in spanwise direction. However, carrying out a spanwise Fourier transformation on the experimental data is not a simple process due to the non-periodic boundary conditions inherent to the experimental setup. 
In order to address this issue, the companion paper \citep{Demange2024arxiv} employs an eigenvalue decomposition of the CSD matrix based on the experimental dataset. The results suggest that eigenmodes are analogous to Fourier modes, and that the acoustic field can be expressed as a superposition of spanwise Fourier modes. Moreover, despite the periodic boundary conditions applied to the simulation, the same methodology can be employed to the simulation data in order to identify the optimal decomposition method in spanwise direction and to compare with the experimental results. A detailed comparison is presented in appendix  \ref{SPOD_line_array}. It shows that the spanwise boundary condition makes insignificant differences for both cases, and the spanwise Fourier decomposition is also identified as the optimal decomposition method for the simulation dataset.

Consequently, the spanwise Fourier transform can be applied to the CSD matrices for both datasets from the experimental and numerical measurements. Limited by the small span width of the airfoil in the simulation, wavenumber differences smaller than $\Delta k_z = 14.36$ cannot be distinguished for the LES data. On the other hand, the span width in the experiment is much larger and thereby contributions of lower wavenumbers can be identified. Further, a larger number of propagative wavenumbers are expected to be observed in the experiments.

\begin{figure}
    \centering
    \begin{subfigure}{0.45\textwidth}
        \centering
        \includegraphics[width = \linewidth]{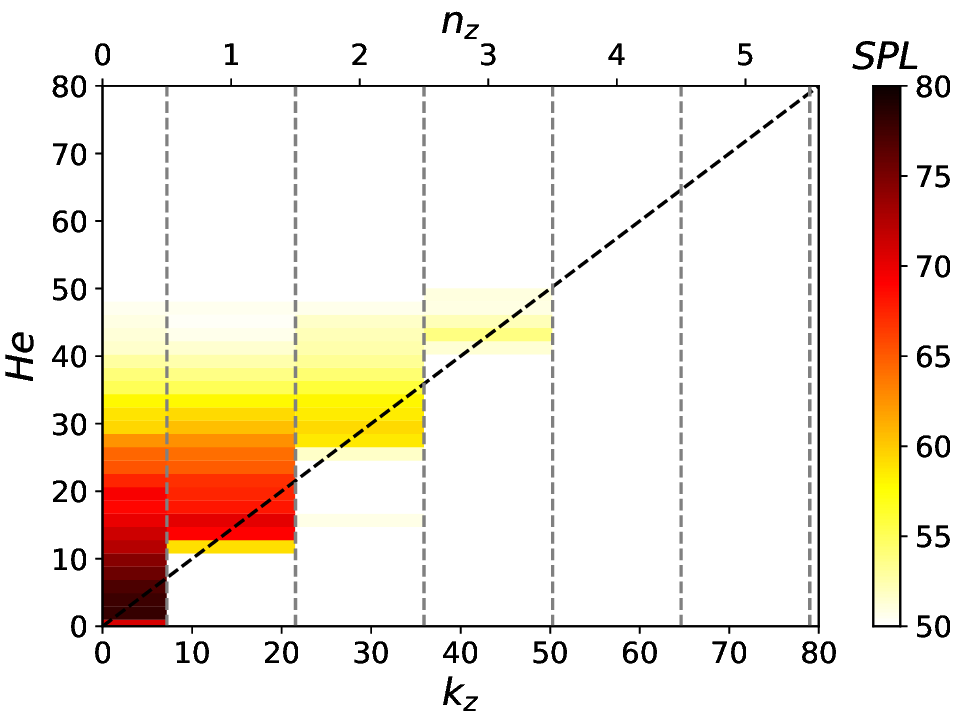}
        \caption{LES, $y/c = -3$}
    \end{subfigure}
    \begin{subfigure}{0.45\textwidth}
        \centering
        \includegraphics[width = \linewidth]{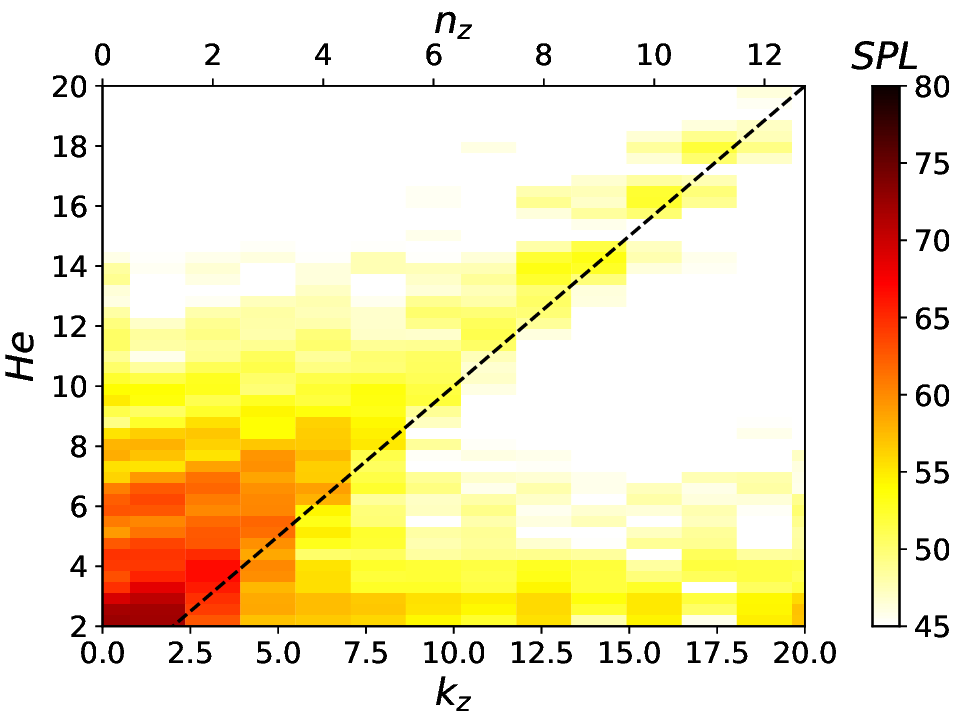}
        \caption{EXP, $y/c = -3$}
    \end{subfigure}
    \begin{subfigure}{0.45\textwidth}
        \centering
        \includegraphics[width = \linewidth]{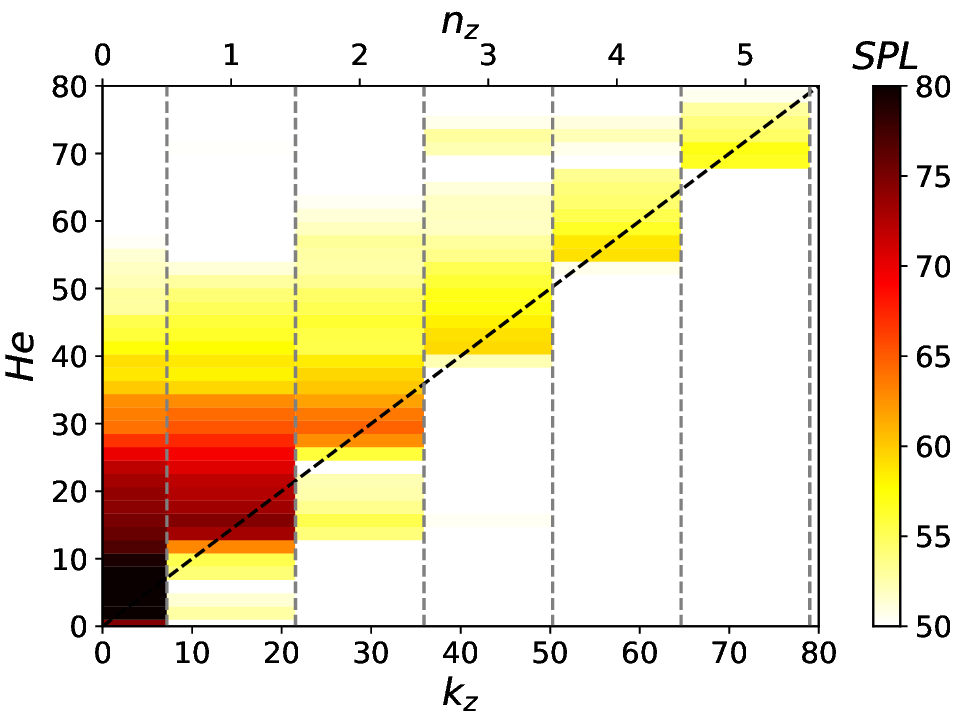}
        \caption{LES, $y/c = -1$}
    \end{subfigure}
    \begin{subfigure}{0.45\textwidth}
        \centering
        \includegraphics[width = \linewidth]{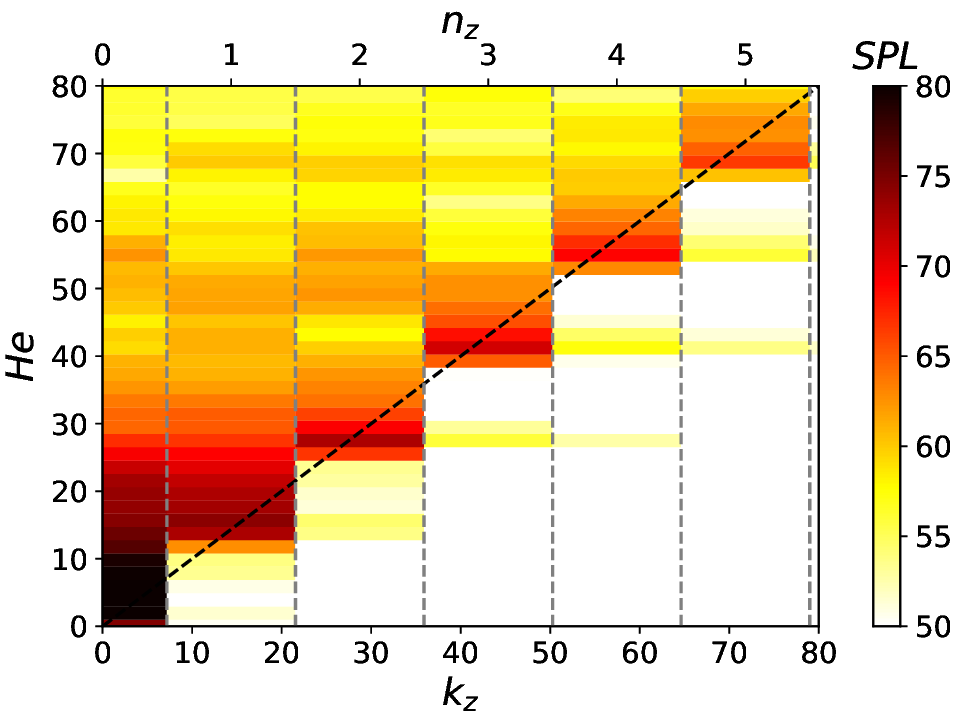}
        \caption{LES, $y/c = 1$}
    \end{subfigure}
    \caption{Frequency-wavenumber spectrum based on spanwise Fourier transform of CSD matrix (with respect to the middle sensor) of acoustic line array. Vertical dashed line indicates scattering condition for each spanwise wavenumber. The dual $x$-axis indicates spanwise wavenumber $k_z = 2\pi n/L_z$, $n_z = 0,1,2\cdots$. The black dashed line indicates acoustic wavenumber $k_0 = He$.}
    \label{fig:kz_He}
\end{figure}

Figure \ref{fig:kz_He} shows magnitude of the Fourier components of CSD matrix for data from LES and experiment in terms of sound pressure level (SPL). The theoretical $n$th propagative wavenumber $n_z$ is indicated by the vertical dashed gray line. The black dashed line indicates the acoustic wavenumber $k_0$ and, with the introduction of $He$, the slope of the scattering line is 1. Both numerical and experimental results are in good agreement with the scattering condition, indicated by the fact that the propagative wavenumbers are at the left of the scattering line ($k_{n_z} < k_0$). However, in the experimental results at low frequencies ($He < 5$), a high level of SPL is observed for wavenumbers higher than the acoustic wavenumber. As suggested in the companion paper, this discrepancy with the theory is due to the spectral leakage in the analysis related to the signal processing and the spanwise width of the domain. Note that simulation data at the $y/c = -3$ shows fewer propagative waves. This is due to the fact that the mesh cannot resolve higher frequency content. To give the reader a broader view of the physics, locations $y/c = [-1, 1]$ are also presented. 
The propagative acoustic waves have higher amplitude above the suction side of the airfoil compared to the measurement from the sensors below the pressure side. This difference is more significant for waves with $n_z \geq 2$.

\subsection{Coherence of surface pressure fluctuations}
\label{coherence}
In this section, we  focus on the coherence of the surface pressure fluctuation and its correlation with farfield acoustics. As discussed in detail in the review work by \citet{LeeSeongkyu2021Tblt}, the surface coherence length is the key input to most acoustic modelling. Thus, the primary objective of this section is to show that the coherence length in the simulation is compatible with the experimental data. Furthermore, correlations between surface pressure fluctuations and farfield acoustics are  presented and compared to experimental results.

\begin{figure}
    \centering
    \begin{subfigure}{ 0.45\textwidth}
    \centering
    \includegraphics[width =\linewidth]{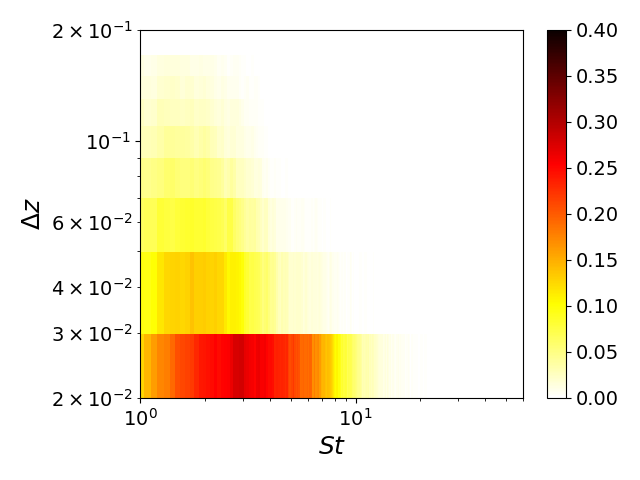}
    \caption{Exp, coherence}
    \label{fig:exp_sur_coherence}
    \end{subfigure}
    \begin{subfigure}{0.45\textwidth}
    \centering
    \includegraphics[width =\linewidth]{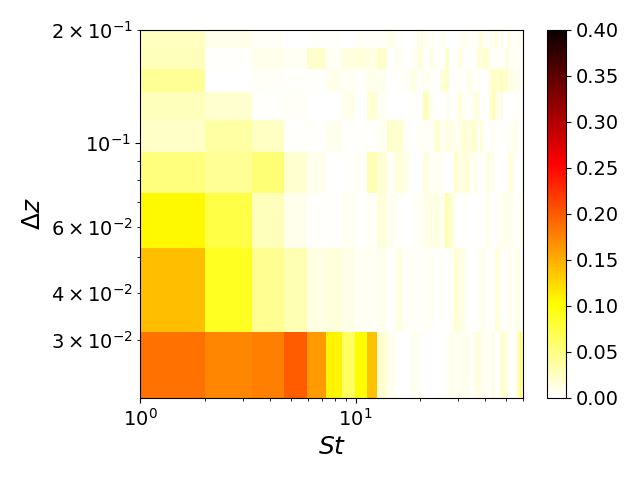}
    \caption{LES, coherence}
    \label{fig:les_sur_coherence}
    \end{subfigure}
    \begin{subfigure}{ 0.45\textwidth}
    \centering
    \includegraphics[width = \linewidth]{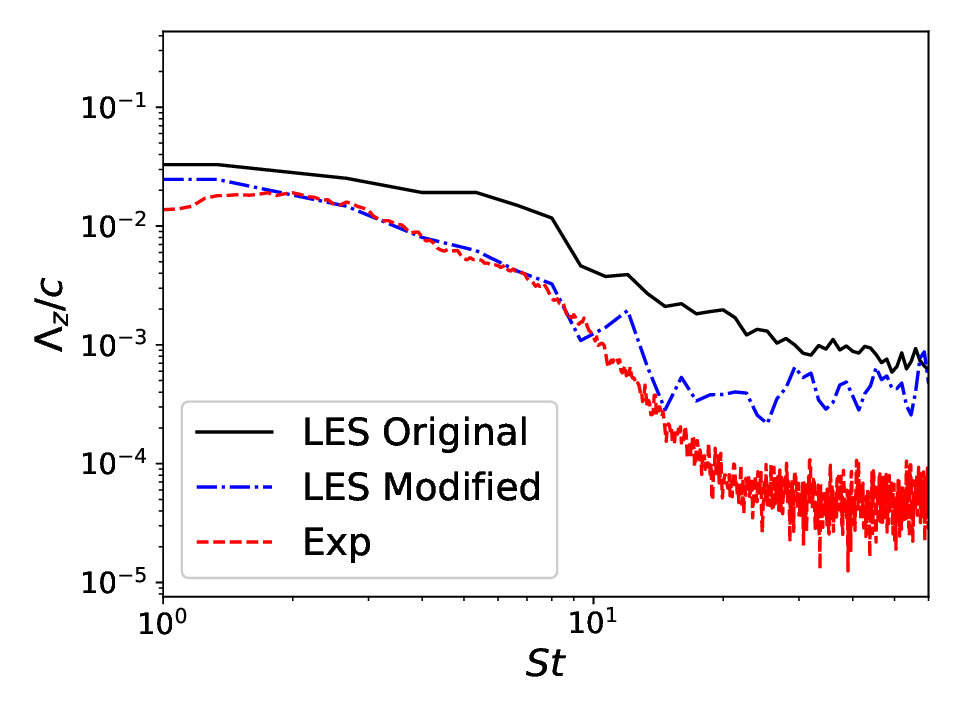}
    \caption{Coherence length}
    \label{fig:coherence_length}
    \end{subfigure}
    \caption{Coherence analysis for the surface pressure fluctuation at $x/c = 88\%$. Coherence is calculated with respect to the mid-span sensor. The coherence length is evaluated with fixed and flexible bin sizes using the Welch method, denoted as \enquote{LES original} and \enquote{LES modified} respectively.}
    \label{fig:surf_coherence}
\end{figure}

The coherence of the surface pressure fluctuations, $\gamma_{XY}$, between two adjacent surface sensors X and Y is defined as:
\begin{equation}
    \gamma_{XY}(\omega) = \frac{|C_{XY}(\omega)|}{\sqrt{P_{XX}(\omega) \times P_{YY}(\omega)}},
    \label{eq:coherence}
\end{equation}
where $P_{XX}$ and $P_{YY}$ are the power spectral density estimates at the locations X and Y, and $C_{XY}$ is the corresponding cross spectral density estimate. The distance between the sensors is $\Delta z = 0.02c$ and the diameter of each sensor is $\Delta = 0.002c$ in the experimental configuration, while the distance between each sensor is $\Delta z = 0.00024c$ in the simulation and each sensor measurement is strictly pointwise. To ensure a direct comparison between the numerical and experimental datasets, the numerical dataset is reprocessed to mimic the experimental conditions. This is done by averaging the simulation signals over an area corresponding to the physical sensors (same location and diameter)  when calculating the CSD. The results are shown in figures \ref{fig:exp_sur_coherence} and \ref{fig:les_sur_coherence}. Overall, the coherence from LES and the experimental datasets show similar values and trends up to $St = 20$. The coherence between surface sensors is concentrated at low frequencies up to $St = 10$ and decays with increasing distance to the reference sensor. 

Since the coherence from the simulations agrees qualitatively well  with experiments, we introduce the coherence length to provide a more quantitative comparison. Similar to the definition from \cite{HerrigAndreas2013BATN}, the coherence length ($\Lambda_z$) can be evaluated by integrating the spanwise coherence with a trapezoidal rule such that
\begin{equation}
    \Lambda_z(\omega) = \int_0^\infty \gamma(\omega, \Delta z) d\Delta z.
\end{equation}
Due to the periodic lateral BC, the coherence is only integrated over half the spanwise width of the numerical domain. This is not the case in experiments where the side plates impose a different boundary condition and the integration is performed over the full width of the airfoil. 
Furthermore, the numerical data have a relatively short time series, making it challenging to fully resolve the lower frequencies using the traditional Welch method with fixed bin sizes. To address this issue, we modified our approach by treating each frequency independently. 
A bin size is selected to ensure that at least four periods of the frequency inspected are contained within each bin. This approach enables the capture of more low-frequency information while mitigating the averaging effect for higher frequencies, which is inherent to the Welch method. 
This analysis is performed for the single point measurements (\enquote{LES original})  and the averaging sensors (\enquote{LES modified}) and presented in the figure \ref{fig:coherence_length}. A close agreement can be observed for the \enquote{LES modified} and experimental datasets. The integrated coherent length analysis shows that the maximum coherent length is around $\Lambda_z/c \approx 2\times10^{-2}$. Additionally, a decay is observed for frequencies $St > 10$.

\subsection{Coherence between surface pressure fluctuations and farfield acoustics}
In the preceding sections, it has been demonstrated that the numerical results  are in close agreement with the experimental measurements for both the farfield acoustics and surface pressure fluctuations. We now proceed to examine the coherence between the two. In both the experimental and simulation studies, the surface pressure and acoustics signals were simultaneously recorded. In the experimental setup, the sensor distribution in the spanwise direction was relatively coarse and non-uniform, limiting the comparison to the spanwise-averaged pressure fluctuations. In any case, focusing on the spanwise-averaged signals is most reasonable for comparisons, since for higher spanwise wavenumbers, the propagative frequency range in the experiment and simulation are different according to the scattering condition.

Due to the relatively short time series in the simulation, direct calculation of the coherence is difficult, especially for the lower frequencies. Therefore, to improve the correlation, the time signal obtained from the acoustic sensors requires a phase shift with respect to the surface sensors, as division of the time series into blocks leads to an artificial loss of coherence for short time series \citep{jaunet2017,blanco2022}. In order to find the correct time delay, here we check the phase delay between two sets of sensors and compare it with two theoretical wave propagation paths. This will also help us to obtain a better understanding of the actual sound generation mechanism.

\begin{figure}
    \centering
    \begin{subfigure}{ 0.45\textwidth}
    \centering
    \includegraphics[width =\linewidth]{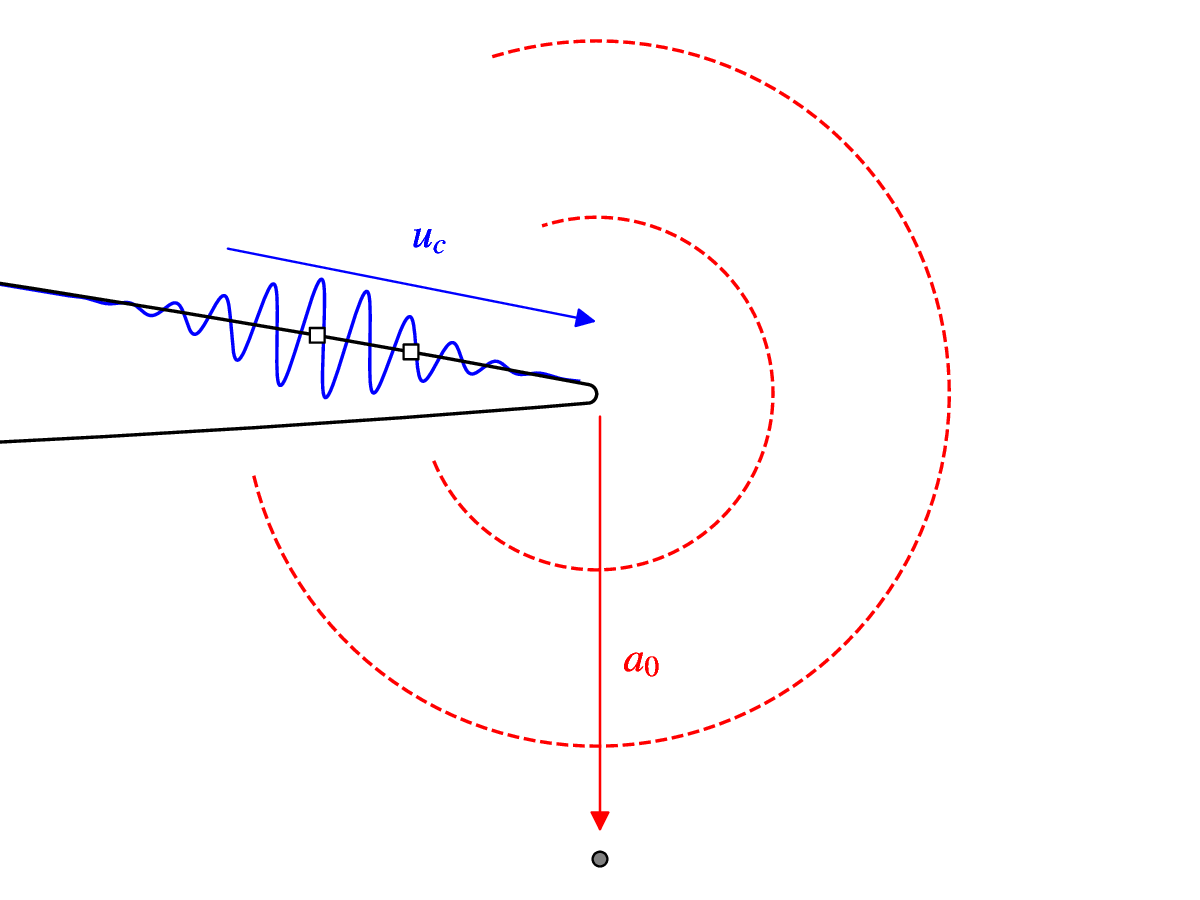}
    \caption{Convection-scattering}
    \label{fig:convectionscatter}
    \end{subfigure}
    \begin{subfigure}{ 0.45\textwidth}
    \centering
    \includegraphics[width =\linewidth]{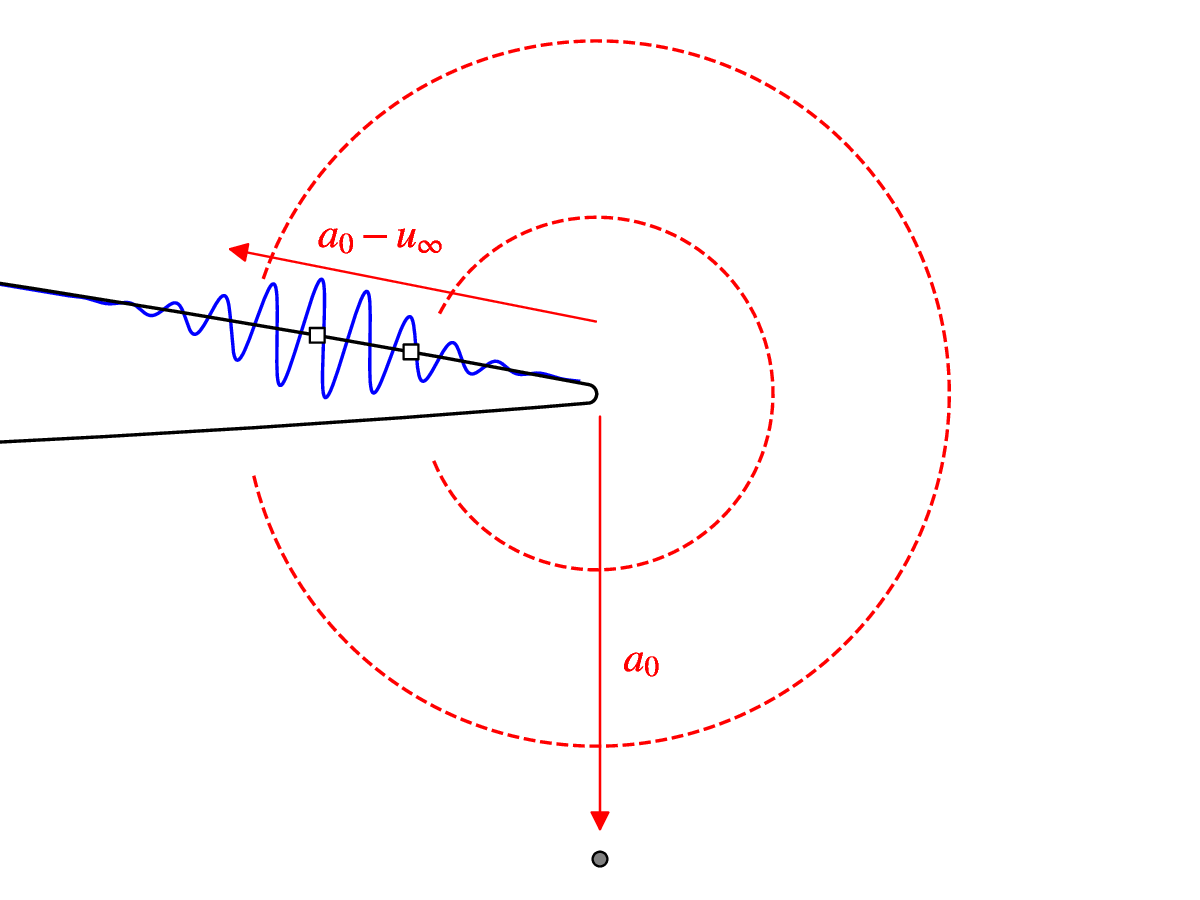}
    \caption{Back-scattering}
    \label{fig:backscatter}
    \end{subfigure}
    \begin{subfigure}{ 0.8\textwidth}
    \centering
    \includegraphics[width =\linewidth]{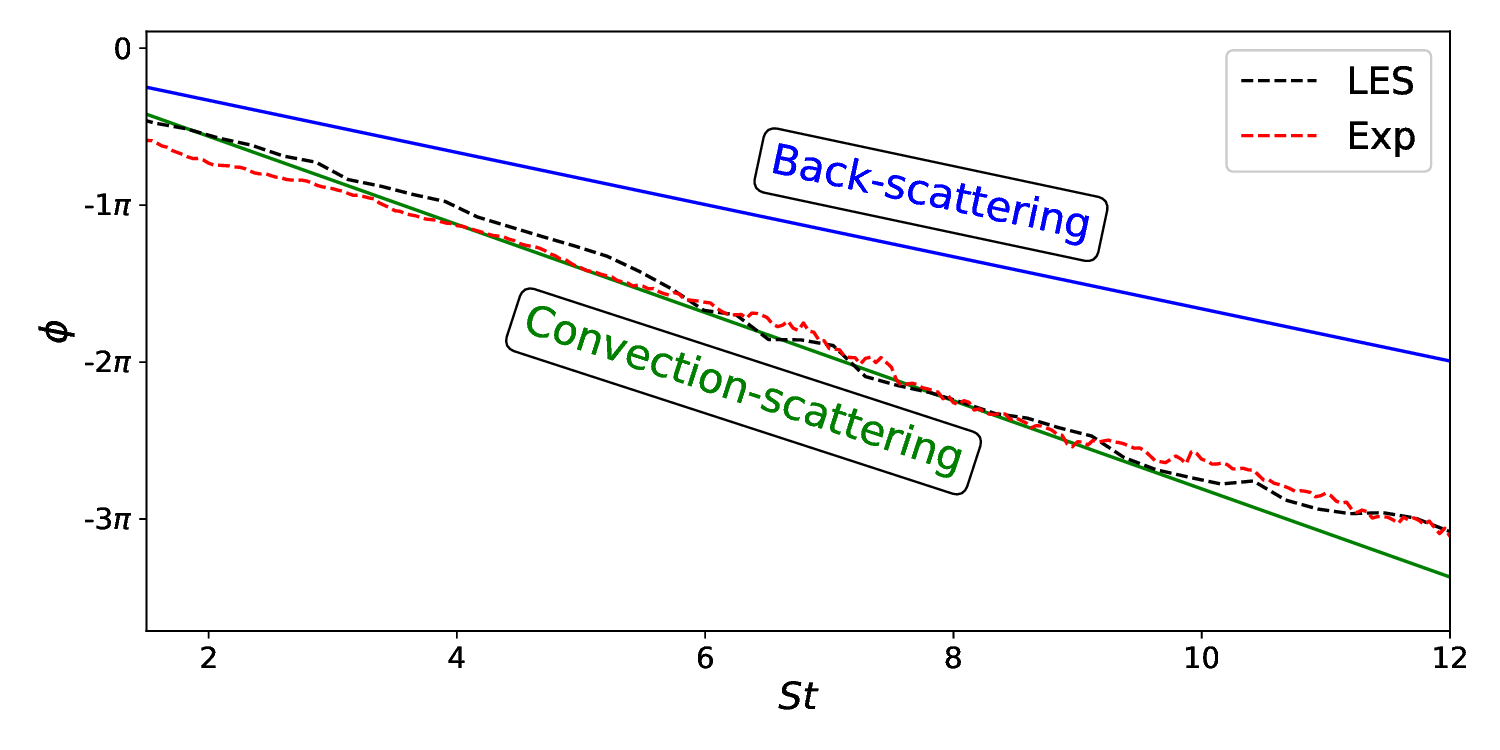}
    \caption{Phase delay}
    \label{fig:coherence_time_delay_phase}
    \end{subfigure}
    \caption{Theoretical phase delays sketches for two scenarios: (a) a hydrodynamic wave is convected from the surface sensor (\opensquare) to the trailing-edge where it scatters an acoustic waves towards the line array (\opencirc); (b) an acoustic wave generated at the trailing-edge is back-scatter to the surface sensor array and scattered towards the line array. (c) Phase delay between spanwise-averaged ($k_z = 0$) surface pressure fluctuation measured at the location $x/c = 0.92$ and acoustics at $y/c = -3$.}
    \label{fig:coherence_phase_path}
\end{figure}

Figure \ref{fig:backscatter} and \ref{fig:convectionscatter} illustrate two possible hypotheses of phenomena that can cause coherence between surface pressure fluctuations and farfield acoustics: (I) hydrodynamic waves being convected over the surface sensors and radiating acoustic waves from the  trailing edge to the farfield (here called `convective-scattering'); (II) acoustic waves generated at the trailing edge that propagate towards the acoustic sensors and the surface sensors simultaneously (here called `back-scattering'). The theoretical time delay is simply calculated as the ratio between the path length and the wave phase velocity. Note here that the hydrodynamic phase velocity is taken to be $c_{ph} = 0.6U_{\infty}$, calculated by correlating signals at different streamwise locations, in agreement with the experimental results reported in the companion paper. The vertical acoustic phase velocity is the speed of sound, and the upstream propagating acoustics is estimated by the difference between the speed of sound and the free stream velocity $U_{\infty}$.

Figure \ref{fig:coherence_time_delay_phase} shows the phase delays, presented as unwrapped phase angle for the experiment and simulations in comparison with the theoretical result. It suggests that both LES and experimental measurements follow the hydrodynamic convection and acoustic radiation route up to $St \approx 10$. 
Also notice that at higher frequencies, the data does not follow any of the two paths mentioned above. As discussed in the following analysis, coherence beyond this frequency is approaching zero and the phase angle calculation is not accurate any more.

\begin{figure}
    \centering
    \begin{subfigure}{ 0.45\textwidth}
    \centering
    \includegraphics[width = \linewidth]{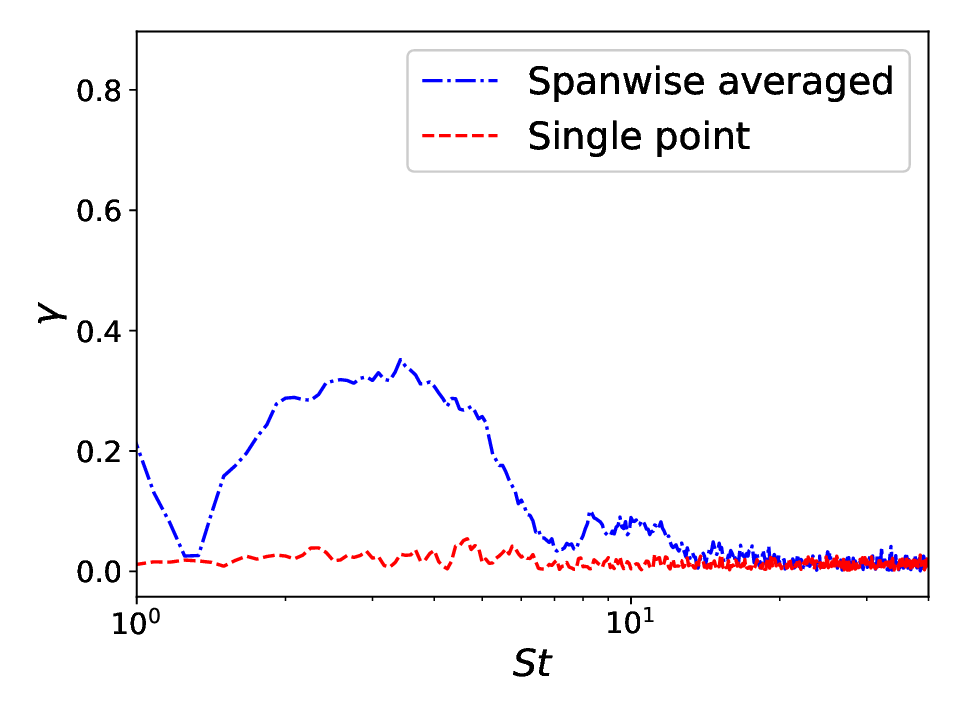}
    \caption{Exp}
    \label{fig:coherence_time_delay_phase_exp}
    \end{subfigure}
    \begin{subfigure}{ 0.45\textwidth}
    \centering
    \includegraphics[width = \linewidth]{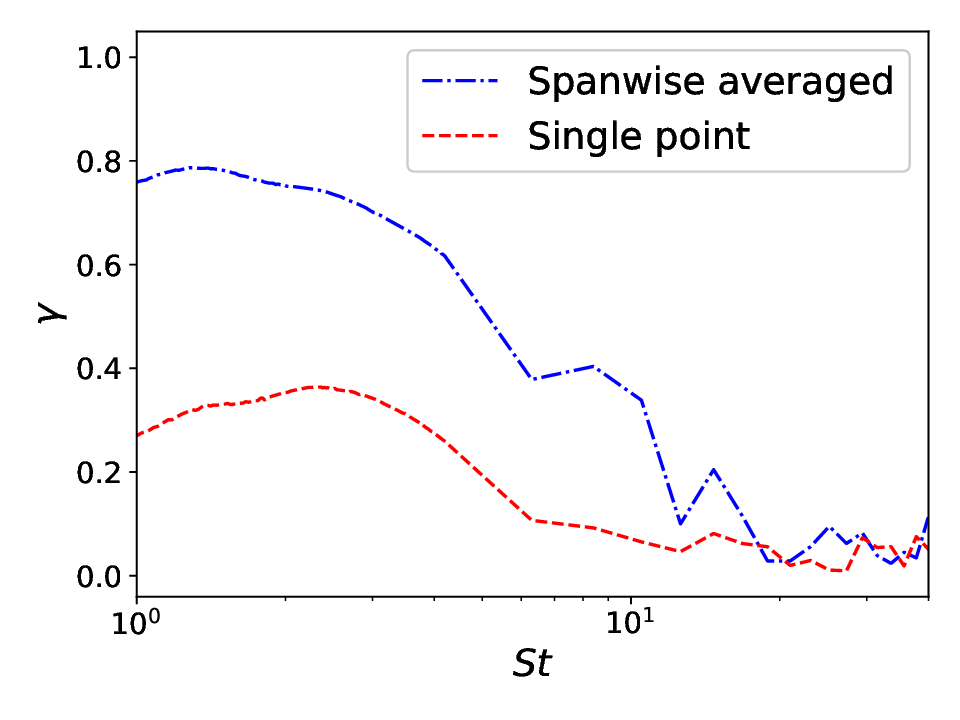}
    \caption{LES}
    \label{fig:coherence_time_delay_phase_les}
    \end{subfigure}
    \caption{Coherence between surface pressure fluctuation at $x/c = 88\%$ and acoustic line array at $x/c = 1, y/c = -3$. Both single pair of sensors and spanwise-averaged $k_z = 0$ signals are presented. Coherence is calculated with time shift.}
    \label{fig:coherence_SPF_acous}
\end{figure}

Figure \ref{fig:coherence_time_delay_phase_exp} shows the coherence between the surface and acoustic signal for the experimental dataset. Accordingly, the  coherence between a single pair of surface and acoustic line array signals remains very low. By taking the span-averaged signals, this coherence is strongly enhanced, with a maximum value of $\gamma \approx 0.4$. Apparently, by isolating the spanwise coherent  part  ($k_z = 0$) of the signals, the coherence is substantially improved within trailing-edge noise frequency range of $1 \leq St \leq 6$, emphasizing the strong role of spanwise coherent structures in trailing-edge noise generation.  This is described in more   detail in the companion paper.

Figure \ref{fig:coherence_time_delay_phase_les} shows the same analysis with the numerical dataset.
Thereby, the time delay determined earlier was taken into consideration to improve the accuracy of the coherent calculations. Similar to the experiment, the spanwise-averaged surface pressure fluctuations are obtained from the location $x/c = 0.92$ and the spanwise-averaged acoustic is measured at $y/c = -3$ below the trailing edge. Analogous to the calculation of the coherence length, flexible bin sizes are used for better resolution of low frequency content.

As shown in figure \ref{fig:coherence_time_delay_phase_les}, the same  trend is observed as in the experiment, namely that coherence is strongly enhanced when considering span-averaged signals instead of point signals. However,  when comparing the numerical and experimental results, higher coherence values can be identified from the simulation data. This is likely due to the much larger span width in the experiment, which leads to significantly more propagative acoustic waves with high spanwise wavenumbers. This leads to more complex dynamics, and thus, each wavenumber has lower energy compared to the numerical results. This leads to lower coherence values in the experiments, especially when considering a single pair of sensors. Another possible reason for the difference is the use of an uneven probe distribution in the experiment, as it is hard to properly distribute sensors in a model that at the same time has a large span and a low coherence length. These issues are absent from calculations using simulation data.  This comparison demonstrates quite clearly the importance of the spanwise domain size when considering noise generated from large spanwise structures.

Note that in this section we have limited ourselves to analysing the correlation between spanwise homogeneous structures in the hydrodynamic and acoustic domain, although current results strongly motivate us to investigate the correlation for higher wavenumbers. From an experimental point of view, performing such an analysis requires a spanwise Fourier decomposition of the surface pressure signal, which is challenging here due to the very limited sensors. On the simulation side, the correct convective time-delay and corresponding signal-shift is very important but difficult to estimate.  Consequently, performing the same analysis with higher wavenumbers is not feasible. Nonetheless, in the subsequent section, we employ a data-driven method to illustrate the relationship between hydrodynamics and acoustics at higher wavenumbers.

\section{Data-driven modelling of wavepackets driving trailing-edge noise}
\label{wavepacket}

In the previous section, a comprehensive validation was conducted between the simulation and experimental datasets. The results demonstrate a strong coherence between spanwise averaged ($k_z = 0$) surface pressure fluctuations and acoustics. 
Nevertheless, the flow structures associated with the turbulent boundary layer which generate the $k_z = 0$ surface pressure fluctuations remain unclear. Furthermore, as mentioned in the previous section, establishing a hydrodynamic-acoustic correlation for higher wavenumbers ($k_z > 0$) is very challenging from the experimental perspective. 
To address this issues, we extend our previous analyses by performing extended spectral proper orthogonal decomposition (ESPOD) analysis. This will help us to identify coherent structures with different spanwise wavenumbers.

%---------------------------------------------------------------------------------
%---------------------------------------------------------------------------------
\subsection{Spectral proper orthogonal decomposition}
Proper orthogonal decomposition (POD) is a data-driven technique that provides a set of optimal orthogonal basis from flow realizations, such as LES snapshots, to describe spatially coherent structures within the flow \citep{lumey_stochastic_1970, berkooz_proper_1993}. In its spectral form, known as SPOD, this basis is established through the eigenvalue decomposition of the cross-spectral density (CSD) matrix of the Fourier-transformed realizations at each frequency, thereby revealing the spatial-temporal coherence of these structures. 
Due to the large number of spatial points in the numerical simulations, it is not feasible to construct the CSD matrix everywhere in the domain. Following the so-called snapshot method \citep{sirovich1987, schmidt_guide_2020}, the decomposition is based on the inner product between different flow realizations $q_{\text{i}}$, given by
\begin{equation}
	\langle q_{\text{i}}, q_{\text{j}}\rangle = \int_{\Omega}q_{\text{j}}^H\mathcal{W}q_{\text{i}} d\mathbf{x}= q_{\text{j}}^H\textbf{W}q_{\text{i}}\text{,}
    \label{eq:inner_product}
\end{equation}
where $\Omega$ is the region of interest, the superscript $H$ denotes the Hermitian transpose, and the discretized weighting operator, $\textbf{W}$, is chosen such that the sum of the eigenvalues defines an energy norm. The SPOD mode shapes have a unit norm, so that the square root of each SPOD eigenvalue represents the amplitude that each mode has in the flow.

A spanwise Fourier transform of the LES snapshots is carried out prior to the SPOD. 
Although there are streaks downstream of the tripping elements, the turbulent structures near the trailing edge are homogeneous, as illustrated in the figure \ref{fig:cf}.  Consequently, the corresponding spanwise Fourier modes emerge as optimal orthogonal basis functions in this direction \citep{berkooz_proper_1993}. Due to the presence of a tripping device and a fully turbulent boundary layer developed at the trailing edge, a significant number of grid points in the spanwise direction is required to avoid spatial aliasing as suggested by \citet{KarbanUgur2022Stai}. Therefore, we interpolated the three-dimensional LES field over 1024 equidistant slices in the spanwise direction. In order to ensure that the interpolation error (spatial distance between the actual and interpolated points) remains below $10^{-8}$, we implemented a spectral interpolation method following the approach of \citet{peplinski2015}. 

As a result of performing SPOD to spanwise Fourier transformed two-dimensional field, the frequency-wavenumber space comprises four quadrants, which correspond to positive and negative wavenumbers and frequencies.
The first and the third, as well as the second and the fourth quadrants, are complex conjugate of each other. Furthermore, positive and negative spanwise wavenumbers indicate waves travelling in the positive and negative spanwise directions, respectively. 
Here, we only present the SPOD modes corresponding to the positive frequencies and wavenumbers.

%---------------------------------------------------------------------------------
%---------------------------------------------------------------------------------

\subsection{ESPOD sub-domains and configuration}
\label{spod_config}

In this work, we rely on the \enquote{extended} version of SPOD (ESPOD) to reveal the correlation between boundary layer hydrodynamics and farfield acoustics. The concept of ESPOD is derived from extended POD (EPOD) \citep{BOREEJ2003Epod}. In contrast to POD, which identifies dominant modes within a single dataset, EPOD extends the analysis to datasets with correlated dynamics and captures the relationship between them. Its spectral version, ESPOD, examines the spatial correlation under the same discrete frequency. This is done by carefully choosing the weighting applied in the equation \ref{eq:inner_product} which can be expressed as:
\begin{equation}
    \textbf{W}_E = M_{\Omega}\textbf{W}
\end{equation}
where $\textbf{W}_E$ is the weight matrix used by the ESPODs and $M_{\Omega}$ describes a mapping coefficient which sets the entries of the quadrature weight matrix $\textbf{W}$ to zero for the points outside a region of interest.
%From a computational point of view, this operation can be performed by setting the entries of the quadrature weight matrix $\mathcal{W}$ to zero for the points outside a region of interest. 
Thus, the energy calculated by the inner product \ref{eq:inner_product} will only be evaluated in the region of interest. However, SPOD modes are reconstructed over the whole domain as linear combinations of flow realisations. Details on the choice of the weighting domain in this study are presented at the end of this section.

In the present work, the flow realizations are defined as the complex Fourier modes of the flow fluctuations $\hat{\mathbf{q}}(\mathbf{x},t) = \left[\hat{\rho},\ \hat{u},\ \hat{v},\ (\hat{w}),\ \hat{T}\right]^{\text{T}}$ around the stationary span-time-averaged flow field $\overline{\mathbf{q}}(\mathbf{x})=\left[\overline{\rho},\ \overline{u},\ \overline{v},\ (\overline{w}),\ \overline{T}\right]^{\text{T}}$.
The norm used to define the optimal orthogonal basis is the compressible energy \citep{ChuBoa-Teh1965Otet,Mack1984,HanifiArdeshir1996Tgic} imposed through the $\mathcal{W}$ operator \eqref{eq:inner_product}, defined as a positive definite weighting tensor: 
%\begin{equation}
%	\textbf{W} =\int_{\Omega}\begin{bmatrix}
%    \frac{\overline{T}}{\gamma\overline{\rho}M^2} &  & & &\\
%    & \overline{\rho} &  & &\\
%    & &\overline{\rho}   & &\\
%    & &  &\overline{\rho}  &\\
%    & &  &  & \frac{\overline{\rho}}{\gamma(\gamma-1)\overline{T}M^2}
%    \end{bmatrix}d\mathbf{x} \text{.}
%    \label{eq:energy_weight_SPOD}
%\end{equation}
\begin{equation}
	\mathcal{W} =\text{diag}\left(
    \frac{\overline{T}}{\gamma\overline{\rho}M^2},\ 
    \overline{\rho},\ 
    \overline{\rho},\ 
    \overline{\rho},\ 
    \frac{\overline{\rho}}{\gamma(\gamma-1)\overline{T}M^2}\right)
   \text{.}
    \label{eq:energy_weight_SPOD}
\end{equation}
where $\gamma$ is the specific heat ratio and the $M$ is the Mach number.

\begin{figure}
    \centering
        \begin{subfigure}{0.49\textwidth}
        \includegraphics[width = \linewidth]{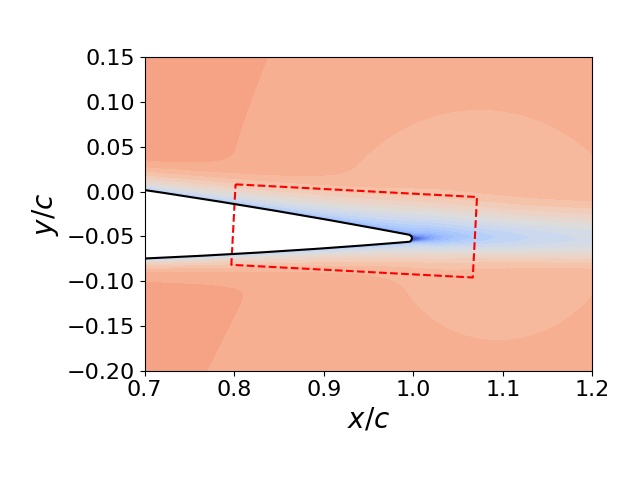}
        \caption{Weighting region for hydrodynamic SPOD}
        \label{fig:weight_TE}
    \end{subfigure}
    \begin{subfigure}{0.49\textwidth}
        \includegraphics[width = \linewidth]{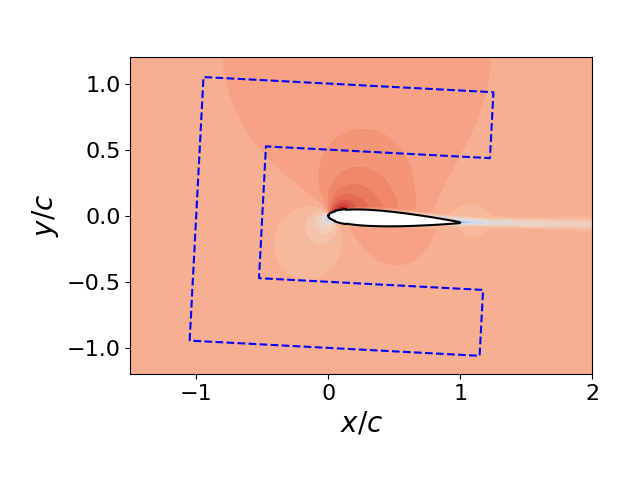}
        \caption{Weighting region for acoustic SPOD}
        \label{fig:weight_acous}
    \end{subfigure}
    \caption{The choice of the subdomain for SPOD weighting matrix $\mathcal{W}$. (a) Red dotted line indicates the domain is restricted near the trailing edge, while (b) the blue dotted line indicates the weighting region in acoustics. The contour plot illustrates the streamwise mean flow $\Bar{u}$.}
    \label{fig:weight_region}
\end{figure}

We consider two sub-domains to conduct the ESPOD analysis in this work, illustrated in figure \ref{fig:weight_region}. Their definition is intended to facilitate the extraction of specific physical phenomena whose contribution dominates the production of compressible energy in different flow regions.

The first sub-domain contains mainly the turbulent boundary layer and wake around the trailing edge, illustrated by the red box in the figure \ref{fig:weight_TE}, where the contribution of (turbulent) hydrodynamics is significantly greater than that of acoustics. Consequently, the basis of SPOD modes obtained from fluctuations in this region will be optimal in terms of turbulent kinetic energy, similar to \citet{Sano2019, Abreu2021}. 
The ESPOD based on this region, referred to as \emph{hydrodynamic ESPOD} (abbreviated to H-SPOD), identifies the hydrodynamic structures that correlate with the farfield acoustics..

The second sub-domain is located in the freestream, illustrated by the blue box in the figure \ref{fig:weight_acous}, where the contribution of acoustics dominates in the energy norm.  This would result in the optimal basis in terms of the acoustic energy, akin to sound power $\Tilde{p}^2$, thereby providing comprehensive validation of the correlation between hydrodynamics and acoustics.
The ESPOD based on this region will be termed \emph{acoustic ESPOD} in the remaining (abbreviated to A-SPOD). In both H/A-SPOD, the mode shapes projected over the whole domain are limited to a box region  extended at least three chords away from the airfoil surface to further reduce the computational cost.

The analysis is performed with the numerical code from \citet{rogowski2023unlocking}, using $N_{\text{FFT}}=512$ frequency bins per block with an overlap of $75\%$, resulting in $N_{\text{blocks}}=57$ blocks. 
The two-sided system is considered since the dataset is complex-valued (see previous sections), which results in the frequency resolution $\Delta He = 0.97$.
A Hanning window is applied to the blocks of flow realizations for the Welch method \citep{WelchP.1967Tuof} in all the analyses considered here.
A validation of the convergence of SPOD is presented in the appendix \ref{SPOD_convergence}.

%---------------------------------------------------------------------------------
%---------------------------------------------------------------------------------

\subsection{Wavepackets in the turbulent boundary layer identified with H-SPOD}
\label{HSPOD}
In this section, the energy spectrum and mode shapes from the H-SPOD analysis are presented to show the dominant hydrodynamic structures and their contribution to the radiating acoustics. Recall that for the H-SPOD, the norm is dominated by the turbulent kinetic energy. When considering the cases $k_z \geq 0$, only $n_z = [0,1,2,3]$ are presented, since the associated propagation frequencies corresponding to these wavenumbers cover up to $He = 57.45$ for the current domain width. This frequency range includes the trailing-edge noise frequencies shown in the figure \ref{fig:spectra_acous}.

\subsubsection{Energy spectrum of H-SPOD}
\begin{figure}
    \centering
    %\includesvg[width = \linewidth]{figs_JFM_wavepackets/SPOD_modes/mode_energy_H.svg}
    \includegraphics[width = \linewidth]{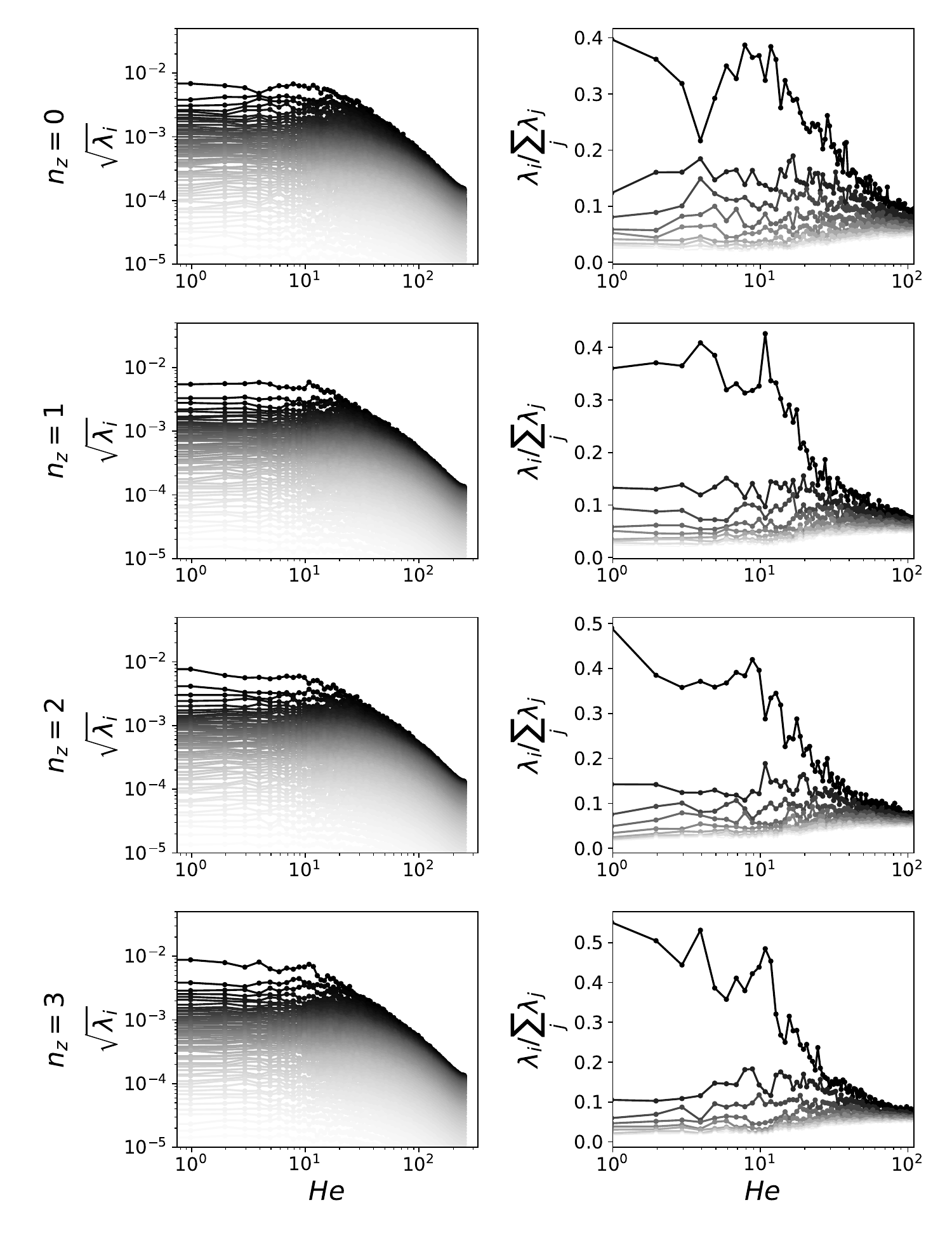}
    \caption{SPOD eigenvalues corresponding to compressible energy for the first five leading spanwise wavenumbers ($n_z = 0,1,2,3$, from top to bottom). Left: the square root of SPOD eigenvalues, right: the energy ratio of each SPOD mode to the summation of all modes.}
    \label{fig:SPOD_energy_uv}
\end{figure}

This section presents the energy spectrum of the H-SPOD, which depicts the distribution of  energy among the coherent structures for each discrete frequencies. Figure \ref{fig:SPOD_energy_uv} (left column)  illustrates the H-SPOD spectrum for the first four wavenumbers. In general, the spectrum exhibits a similar shape to that observed in the surface pressure spectrum illustrated in  figure  \ref{fig:spectra_hydro}.  At all wavenumbers the spectrum shows a plateau up to $He = 10$ and then decays rapidly at higher frequencies. The rapid decay within $He > 10$ suggests that the temporal aliasing of the energy near the trailing edge is insignificant.
However, because of the relatively shorter time series obtained for the SPOD analyses, it is difficult to achieve a good convergence in the very low-frequency regime in comparison with the surface pressure spectrum, which demonstrates a gradual decrease in energy.

Additionally, the ratio of H-SPOD mode energy to total energy ($\lambda_i/\sum_j \lambda_j$) is also presented in figure \ref{fig:SPOD_energy_uv} (right column) to check the contribution of each H-SPOD mode. Within the low-frequency regime ($1 \le He \le 20$), the leading H-SPOD mode constitutes a significant portion of the total energy.
In the case of $n_z = 0$, the leading H-SPOD mode accounts for approximately 40\% of the total energy, while the sub-leading H-SPOD modes account for less than 20\% of the total compressible energy. As the wavenumber increases, the leading H-SPOD mode gains a greater proportion of energy. In the case of $n_z = 3$, the leading H-SPOD mode takes up to 54\% of the total energy. Such proportions in the energy distribution indicates that the trailing edge hydrodynamics are mostly low rank for frequencies up to $He = 20$.
In the following analyses, the leading H-SPOD mode is employed for the presentation of mode shapes. In the higher frequency regime, the contributions of the leading and sub-leading modes become less distinguishable, and thus we restrict our analyses within the low frequency regime.

%---------------------------------------------------------------------------------
%---------------------------------------------------------------------------------

\subsubsection{Mode shapes from H-SPOD}

\begin{figure}
    \centering
    \begin{subfigure}{0.49\textwidth}
    \centering
    \includegraphics[width = \linewidth]{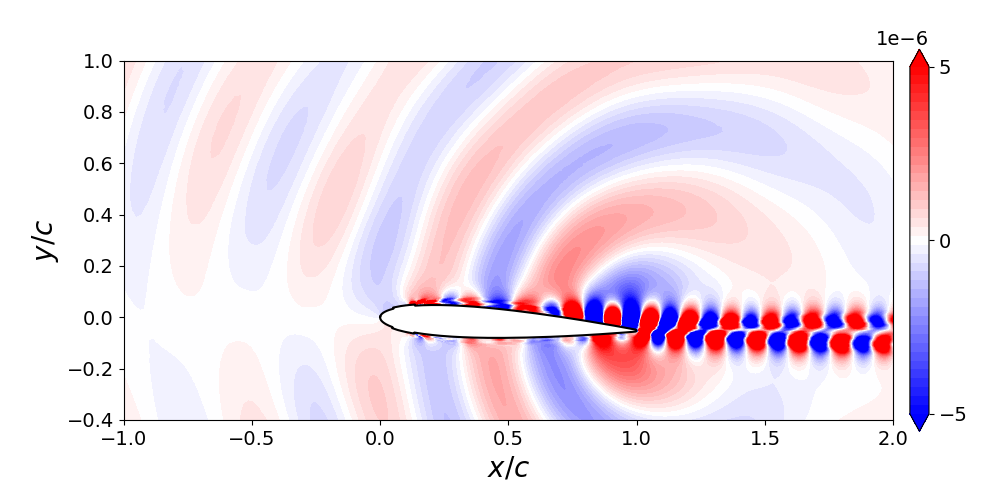}
    \caption{$n_z = 0, ~ \Tilde{p}, ~ He = 9.82$}
    \end{subfigure}
    \begin{subfigure}{0.49\textwidth}
    \centering
    \includegraphics[width = \linewidth]{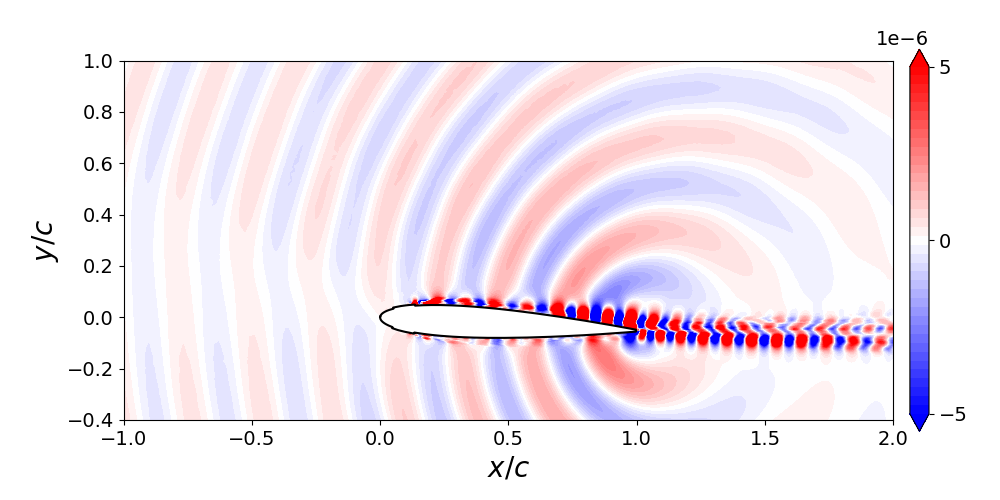}
    \caption{$n_z = 0, ~ \Tilde{p}, ~ He = 16.69$}
    \label{fig:SPOD_p_16.69_0}
    \end{subfigure}
    \begin{subfigure}{0.49\textwidth}
    \centering
    \includegraphics[width = \linewidth]{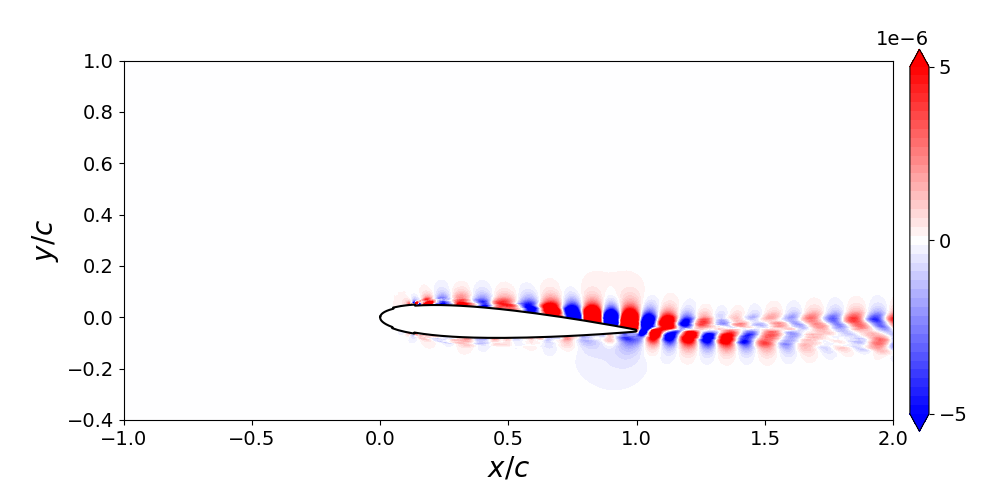}
    \caption{$n_z = 1, ~ \Tilde{p}, ~ He = 9.82$}
    \end{subfigure}
    \begin{subfigure}{0.49\textwidth}
    \centering
    \includegraphics[width = \linewidth]{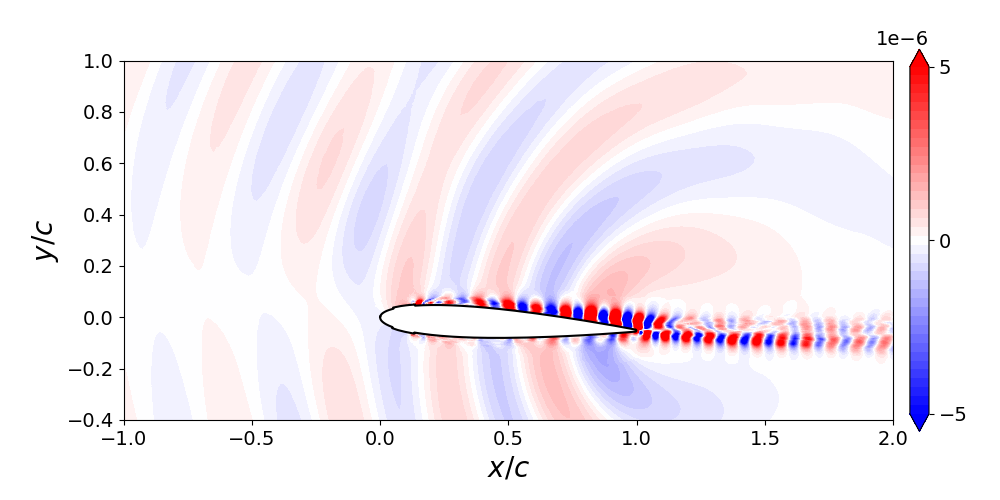}
    \caption{$n_z = 1, ~ \Tilde{p}, ~ He = 16.69$}
    \end{subfigure}
    \begin{subfigure}{0.49\textwidth}
    \centering
    \includegraphics[width = \linewidth]{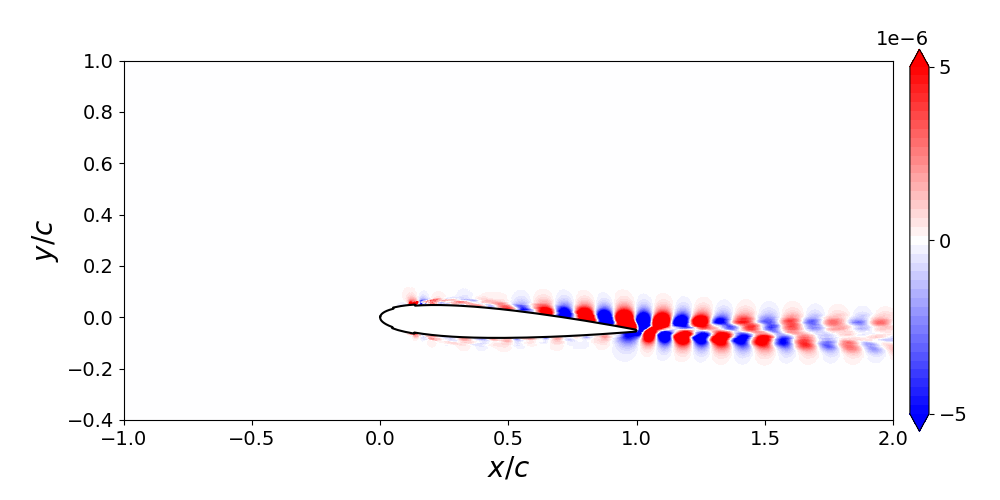}
    \caption{$n_z = 2, ~ \Tilde{p}, ~ He = 9.82$}
    \end{subfigure}
    \begin{subfigure}{0.49\textwidth}
    \centering
    \includegraphics[width = \linewidth]{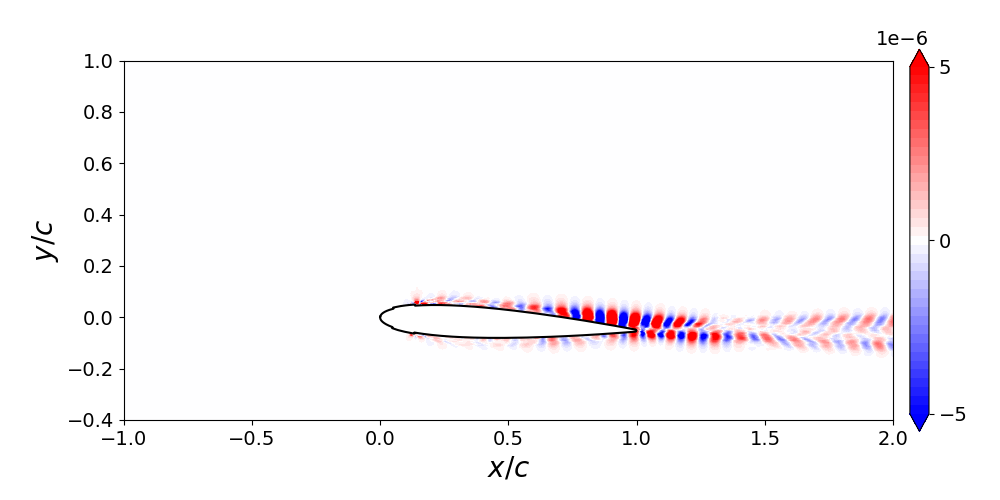}
    \caption{$n_z = 2, ~ \Tilde{p}, ~ He = 16.69$}
    \label{fig:SPOD_p_16.69_2}
    \end{subfigure}
    \caption{The \textit{leading} H-SPOD mode shape $\Tilde{p}$ for the first three spanwise wavenumbers ($n_z = 0, 1, 2$) at the frequency $He = 9.82$ (a, c, e) and $He = 16.69$ (b, d, f), respectively. The $\Tilde{p}$ is premultiplied by the square root of its eigenvalue to give the real amplitude.}
    \label{fig:leading_SPOD_mode_kz012_p_He_9.82_16.69}
\end{figure}

% Subsequently, the pressure fluctuations are presented for analysis of the most energetic hydrodynamic structures and associated acoustic radiation.
Figure \ref{fig:leading_SPOD_mode_kz012_p_He_9.82_16.69} illustrates the pressure distribution of the leading H-SPOD mode for the three leading wavenumbers, $n_z = 0, 1, 2$, at $He = 9.82, 16.69$.  
For the wavenumbers and frequencies considered here, the mode shapes reveal a wavepacket extending from the trip on both the suction and pressure sides to the wake of the airfoil. As a consequence of the adverse pressure gradient, the fluctuations on the suction side have a significantly larger amplitude than those on the pressure side.
In the case of $n_z = 0$, strong radiated acoustic waves can be identified for both frequencies presented here. The origin of the acoustics is located at the trailing edge. 

The SPOD mode shapes further show that acoustic waves are evanescent for the case of  $n_z = 1$ and $He = 9.82 < k_0$, but are propagative for $He = 16.69 > k_0$, which is in agreement with the scattering condition (and as observed in \S \ref{scattering_condition}). For $n_z = 2$ the acoustic waves are evanescent for both frequencies. Note that the amplitudes of evanescent waves decrease exponentially and are not visible for the used colour-scale.
The presence of the propagative acoustic waves in the H-SPOD mode shapes further indicates that there is a strong correlation between the surface pressure fluctuations and the farfield acoustics, as will be quantified by the ESPOD.

These present observations are consistent with previous analyses of broadband trailing-edge noise \citep{Sano2019, Abreu2021}. These studies find SPOD modes with $n_z = 0$ to be most energetic with  wavepacket-like structures extending from the position where the boundary layer is tripped to the near wake. Likewise the present study, strong acoustic radiation from the trailing edge was identified.  However, due to the limited span in the previous studies, the authors limited their analyses to $n_z = 0$, only. In the present work we show that the wavepacket structures exist also for higher wavenumbers with the acoustic radiation following the scattering condition.

\begin{figure}
    \centering
    \begin{subfigure}{0.49\textwidth}
    \centering
    \frame{\includegraphics[width = 0.9\linewidth]{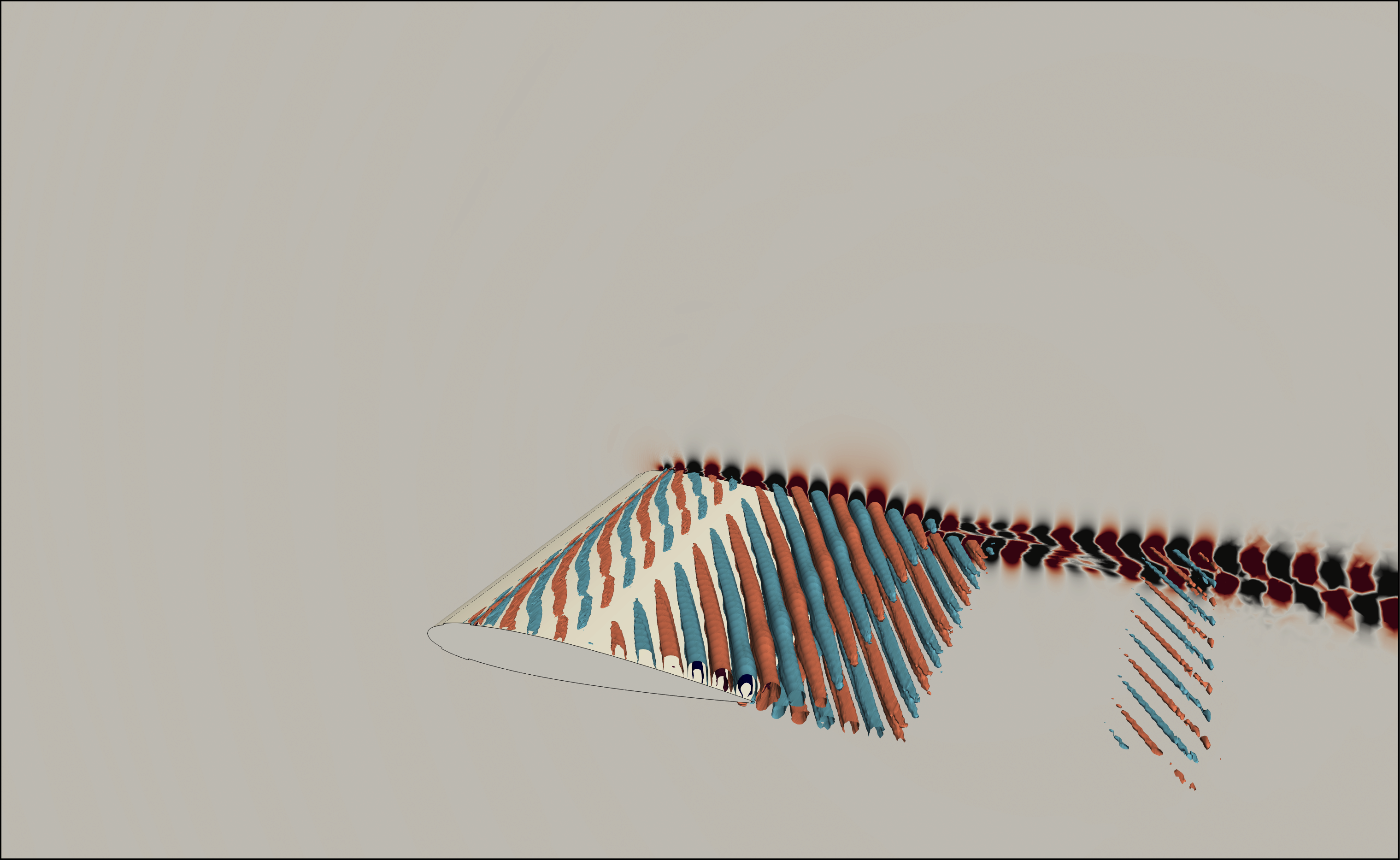}}
    \caption{$n_z = 1, ~ He = 9.82$}
    \end{subfigure}
    \begin{subfigure}{0.49\textwidth}
    \centering
    \frame{\includegraphics[width = 0.9\linewidth]{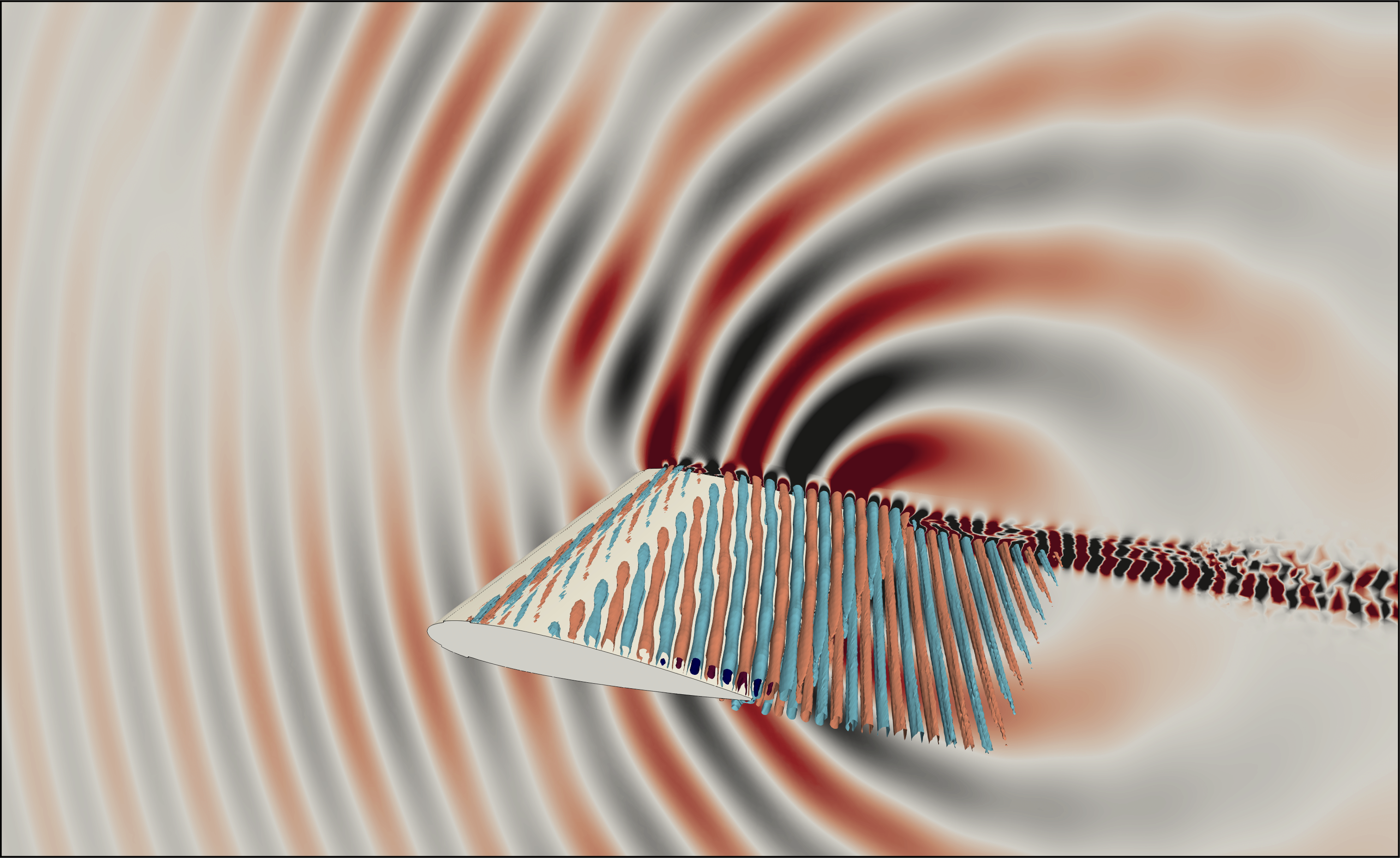}}
    \caption{$n_z = 1, ~ He = 16.69$}
    \end{subfigure}
    \caption{Pressure iso-surface of three dimensional SPOD mode shapes  $\Tilde{p}$ for evanescent (a) and propagative (b) conditions. The colour scale for the acoustic pressure fluctuation is saturated in the range $\left[-5\times10^{-6},5\times10^{-6}\right]$ from black to red, and the hydrodynamic pressure fluctuation is in the range $\left[-0.003,0.003\right]$ from blue to orange.}
    \label{fig:3D_SPOD_modes}
\end{figure}

It should be noted that due to the non-zero wavenumbers, the SPOD modes are three-dimensional, whereby the wavepackets and scattered acoustic waves are no longer parallel to the trailing edge.  To enhance our understanding of these dynamics, the three-dimensional pressure field of the SPOD modes is  reconstructed, following:
\begin{equation}
    \tilde{p}_{3D}(\omega) = \tilde{p}_{2D}(\omega)e^{ik_zz}
\end{equation}
where $\omega$ is the angular frequency of the H-SPOD modes. The results associated with $n_z = 1$ at the evanescent and propagative frequencies are visualized in the figure \ref{fig:3D_SPOD_modes}. In both cases, the hydrodynamic wavepackets  are oblique with respect to the trailing edge. The amplitude of the wavepacket structures decay shortly downstream of the tripping element and then grow again close to the trailing edge. This may indicate that the tripping device might also contribute to the noise generation.
To better visualize the three-dimensional spatio-temporal shapes of the acoustic and hydrodynamic fluctuations, animations corresponding to the three-dimensional reconstructions of the SPOD modes for more wavenumbers are provided as supplementary material.

%---------------------------------------------------------------------------------
%---------------------------------------------------------------------------------

\subsubsection{Low-order reconstruction of the far-field acoustics}
\begin{figure}
    \centering
    \begin{subfigure}{0.39\textwidth}
    \centering
    \includegraphics[width = \linewidth]{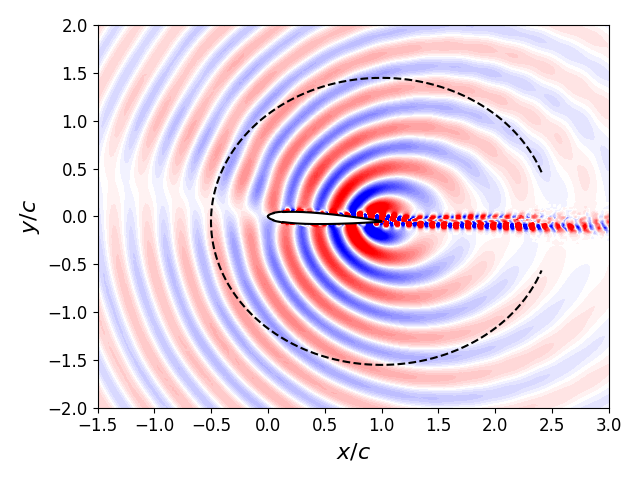}
    \caption{}
    \label{subfig:acous_region}
    \end{subfigure}
    \begin{subfigure}{0.6\textwidth}
    \centering
    \includegraphics[width = \linewidth]{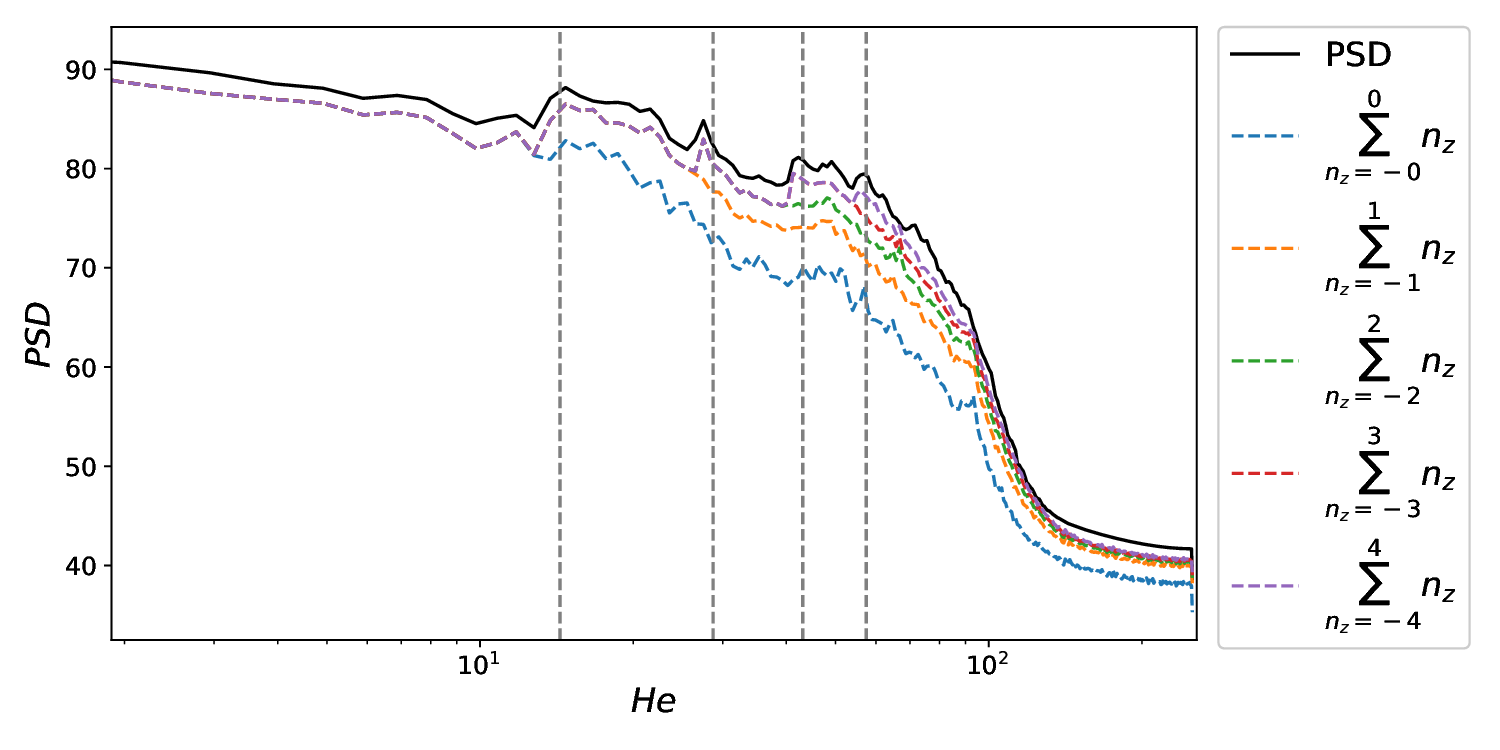}
    \caption{}
    \label{fig:psd_spod}
    \end{subfigure}
    \caption{(a) Region of the acoustic field (dashed circle arc) used to integrate the power spectra density representation of the pressure. The wake is avoided from the integration patch. (b) The contribution of the accumulated spanwise wavenumbers $n_z$ to the radiated pressure. PSD calculated directly from the three-dimensional  field. Both are integrated along the arc-circle as in (a). Vertical dashed lines indicate the  scattering condition \eqref{eq:scatter}.}
    \label{fig:psd_spod_and_int_region}
\end{figure}

We have shown that wavepackets in the turbulent boundary layer generate the trailing-edge noise, as long as their wavenumbers fulfill the scattering condition. However, it is still unclear whether the acoustic field represented by the leading H-SPOD modes can offer a good approximation of the LES or experimental data. 
Thus, we compare the integrated acoustic spectrum obtained from the H-SPOD modes with the direct measurement from the simulation. The integration is carried out along an arc-circle in the freestream, as presented with the dashed circle arc in the figure \ref{subfig:acous_region}. The radius of the arc is 1.5 chords length, with the origin located at the trailing edge. Integration over the wake is avoided.
Furthermore, to highlight the contribution of each wavenumber, the acoustics are reconstructed by accumulating the leading twenty SPOD modes from the first five wavenumbers. The results are presented in the figure \ref{fig:psd_spod}. Note that the large number of H-SPOD modes chosen here is due to the slow convergence of acoustic energy in H-SPOD. This is discussed in more detail in \S \ref{A_H_SPOD_comapre}. 

Figure \ref{fig:psd_spod} illustrates the dominant role played by each wavenumber, $n_z$, within its corresponding frequency range.
In the frequency range $He \le 11.78$, only the leading wavenumber contributes to the acoustic energy. In the higher frequency range, $11.78 \le He \le 25.52$, the second wavenumber contributes a larger proportion of the acoustic radiation, with the acoustic energy contained in the second wavenumber representing approximately 60\% of the total acoustic energy.  As we move to higher frequencies, more wavenumbers begin to contribute to the acoustic radiation, which links back to the scattering condition. Therefore, the figure shows the theoretical scattering condition calculated from  equation \eqref{eq:scatter} for the five wavenumbers, indicated by the vertical dashed gray line. 
It can be seen that the scattering conditions (cut-off frequencies) obtained from the LES dataset are  somewhat lower than the theoretical values. This is likely due to the  negligence of the mean flow velocity profile, in the derivations of the scattering conditions. However, the difference is small, and we still consider the prediction from equation \eqref{eq:scatter} as reasonable.

%---------------------------------------------------------------------------------
%---------------------------------------------------------------------------------
%---------------------------------------------------------------------------------
%---------------------------------------------------------------------------------

\subsection{Trailing-edge noise reduced-order model based on A-SPOD modes}
\label{ASPOD}
The H-SPOD analysis indicates that the surface hydrodynamic wavepackets are a source of the acoustic radiation. However, twenty leading SPOD modes are required to reconstruct the farfield acoustics with maximum 2 dB error. Such slow convergence suggest that H-SPOD may not be optimal for the reconstruction of the acoustics. It should be recalled that H-SPOD considers the optimal SPOD basis with respect to the compressible energy inside the turbulent boundary layer rather than acoustics.
Thus, the objective of this section is to provide a more efficient (low-ranked) reduced-order model for the acoustic prediction and to determine the relevance of hydrodynamic energy in the context of acoustic radiation.

%---------------------------------------------------------------------------------
%---------------------------------------------------------------------------------

\subsubsection{A-SPOD Energy spectrum}
In order to identify an optimal SPOD basis for the acoustic field, it is beneficial to apply the ESPOD based on the compressible energy measured solely within the acoustic farfield. As the pressure fluctuations in this region are much more significant than the hydrodynamic components, the compressible energy presented here is equivalent to acoustic energy ($\tilde{p}^2$). And the same compressible norm from H-SPOD is employed in A-SPOD computations.  

\begin{figure}
    \centering
    \begin{subfigure}{0.49\textwidth}
    %\includesvg[width = \linewidth]{figs_JFM_wavepackets/SPOD_modes_weight_acous/kz0_p_energy.svg}
    \includegraphics[width = \linewidth]{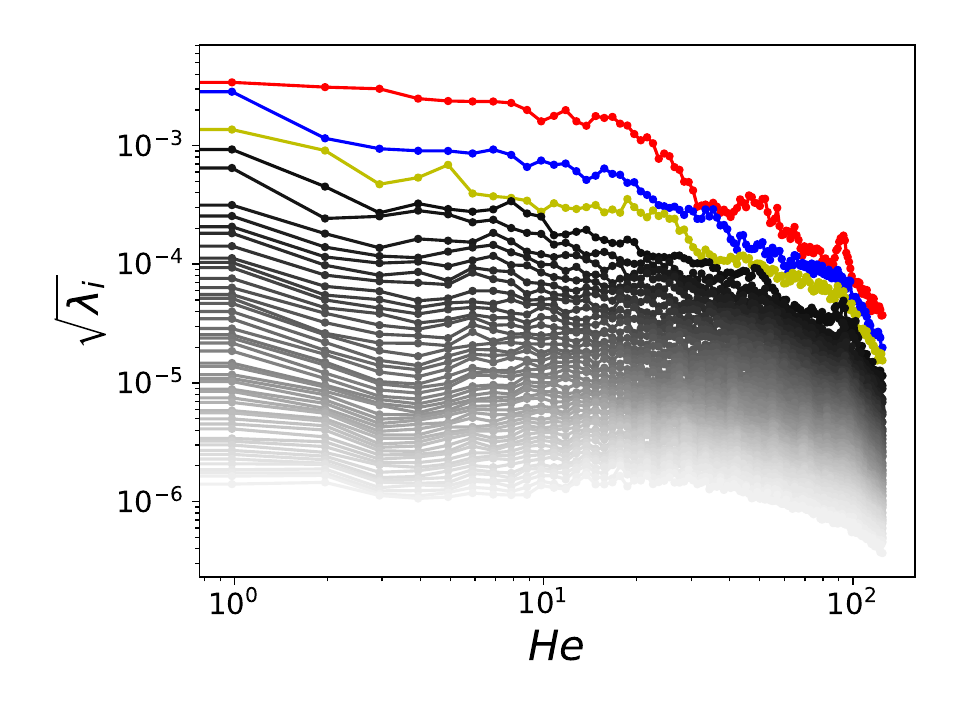}
    \caption{$n_z = 0$}
    \end{subfigure}
    \begin{subfigure}{0.49\textwidth}
    %\includesvg[width = \linewidth]{figs_JFM_wavepackets/SPOD_modes_weight_acous/kz1_p_energy.svg}
    \includegraphics[width = \linewidth]{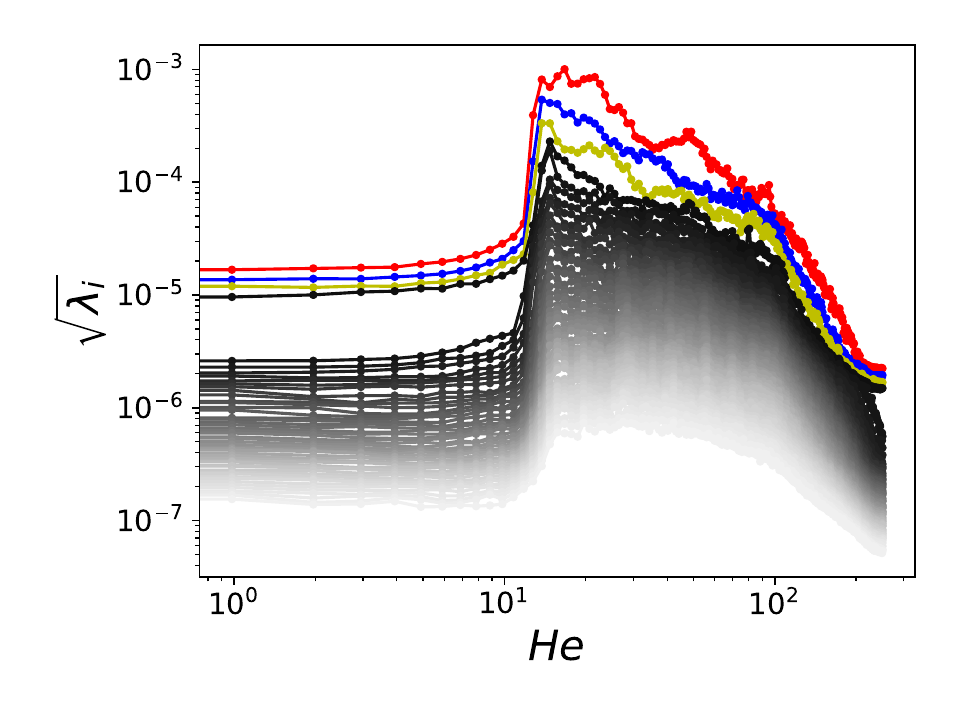}
    \caption{$n_z = 1$}
    \end{subfigure}
    \begin{subfigure}{0.49\textwidth}
    %\includesvg[width = \linewidth]{figs_JFM_wavepackets/SPOD_modes_weight_acous/kz2_p_energy.svg}
    \includegraphics[width = \linewidth]{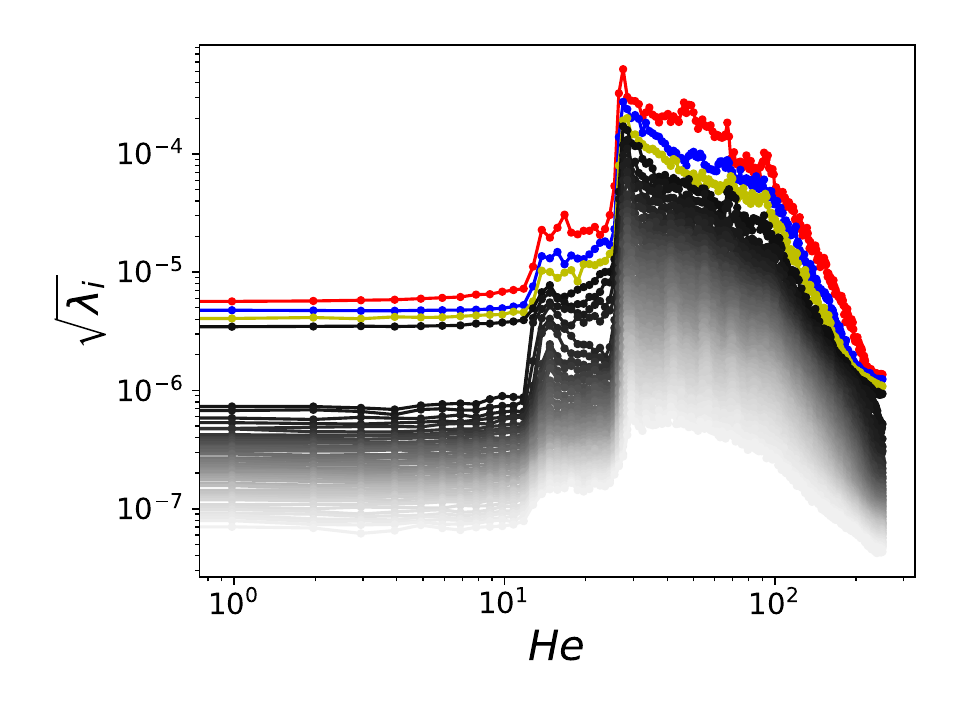}
    \caption{$n_z = 2$}
    \end{subfigure}
    \begin{subfigure}{0.49\textwidth}
    %\includesvg[width = \linewidth]{figs_JFM_wavepackets/SPOD_modes_weight_acous/kz3_p_energy.svg}
    \includegraphics[width = \linewidth]{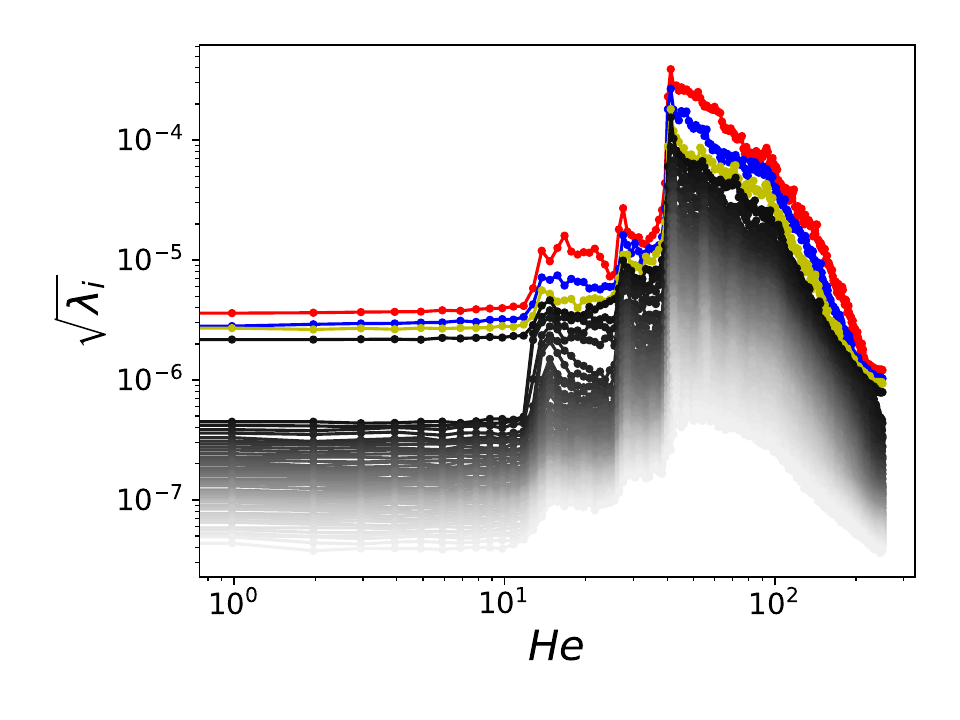}
    \caption{$n_z = 3$}
    \end{subfigure}
    \caption{Spectrum of the first four spanwise wavenumbers from the A-SPOD analyses. The first three SPOD modes are coloured with red, blue and yellow, respectively.}
    \label{fig:spectrum_weight_acous}
\end{figure}

The A-SPOD spectrum for the first four wavenumbers is presented in figure \ref{fig:spectrum_weight_acous}. In the case of $n_z = 0$, the spectrum is broadband. The energy contained in the leading mode exceeds 80\% in the frequency range $3 \le He \le 25$. The ratio between the energy contained in the leading SPOD mode and the total energy is more significant in the A-SPOD than in the H-SPOD, which is expected given that turbulent dynamics are a much more complex phenomenon than acoustic dynamics. 
In the case of $n_z = 1$, the spectrum demonstrates a notable reduction in total energy within the frequency range up to $He \approx 12$, where a significant step is observed. After this threshold frequency, the trend is similar to that observed for $n_z = 0$. The SPOD energy at the lower frequencies is at least one order of magnitude lower, indicating the non-propagative condition. As the wavenumbers are increased further, the spectrum exhibits the appearance of multiple steps, which is consistent with the expectation that more propagative waves will be present in the higher frequency regime. The frequency thresholds follow the scattering conditions obtained from H-SPOD, as illustrated in figure \ref{fig:psd_spod}. As with the case of $n_z = 0$, the low-rank feature also persists in the higher wavenumber contents, and the leading A-SPOD mode is an effective representation of the flow. Therefore, A-SPOD may offer potential benefits for reduced-order modelling.

%---------------------------------------------------------------------------------
%---------------------------------------------------------------------------------

\subsubsection{A-SPOD mode shapes}
\begin{figure}
    \centering
    \begin{subfigure}{0.49\textwidth}
    \includegraphics[width = \linewidth]{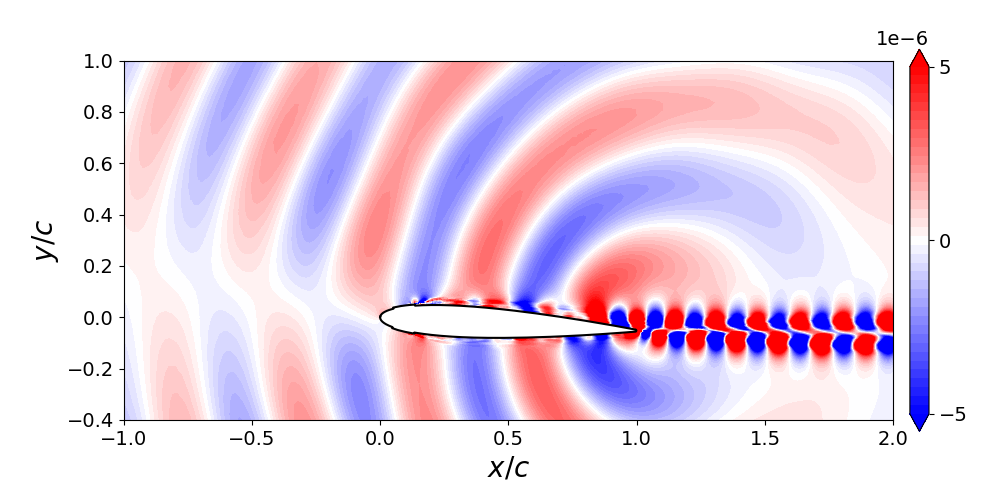}
    \caption{$n_z = 0, ~ He = 9.82$}
    \end{subfigure}
    \begin{subfigure}{0.49\textwidth}
    \includegraphics[width = \linewidth]{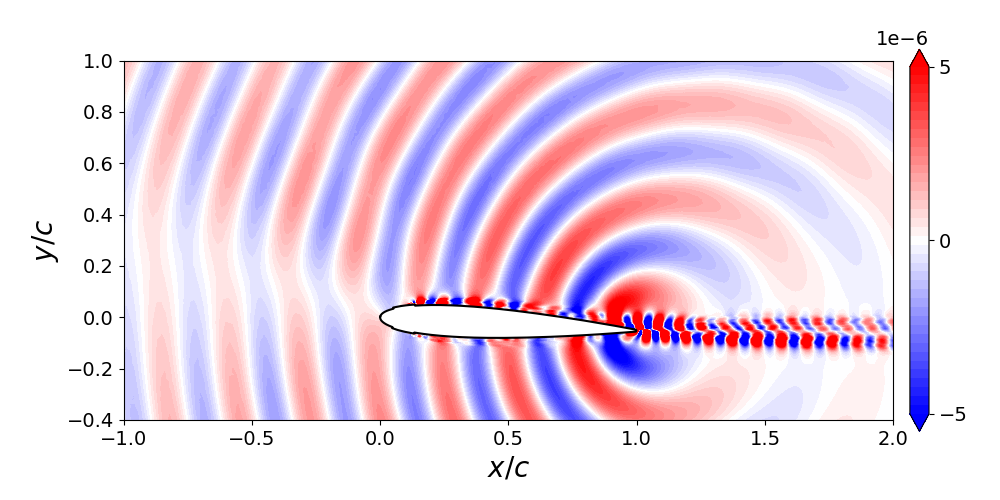}
    \caption{$n_z = 0, ~ He = 16.69$}
    \end{subfigure}
    \begin{subfigure}{0.49\textwidth}
    \includegraphics[width = \linewidth]{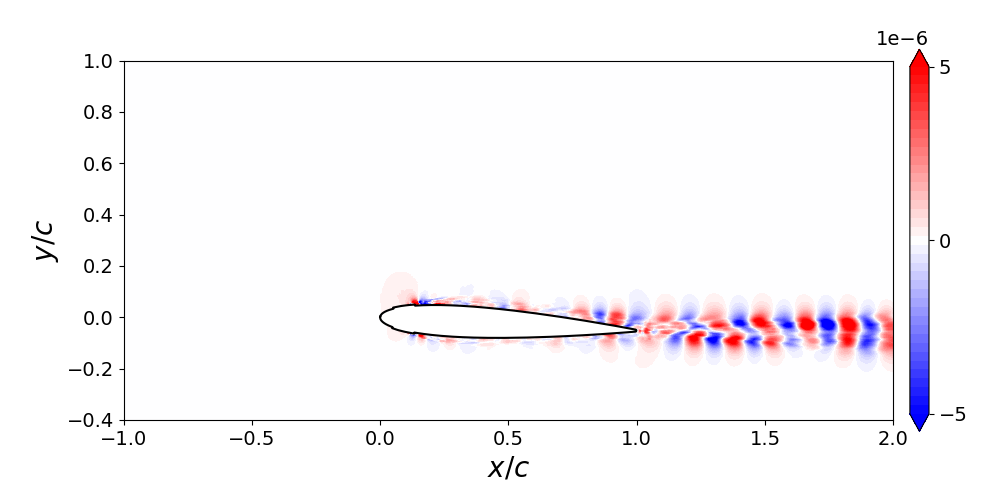}
    \caption{$n_z = 1, ~ He = 9.82$}
    \end{subfigure}
    \begin{subfigure}{0.49\textwidth}
    \includegraphics[width = \linewidth]{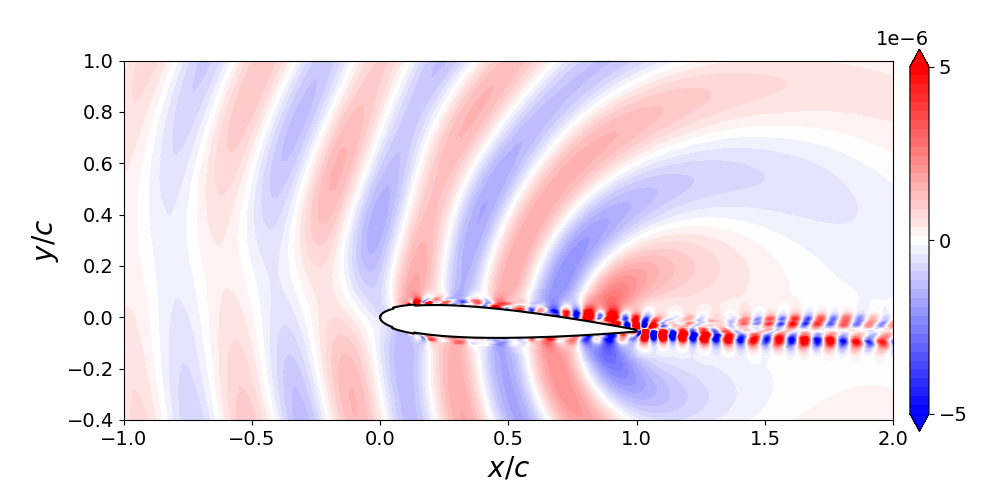}
    \caption{$n_z = 1, ~ He = 16.69$}
    \end{subfigure}
    \caption{The \textit{leading} A-SPOD mode shapes $\tilde{p}$ of wavenumbers $n_z = 0, 1$ for frequencies $He = 9.82$ and $16.69$.}
    \label{fig:A-spod_modes_p}
\end{figure}

The pressure component of A-SPOD modes are presented in figure \ref{fig:A-spod_modes_p}. The frequencies and wavenumbers presented here are the same as those shown in figure \ref{fig:leading_SPOD_mode_kz012_p_He_9.82_16.69} to allow for a direct comparison of the H-SPOD and A-SPOD approaches. In general, the A-SPOD modes have similar wavepacket structures in the boundary layer and acoustic farfield   compared to the H-SPOD modes. Moreover, the A-SPOD modes obey the scattering condition, with evanescent  acoustic waves for $n_z = 1$, $He = 9.82$ and propagative waves for the other cases. 

However, some differences between the A-SPOD and H-SPOD modes can be observed. The acoustic waves in the A-SPOD have higher pressure amplitudes, the wavepacket structures immediately downstream of the tripping element have lower amplitudes compared to the H-SPOD modes, and the wavepackets are more concentrated in the vicinity of the trailing edge and wake. It is also very interesting to see wavepacket structures  in the boundary layer for  the case of $n_z = 1$ and $He = 9.82$, which corresponds to non-propagative conditions. It is likely that the noise generated by the tripping element generates these wave-like structures. The noise level in this case is very low and cannot be observed under the current colour-scales. In the appendix  \ref{compare_ASPO_HSPOD_mode_shape}, the A- and H-SPOD modes  are further compared in terms of the velocity components $\Tilde{u}$ and $\Tilde{v}$, showing good agreement.

Overall, A-SPOD provide a very efficient basis for the low-dimensional representation of the farfield acoustics, and it also identifies wavepackets in the turbulent boundary layer as the driving mechanisms of the trailing-edge noise. These flow structures have a similar shape as the most energetic flow structures obtained via H-SPOD.

%---------------------------------------------------------------------------------
%---------------------------------------------------------------------------------

\subsubsection{Comparison of low-rankness between A-SPOD and H-SPOD}
\label{A_H_SPOD_comapre}

From the  A-SPOD and H-SPOD analysis, we can conclude that the hydrodynamic wavepackets are the sources that drive the trailing-edge noise. However, the two approaches yield significantly different levels of efficiency for low-order modelling of  the farfield acoustics. One of the objects of this paper is to provide guidance for new acoustic modelling. Therefore, in this section we will focus on the comparison between these two approaches for acoustic reconstructions.

\begin{figure}
    \centering
    \begin{subfigure}{0.49\textwidth}
    \includegraphics[width = \linewidth]{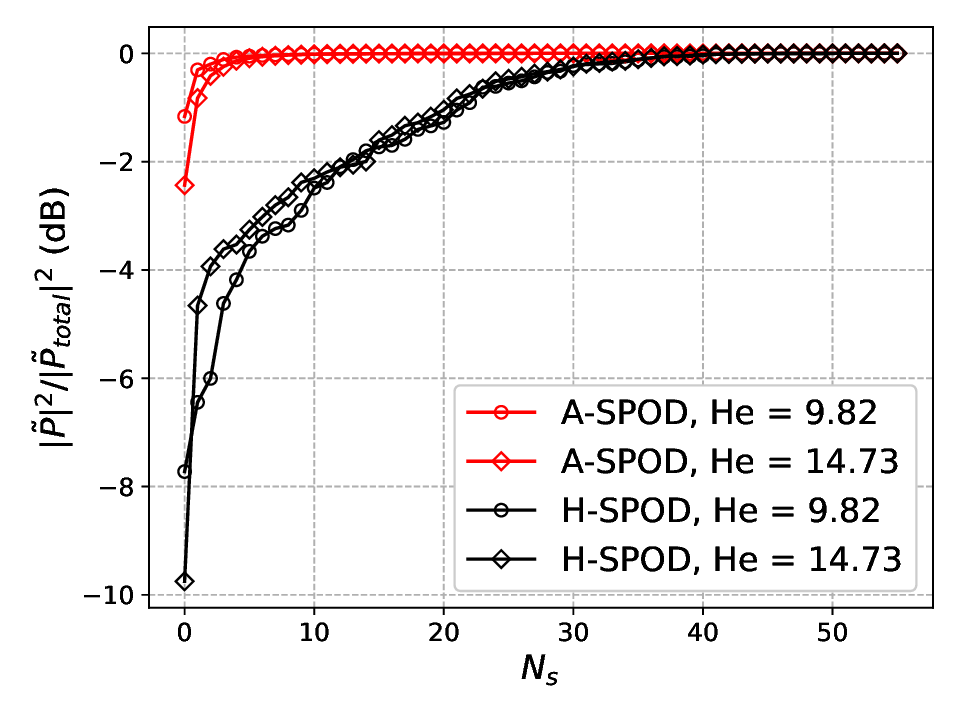}
    \caption{}
    \label{subfig:p_accum}
    \end{subfigure}
    \begin{subfigure}{0.49\textwidth}
    \includegraphics[width = \linewidth]{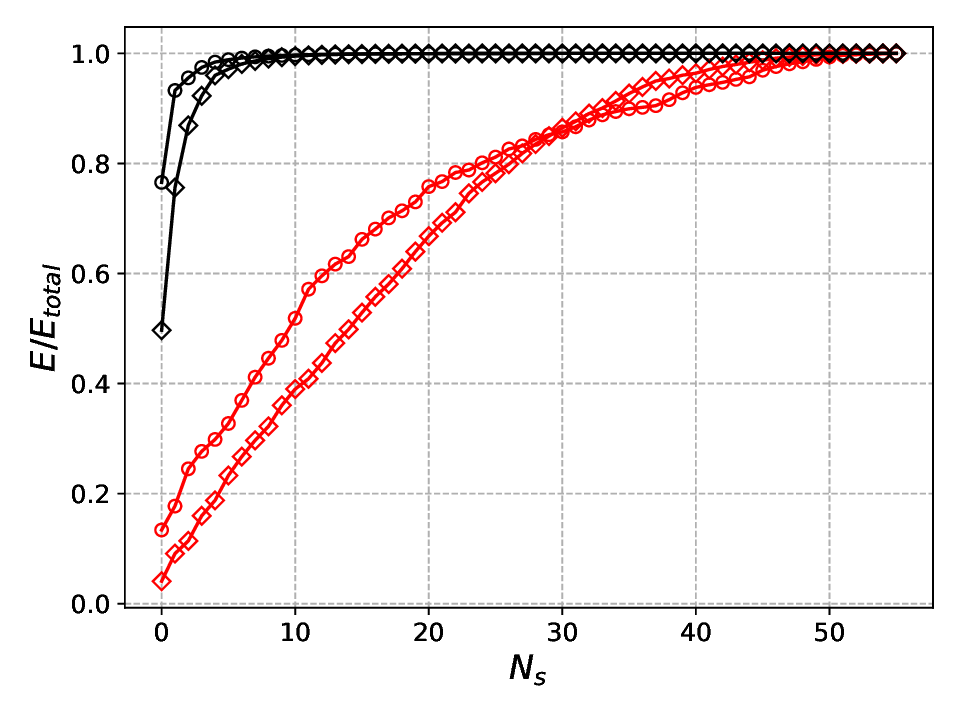}
    \caption{}
    \label{subfig:E_accume}
    \end{subfigure}
    \begin{subfigure}{0.49\textwidth}
    \includegraphics[width = \linewidth]{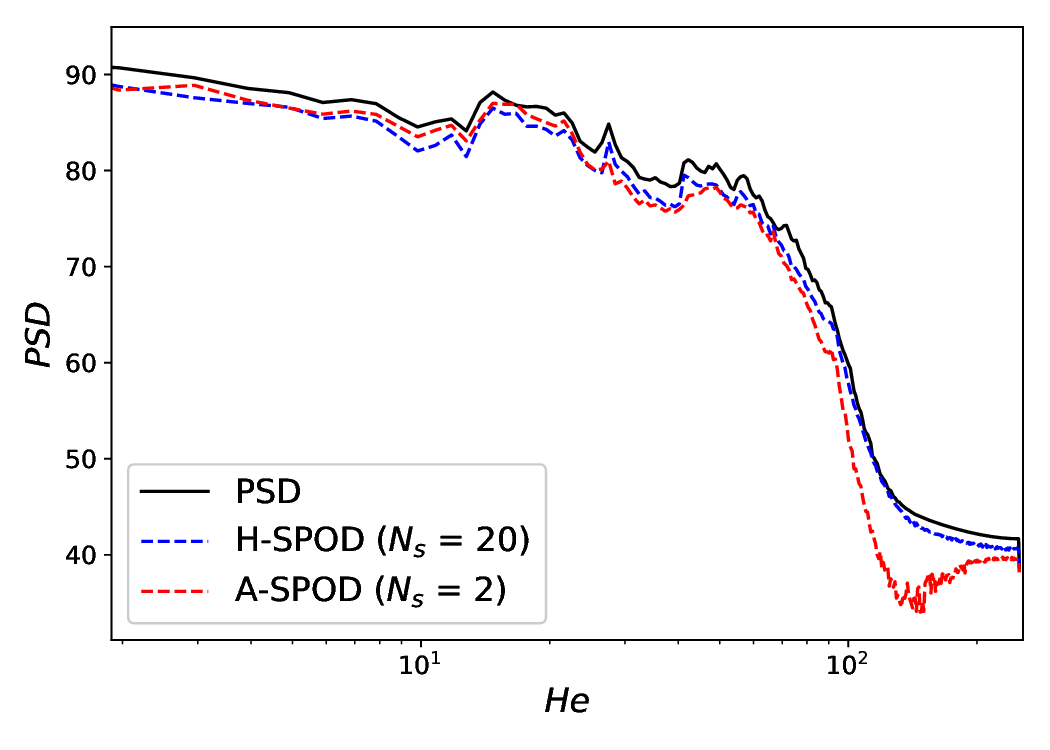}
    \caption{}
    \label{fig:compare_Aspod_Hspod_PSD}
    \end{subfigure}
    \caption{(a) Ratio of acoustic energy between accumulated SPOD modes ($N_s$) to total acoustic energy integrated in the free stream (see figure \ref{fig:weight_acous}). (b) Ratio of compressible energy between accumulated SPOD modes to total energy integrated near the trailing edge (see figure \ref{fig:weight_TE}). (c) Comparison of the integrated PSD obtained from reconstructed H-SPOD, A-SPOD and the original dataset. Note that \textit{twenty} H-SPOD modes and \textit{two} A-SPOD modes are used for the reconstructions.}
    \label{fig:SPOD_kth_mode_contribution}
\end{figure}

As mentioned above, twenty H-SPOD modes are used to reconstruct the acoustic field, ensuring a maximum difference of less than 2 dB compared to the LES measurement. Here, we complete the story by checking the convergence of the farfield acoustic energy with respect to the increasing number of H-SPOD modes. This is done by evaluating the sound power $|\Tilde{p}|^2$ in a freestream sub-domain as illustrated in  figure \ref{fig:weight_acous}. The results are presented in figure \ref{subfig:p_accum}. 
For both frequencies examined here, the difference in SPL between the leading H-SPOD mode and the total acoustic energy is above 7.8 dB. In sharp contrast, the difference between the leading A-SPOD mode and total acoustic energy is below 2.4 dB. Furthermore, a much faster convergence with increasing mode number for the  A-SPOD modes can be identified. Comparing with H-SPOD, A-SPOD requires only two modes to  reconstruct the acoustics within 1 dB difference, while the H-SPOD requires 24 modes. Together with the significant difference between the energy contained within the leading modes, these results clearly show that the A-SPOD approach is a more appropriate candidate for the reduced-order modelling of acoustics.

Next, we build the three-dimensional acoustic field using A-SPOD to compare with H-SPOD reconstruction and the measurement directly from the simulation. We perform the same analysis on the integrated acoustics along the circle arc indicated in figure \ref{subfig:acous_region}. 
We reuse the spectrum reconstructed by the leading twenty H-SPOD modes while  only two first A-SPOD modes are used for integration. Results are shown in figure \ref{fig:compare_Aspod_Hspod_PSD} where the contributions of all wavenumbers are considered. As can be seen there, two A-SPOD modes can give a better reconstruction than twenty H-SPOD modes up to $He \approx 25$. For the higher frequency range, more A-SPOD modes are required to achieve  better results. However, reconstructions by only two A-SPOD modes still have compatible amplitudes with obtained using twenty H-SPOD modes. This favourable comparison is suggesting that A-SPOD can provide an efficient low-rank reduced-order model of the acoustic field.

We hypothesize that A-SPOD is more suitable for low-order modelling than  H-SPOD, because the latter, being  optimal in terms of turbulent kinetic energy, might recover structures that are most energetic but may not cause the strongest acoustic radiation. 
To validate this hypothesis, we integrate the weighted energy of A-SPOD near the trailing edge, in the region marked by a red box in  figure \ref{fig:weight_TE}. This allows the quantification of the energy content of the leading SPOD mode that generates most of the acoustics with respect to the total energy. The result is shown in figure \ref{subfig:E_accume}. Here, $E_{total}$ stands for the total energy contained with respect to a single frequency and wavenumber. 
In contrast to the case of acoustic energy (figure \ref{subfig:p_accum}), the H-SPOD shows a faster convergence of the accumulated energy with increasing number of modes. And the leading H-SPOD mode contains a significantly larger proportion of energy compared to the A-SPOD. The energy contained within each A-SPOD is small and almost equal.
This suggests that a minor proportion of the hydrodynamic energy contributes significantly to the acoustic radiation. Recall that by performing a Fourier transform in the spanwise direction, turbulent flow structures are understood as a superposition of many Fourier modes with similar amplitudes.  Consequently, the energy contained in the structures contributing to the acoustic radiation is  insignificant among the turbulent flow. Focusing on the correlation between the turbulent structures and the acoustics could be challenging for accurate modelling of the acoustics. 
However, concentrating on the acoustic farfield provides a better set of modes for reduced-order modelling of generated noise.

\section{Conclusions}
\label{conclusion}
In this paper and the companion one \citep{Demange2024arxiv}, we investigate the generation mechanisms of broadband trailing-edge noise through numerical simulation and experiment.
The work presented here focuses on the numerical investigation of the problem, in which we perform a high-fidelity wall-resolved large eddy simulation with the similar geometry and flow conditions as in the experiments. The simulation was performed using the compressible high-order flux reconstruction framework, PyFR. The geometry under investigation is identical to that used in the experimental study, which incorporated zig-zag tripping elements in the vicinity of the leading edge. The simulation is designed to be comparable to the experiment having the same chord-based Reynolds number, Re = 200,000. However, there are some differences between the numerical and the experimental setup (e.g. span width, installation effects, and freestream Mach number). Following the scattering condition proposed by \citet{Nogueira2017Ampf}, the span extension of the numerical domain is chosen to be 43.75\% of the chord length, allowing for several propagative spanwise wavenumbers.

An in-depth cross-validation of the numerical and experimental datasets is performed. Velocity profiles obtained at the middle of the span on the airfoil surface and in the wake show good agreement with experimental results. Results of spectral analysis of farfield and surface pressure sensors also show good alignment with experiment when using appropriate scaling methods. Analysis of data for different spanwise Fourier modes showed that the acoustic modes are propagative only if $k_z < k_0$, where $k_z$ is the spanwise wavenumber and $k_0$ is the acoustic wavenumber, respectively. This confirms the work by \citet{Nogueira2017Ampf}.

Furthermore, we identified a strong coherence between spanwise-averaged surface pressure fluctuations and farfield acoustics, with up to 80\%  coherence, in the broadband noise frequency regime. This is in an apparent contrast to underlying ideas in  acoustic prediction models, where small surface coherent lengths rather than large spanwise coherent structures play an important role. As discussed in the companion paper, this contrast is indeed only apparent: low coherence lengths in turbulent boundary layers show that many spanwise wavenumbers are present in the field, but the scattering condition indicates that among these wavenumbers, only the lowest ones contribute to sound radiation.

To further reveal the role of spanwise coherent structures with low spanwise wavenumbers on trailing-edge noise, we employ the spectral proper orthogonal decomposition (SPOD) method on the spanwise Fourier transformed two-dimensional fields. In particular, we use the extended SPOD focused on a region in the vicinity of the trailing edge to identify wavepackets that are correlated with the acoustics. We find that the dominant hydrodynamic structures are spanwise coherent wavepackets concentrated around the trailing edge. Inspecting the pressure component of these modes, we find strong acoustic waves scattered at the trailing edge and propagating into the farfield, thereby following the scattering condition mentioned above.

The SPOD low-order model is validated by integrating the acoustic amplitude along a circular arc and comparing it with the LES dataset. Using the leading 20 H-SPOD modes achieves a maximum difference of 2 dB, but slow convergence indicates inefficiency. An acoustic-based A-SPOD addresses this issue by ranking modes based on farfield sound power, capturing up to 80\% of total acoustic energy with the leading mode. A-SPOD modes resemble H-SPOD modes, confirming wavepackets as the noise source. Notably, A-SPOD reconstruction  achieves a close agreement with LES with only two modes, enabling a low-rank, physics-based reduced-order model via resolvent analysis.

Finally, we also use the A-SPOD modes to recompute the  energy near the trailing edge. The results indicate that the hydrodynamic energy associated with the acoustic radiation is  small and that a minor proportion of the hydrodynamic energy contributes significantly to the acoustic radiation. 

In summary, this work, together with the companion paper \citep{Demange2024arxiv}, completes a joint numerical and experimental investigation of the broadband trailing-edge noise generation mechanisms. In contradiction to most acoustic models, we identify the importance of spanwise coherent structures with large spanwise wavelength for trailing-edge noise generation. The datasets validate the scattering condition and further identify spanwise coherent wavepacket-like structures as the driver of noise generation. Furthermore, we propose an efficient reduced-order acoustic model using acoustic-weighted SPOD analysis. The current work stats the ground for future resolvent analysis and adjoint-based shape optimisation for physics-based tailored noise control.

\section*{Acknowledgments}
Zhenyang Yuan would like to acknowledge Swedish Research Council for supporting current work under Grant 2020-04084. The computations were performed on resources by the National Academic Infrastructure for Supercomputing in Sweden (NAISS) at the LUMI supercomputer cluster in Finland. 
Simon Demange would like to acknowledge Deutsche Forschungsgemeinschaft for supporting current work under grant number 458062719.

\appendix

\section{Eigenvalue decomposition to the CSD matrix}
\label{SPOD_line_array}

As discussed in \S \ref{scattering_condition}, checking the scattering condition with respect to isolated spanwise wavenumbers and frequencies relies on the Fourier decomposition in both space and time. The companion paper \citep{Demange2024arxiv} shows that the dominant spanwise structures are Fourier modes despite non-periodic boundary conditions in the spanwise direction. In the simulation, a periodic boundary condition allows the spanwise Fourier decomposition. However, it is unclear whether Fourier modes are also optimal structures in the spanwise direction compared to the experimental results, specially given that the current configuration is not strictly homogeneous in the spanwise direction due to the zig-zag trip. Hence, the differences introduced by the different spanwise boundary conditions remain a question. Thus, motivated by these questions, we perform the same analysis that was performed on the experimental dataset.

\begin{figure}
    \centering
    \begin{subfigure}{\textwidth}
        \centering
        \includegraphics[width = \linewidth]{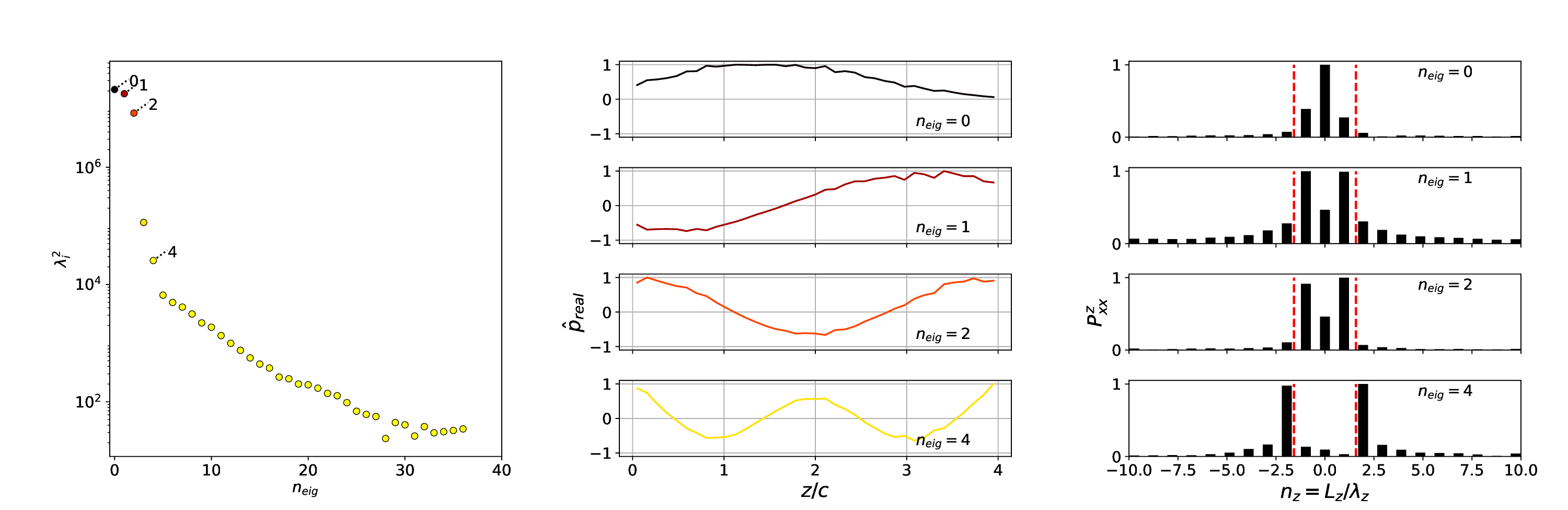}
        \caption{Exp, $He = 2$}
        \label{fig:SPODLinearray_exp}
    \end{subfigure}
    \begin{subfigure}{\textwidth}
        \centering
        \includegraphics[width = \linewidth]{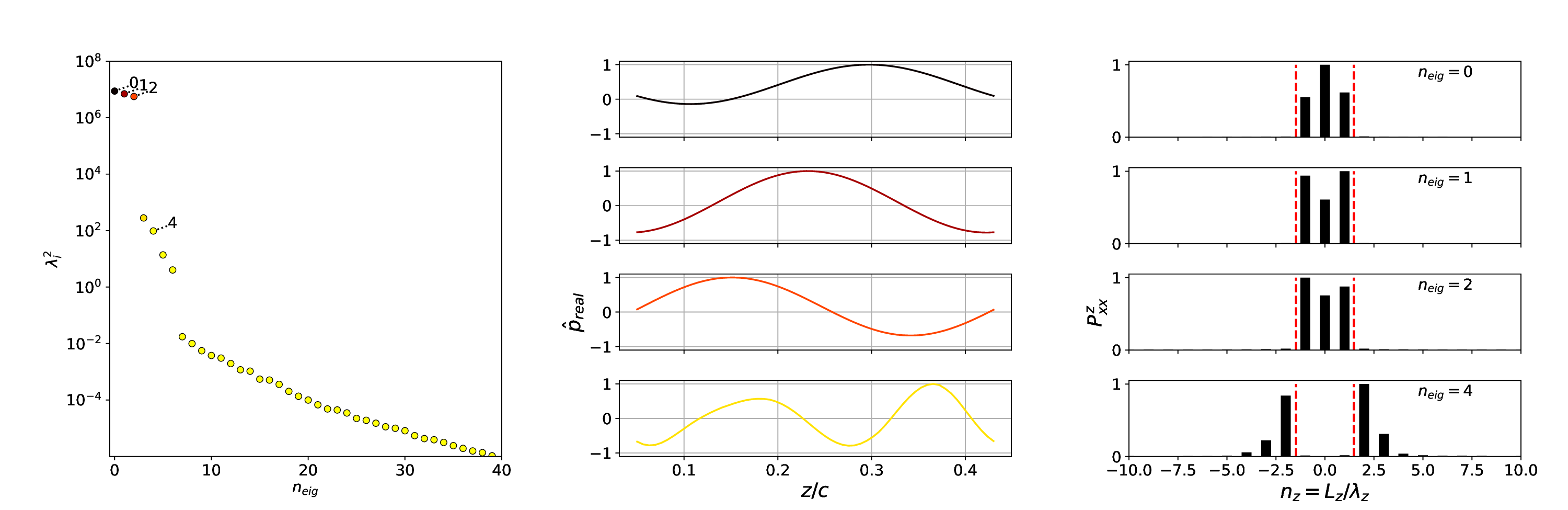}
        \caption{LES, $He = 16$}
        \label{fig:SPODLinearray_les}
    \end{subfigure}
    \caption{Eigenvalue analysis on the CSD matrices constructed from the acoustic line array signals from the simulation and experimental datasets. Eigenvalues (left), eigenmodes (mode) and the spanwise Fourier decomposition of the eigenmodes (right) are shown. The scattering condition ($-k_0 < k_z < k_0$), indicated by the vertical dashed red line, is shown to illustrate the wavenumbers that generate the radiated sound.}
    \label{fig:SPOD_line_array}
\end{figure}

The results at frequency $He = 2$ for the experimental signals and $He = 16$ for the numerical signals are shown in figure \ref{fig:SPOD_line_array}. The frequencies are chosen to have two propagative waves according to the scattering condition \ref{eq:scatter} for both datasets. In both cases, the eigenvalues (shown in the \textit{left} column) show that the leading modes have significant energy compared to the others. This indicates the low-rank feature of the acoustics. 
In both cases, there are three eigenvalues with significant amplitudes compared to the others. A rapid decrease in amplitude can be seen after $n_{eig} = 2$.

The shape of the dominant SPOD modes are shown in the \textit{middle} column of figure \ref{fig:SPOD_line_array}. The real part of the leading four modes are presented, indicated by the numbers labelled in the spectra. The results show that the mode shape is similar to Fourier modes. Further insight into the similarity of the mode shapes compared to the Fourier modes can be obtained by performing a spatial Fourier transform of the SPOD modes, as shown in the right column. It can be  seen 
 that the SPOD modes are dominated by very few   spanwise Fourier modes. The first SPOD mode consists mainly of a mode with zero spanwise wavenumber, while the next lower modes consists of two modes with same wavenumber but opposite sign. 
This observation suggests that the SPOD modes represent a  superposition of left and right travelling acoustic waves identified by positive and negative wavenumbers $n_z$. 
these observations are consistent with experimental findings shown in the companion paper, indicating that the periodic boundary condition has less influence on the mode shapes. Thus, a fair comparison can be made for both datasets. 

In figure \ref{fig:SPOD_line_array}, the scattering condition (given by the acoustic wavenumber $k_0$) is indicated as vertical dashed lines. The region where $-k_0 < k_z < k_0$ indicates the propagative waves, while the rest are evanescent waves.
As can be seen, only the first three SPOD modes are propagative, which explains their significant  energy shown in the eigenvalue spectrum. 

Overall, the  SPOD of the CSD matrix reveals modes that are very similar to spanwise Fourier modes, which  justifies the Fourier  decomposition of the acoustic data.

%--------------------------------------------------------------------
%--------------------------------------------------------------------

\section{SPOD convergence analysis}
\label{SPOD_convergence}
In order to illustrate the convergence and robustness of the computed SPOD modes, a convergence analysis was conducted. The methodology employed is similar to that described by \citet{Abreu2021}, where the entire time signal is split into two equal parts, each comprising 75\% of the original dataset. Subsequently, SPOD was performed on each part separately. Similar to \eqref{eq:coherence}, a correlation relation between the modes from each dataset can be defined to determine the convergence:
\begin{equation}
    \sigma_{i,k} = \frac{\left<\Psi_{k}, \Psi_{i,k}\right>}{\sqrt{\left<\Psi_{k}, \Psi_{i,k}\right>}\sqrt{\left<\Psi_{i,k}, \Psi_{i,k}\right>}}
\end{equation}
where $\sigma$ stands for the correlation coefficient, $i \in \{1,2\}$ are two subsets of the original LES data and $k \in \{1,2,3\dots, N_b\}$ are the indices of the SPOD modes. Note that $\left<\cdot,\cdot\right>$ indicates the non-weighted inner product, which represents the projection of the partial dataset $\Psi_{i,k}$ to the full dataset $\Psi_{k}$. It is anticipated that, in the event of a sufficiently long time series, the correlation coefficient will approach a value of one. 

\begin{figure}
    \centering
    \begin{subfigure}{0.49\textwidth}
        \centering
        %\includesvg[width = \textwidth]{figs_JFM_wavepackets/SPOD_convergence_rho_H.svg}
        \includegraphics[width = \linewidth]{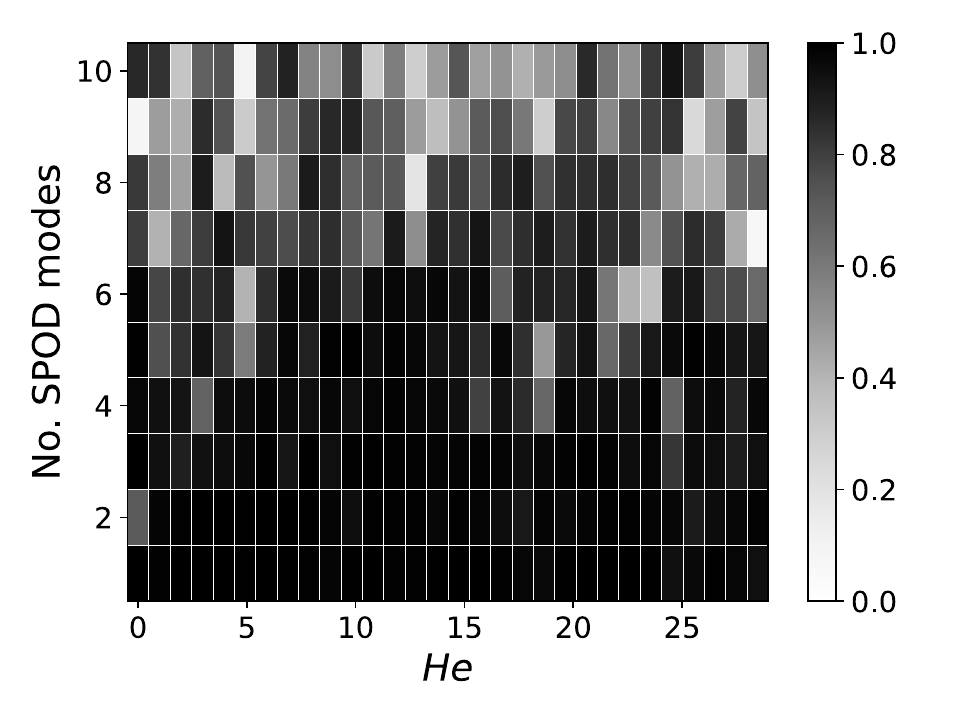}
        \caption{Hydrodynamic SPOD}
        \label{fig:conv_H}
    \end{subfigure}
    \begin{subfigure}{0.49\textwidth}
        \centering
        %\includesvg[width = \textwidth]{figs_JFM_wavepackets/SPOD_convergence_rho_A.svg}
        \includegraphics[width = \linewidth]{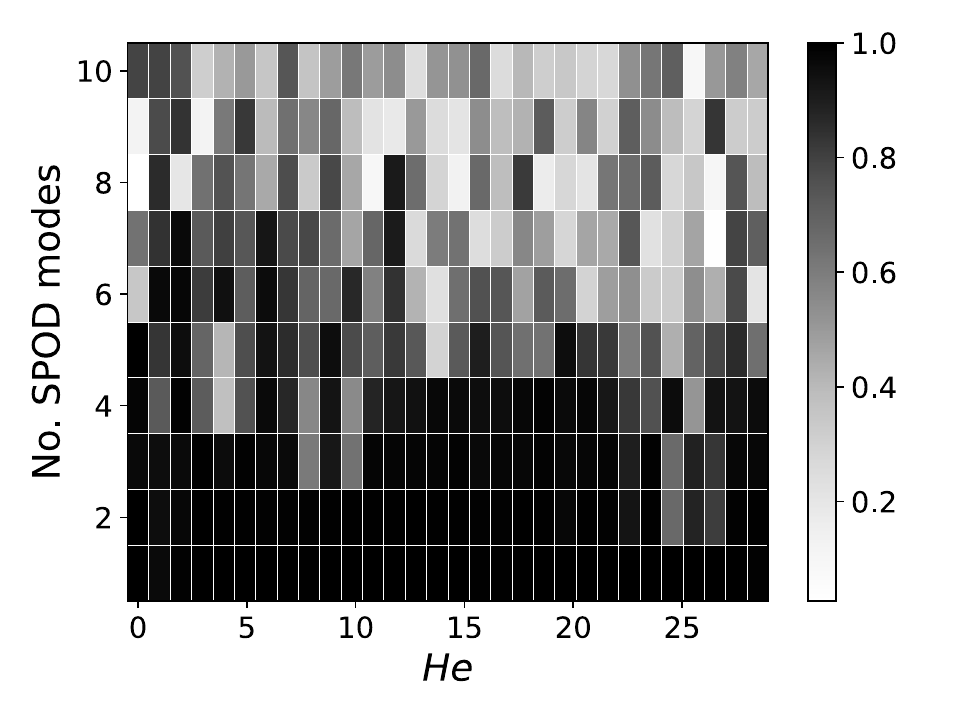}
        \caption{Acoustic SPOD}
        \label{fig:conv_A}
    \end{subfigure}
    \caption{Correlation coefficient $\sigma_{1,k}$ for the leading 10 SPOD modes with $He \leq 30$. Here shows the dataset $i = 1, 2$ with 75\% overlapping with the original LES dataset.}
    \label{fig:convergence_spod}
\end{figure}

Figure \ref{fig:convergence_spod} illustrates the correlation coefficient $\sigma_{1,k}$ for the leading 10 SPOD modes within the frequency range $He < 30$ for both hydrodynamic SPOD and acoustic SPOD. Accordingly, the leading SPODs are very well converged, with all leading SPOD modes having $\sigma_{1,k} \geq 0.98$ for $He < 25$. For the majority of analyses, the leading SPOD modes are presented, and therefore, it can be concluded that the SPOD analysis is sufficiently well converged.

%--------------------------------------------------------------------
%--------------------------------------------------------------------
\section{H-SPOD mode shapes of the velocity components}
\begin{figure}
    \centering
    \begin{subfigure}{0.49\textwidth}
    \centering
    \includegraphics[width = \linewidth]{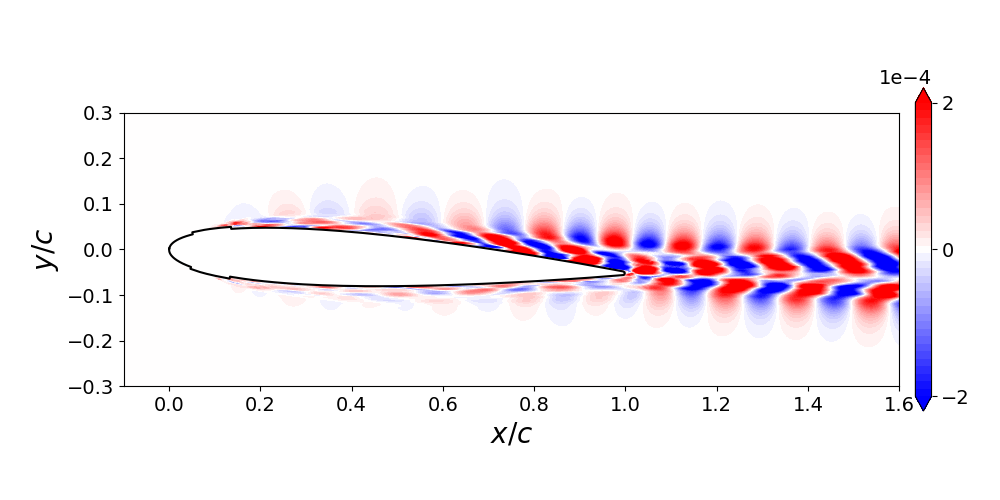}
    \caption{$n_z = 0, ~ \Tilde{u}$}
    \label{fig:SPOD_u_9.82_0}
    \end{subfigure}
    \begin{subfigure}{0.49\textwidth}
    \centering
    \includegraphics[width = \linewidth]{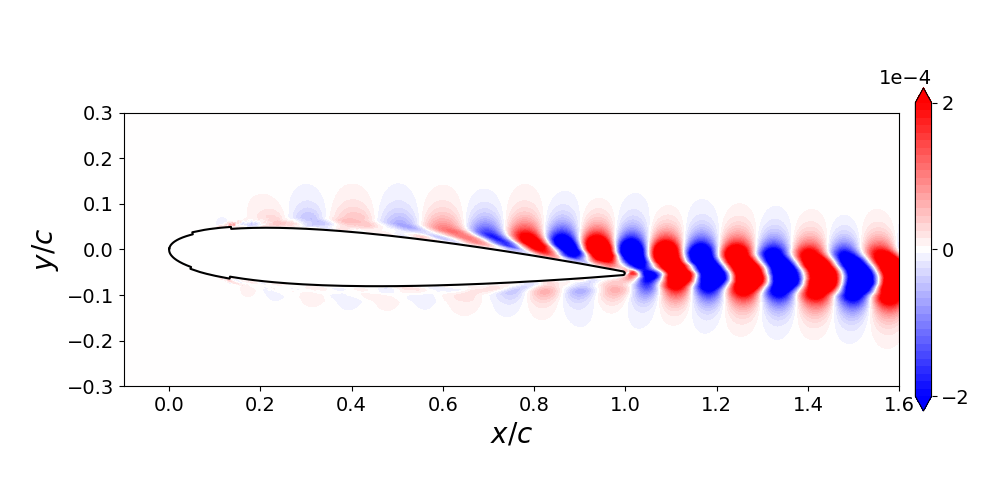}
    \caption{$n_z = 0, ~ \Tilde{v}$}
    \label{fig:SPOD_v_9.82_0}
    \end{subfigure}
    \begin{subfigure}{0.49\textwidth}
    \centering
    \includegraphics[width = \linewidth]{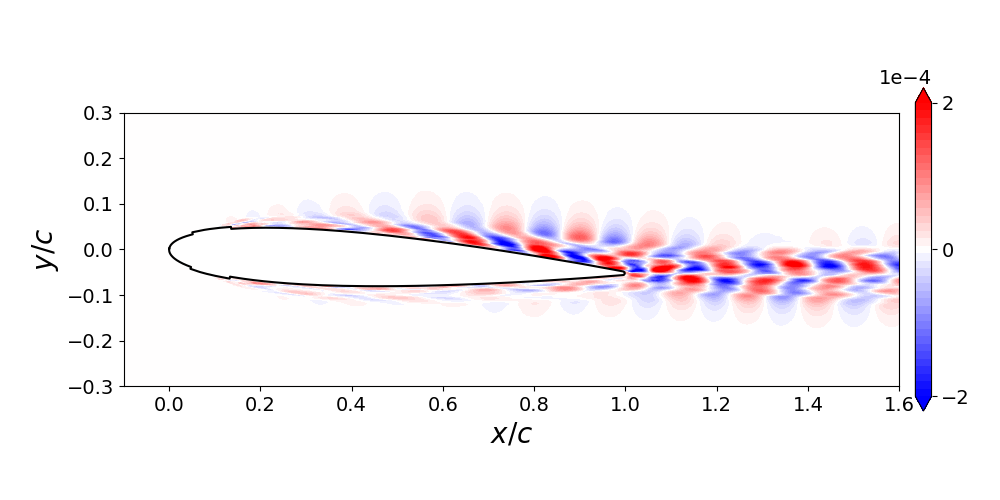}
    \caption{$n_z = 1, ~ \Tilde{u}$}
    \label{fig:SPOD_u_9.82_1}
    \end{subfigure}
    \begin{subfigure}{0.49\textwidth}
    \centering
    \includegraphics[width = \linewidth]{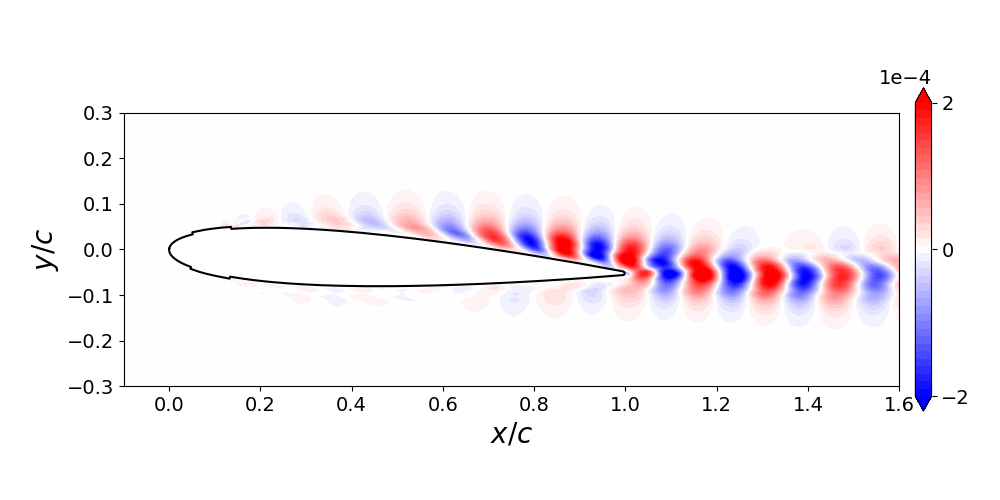}
    \caption{$n_z = 1, ~ \Tilde{v}$}
    \label{fig:SPOD_v_9.82_1}
    \end{subfigure}
    \begin{subfigure}{0.49\textwidth}
    \centering
    \includegraphics[width = \linewidth]{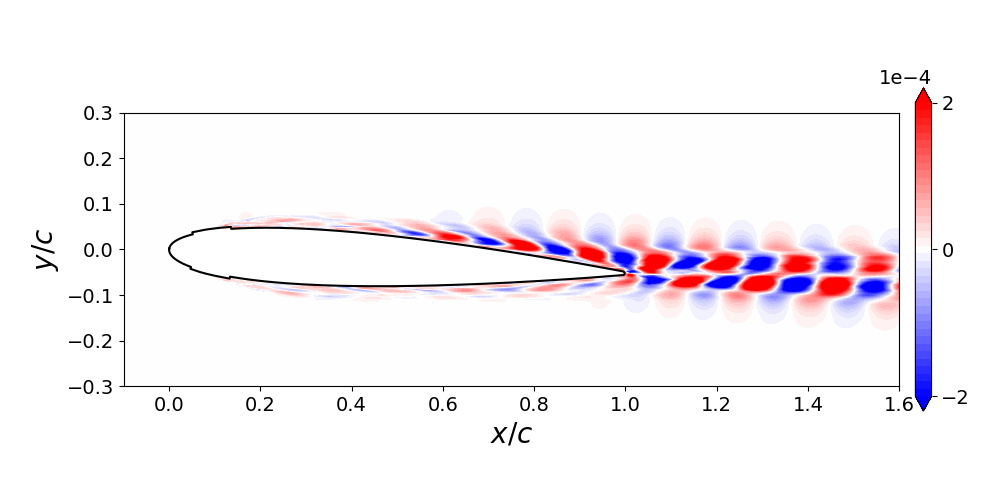}
    \caption{$n_z = 2, ~ \Tilde{u}$}
    \label{fig:SPOD_u_9.82_2}
    \end{subfigure}
    \begin{subfigure}{0.49\textwidth}
    \centering
    \includegraphics[width = \linewidth]{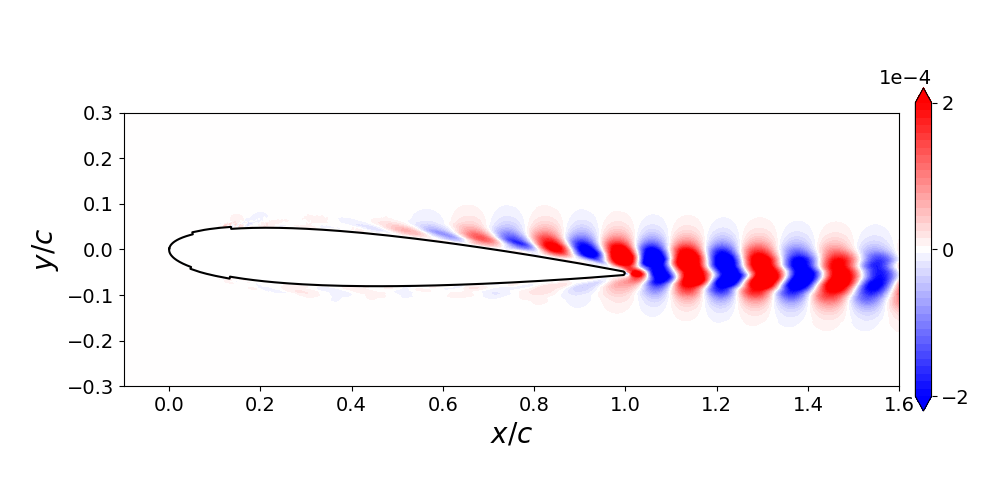}
    \caption{$n_z = 2, ~ \Tilde{v}$}
    \label{fig:SPOD_v_9.82_2}
    \end{subfigure}
    \begin{subfigure}{0.49\textwidth}
    \centering
    \includegraphics[width = \linewidth]{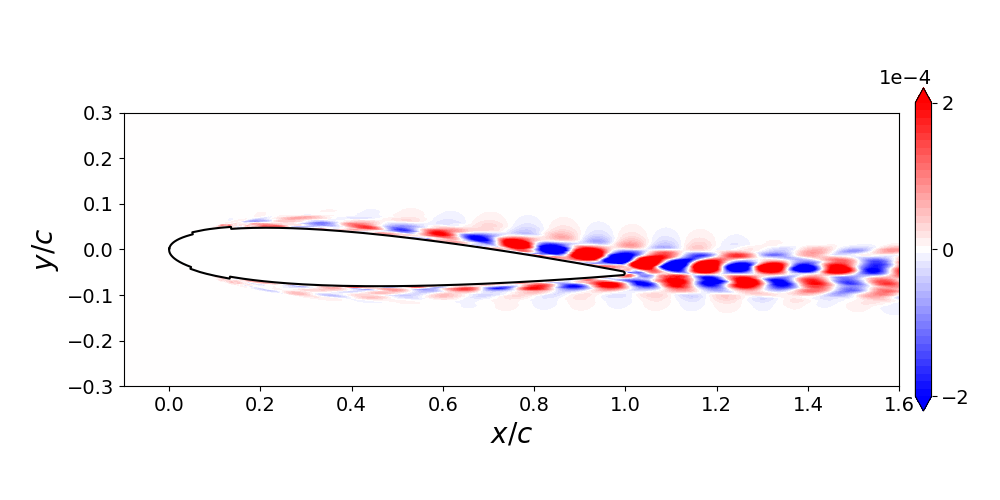}
    \caption{$n_z = 3, ~ \Tilde{u}$}
    \label{fig:SPOD_u_9.82_3}
    \end{subfigure}
    \begin{subfigure}{0.49\textwidth}
    \centering
    \includegraphics[width = \linewidth]{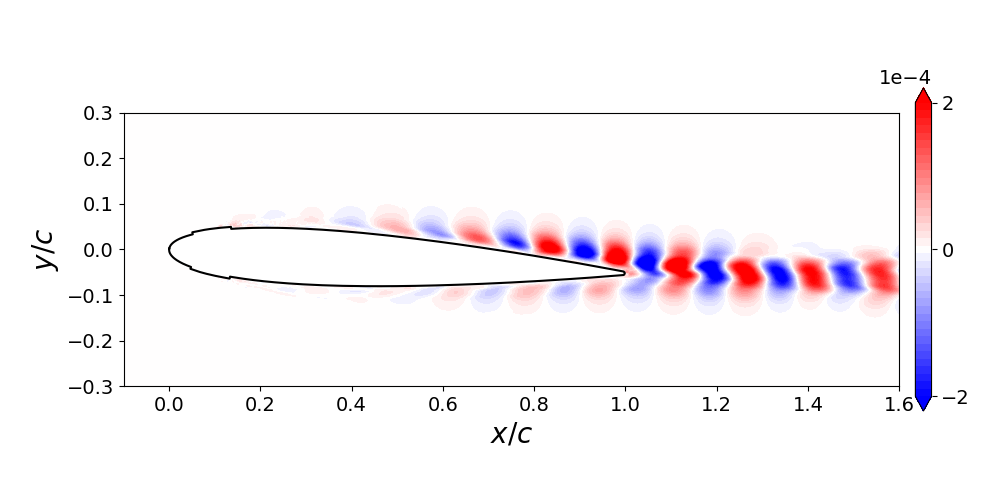}
    \caption{$n_z = 3, ~ \Tilde{v}$}
    \label{fig:SPOD_v_9.82_3}
    \end{subfigure}
    \caption{The \textit{leading} SPOD mode shape for the first three leading spanwise wavenumbers ($n_z = 0, 1, 2, 3$) at the frequency $He = 9.82$. (a, c, e, g) and (b, d, f, h) show mode shapes corresponding to $\Tilde{u}$, and $\Tilde{v}$, respectively.}
    \label{fig:leading_SPOD_mode_kz012_He_9.82}
\end{figure}

Figure \ref{fig:leading_SPOD_mode_kz012_He_9.82} shows velocity components $\Tilde{u}$ and $\Tilde{v}$ of the leading SPOD mode for the first four spanwise wavenumbers at frequency $He = 9.82$, where the relative energy share of the leading SPOD mode is maximum. 
For the wavenumbers and frequency considered here, the mode shapes reveal a wavepacket extending from the tripping elements on both the suction and pressure sides to the wake of the airfoil. 
The maximum amplitude of the wavepacket is found in the region close to the trailing edge. 
Similar to the pressure mode presented in figure \ref{fig:leading_SPOD_mode_kz012_p_He_9.82_16.69}, the fluctuations on the suction side have a significantly larger amplitude than those on the pressure side. Furthermore, over the suction side of the airfoil, the streamwise velocity fluctuations $\Tilde{u}$ show a phase change in the wall normal direction. 
These structures are typically found in the shear layer induced instabilities, such as the Kelvin-Helmholtz instability, especially in the region where an adverse pressure gradient acts \citep{Crighton_Gaster_1976, MICHALKE1972213, heningson2001}.

In the case of $n_z = 2$ the wavepacket structures move further downstream and into the trailing edge and near-wake region and almost disappear on the pressure side. This observation is particularly pronounced following the evolution of the $\Tilde{v}$ component.  This is likely due to the fact that the $n_z = 2$ case picks up the streaks generated by the tripping elements, which delay the growth of the turbulent boundary layer.

In general, the wavepacket structures presented here are very similar to those found in the work of \citet{Sano2019, Abreu2021}, despite the different tripping techniques employed. Streaks from tripping could have influence downstream of the tripping elements for specific wavenumbers, but the influence is less significant in the region close to the trailing-edge.

%--------------------------------------------------------------------
%--------------------------------------------------------------------
\section{Comparison between the acoustic and hydrodynamic SPOD mode shapes}
\label{compare_ASPO_HSPOD_mode_shape}
\begin{figure}
    \centering
    \begin{subfigure}{0.49\textwidth}
    \centering
    \includegraphics[width = \linewidth]{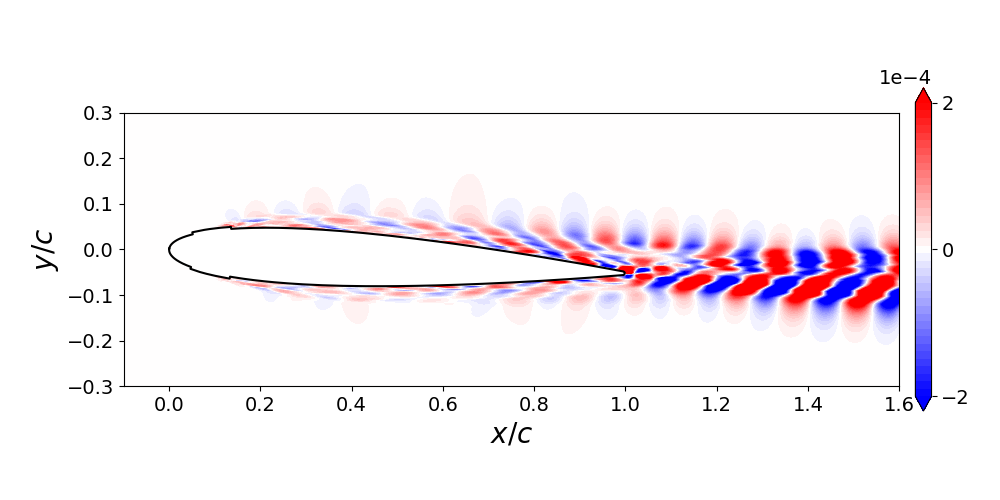}
    \caption{$n_z = 0, ~ \Tilde{u}, $ ASPOD}
    \end{subfigure}
    \begin{subfigure}{0.49\textwidth}
    \centering
    \includegraphics[width = \linewidth]{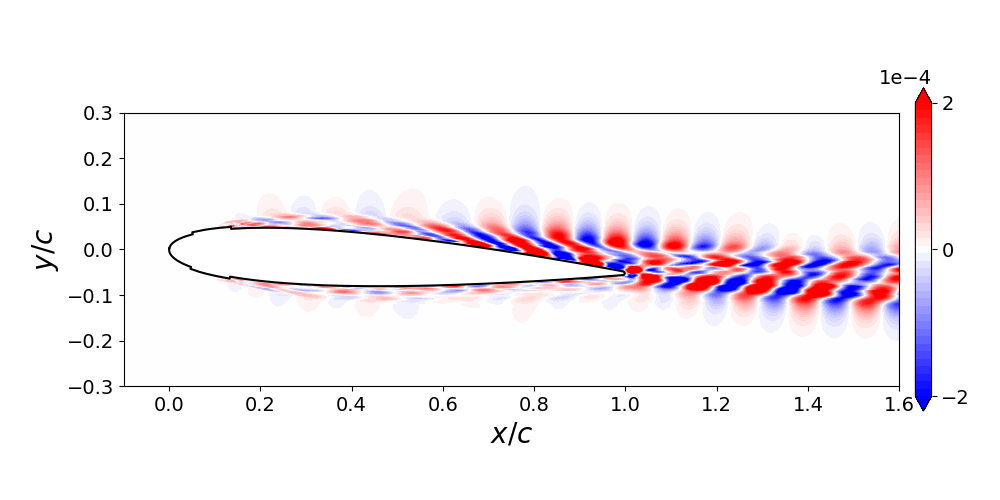}
    \caption{$n_z = 0, ~ \Tilde{u}, $ HSPOD}
    \end{subfigure}
    \begin{subfigure}{0.49\textwidth}
    \centering
    \includegraphics[width = \linewidth]{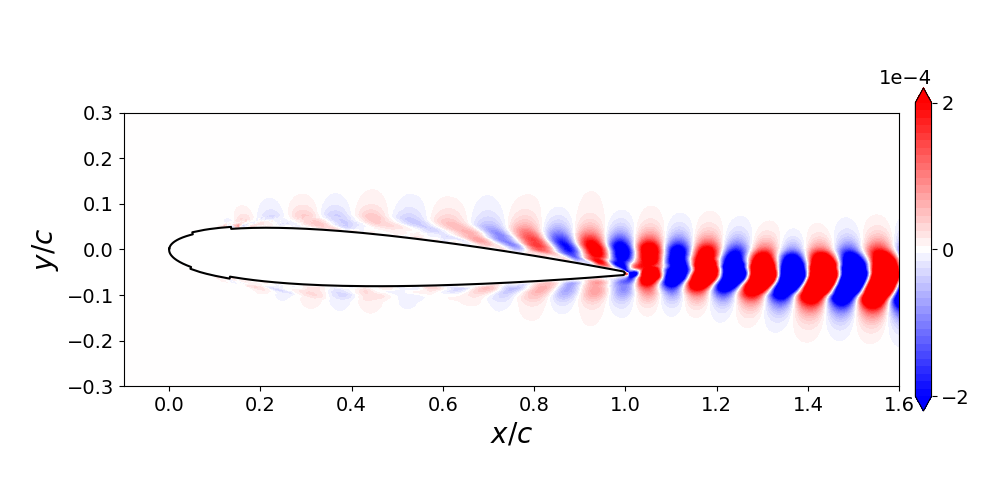}
    \caption{$n_z = 0, ~ \Tilde{v}, $ ASPOD}
    \end{subfigure}
    \begin{subfigure}{0.49\textwidth}
    \centering
    \includegraphics[width = \linewidth]{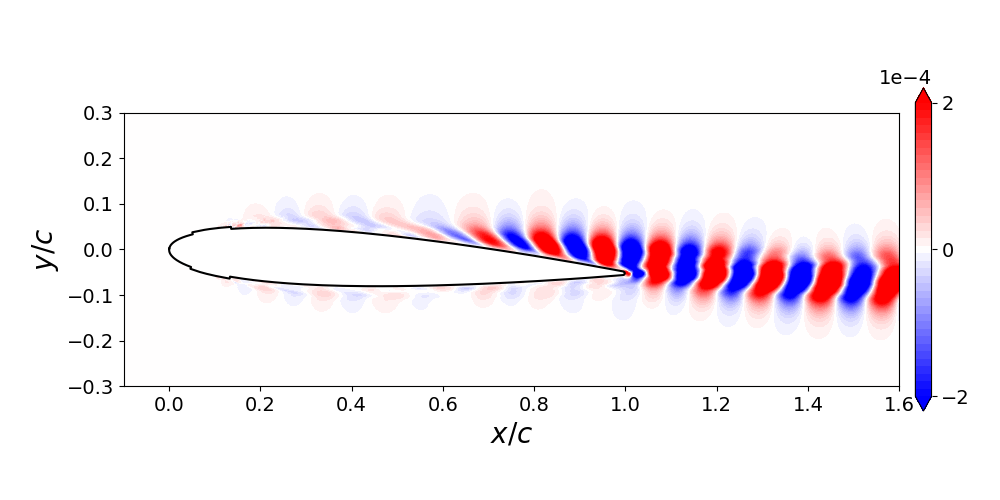}
    \caption{$n_z = 0, ~ \Tilde{v}, $ HSPOD}
    \end{subfigure}
    \begin{subfigure}{0.49\textwidth}
    \centering
    \includegraphics[width = \linewidth]{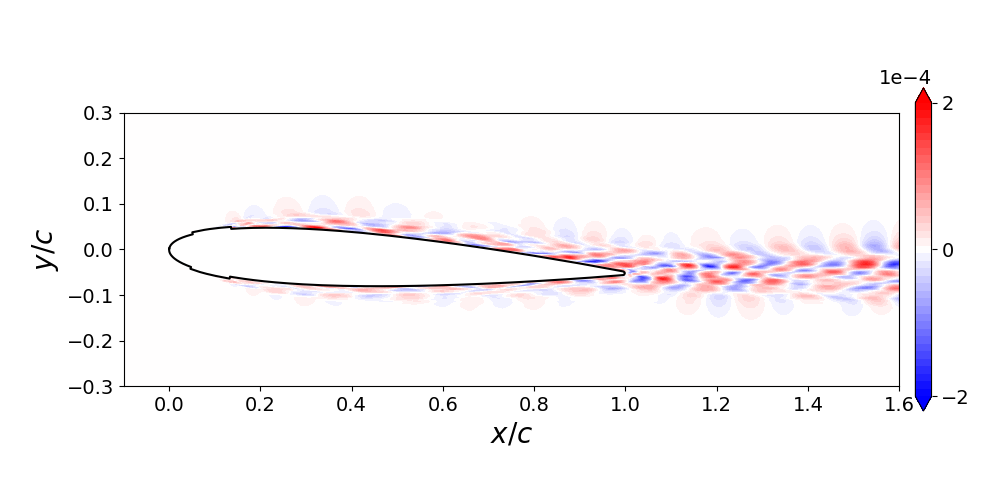}
    \caption{$n_z = 1, ~ \Tilde{u}, $ ASPOD}
    \end{subfigure}
    \begin{subfigure}{0.49\textwidth}
    \centering
    \includegraphics[width = \linewidth]{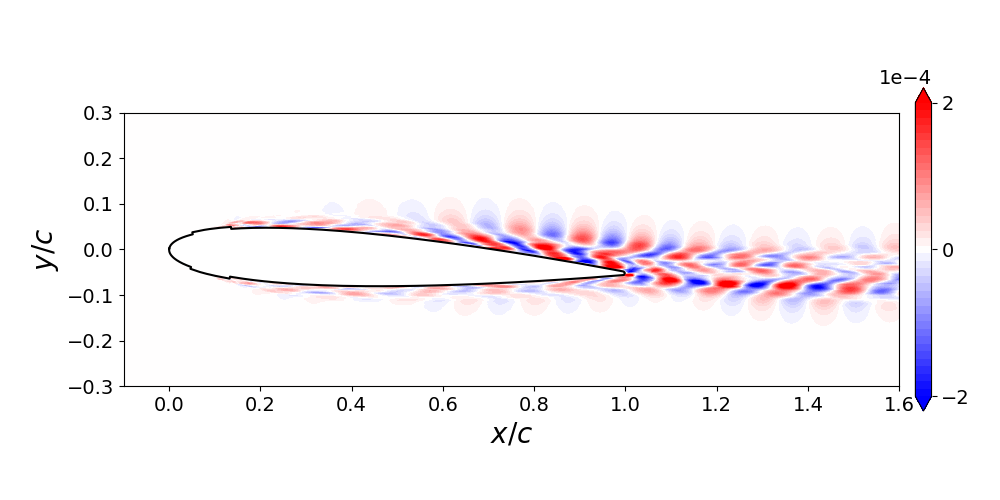}
    \caption{$n_z = 1, ~ \Tilde{u}, $ HSPOD}
    \end{subfigure}
    \begin{subfigure}{0.49\textwidth}
    \centering
    \includegraphics[width = \linewidth]{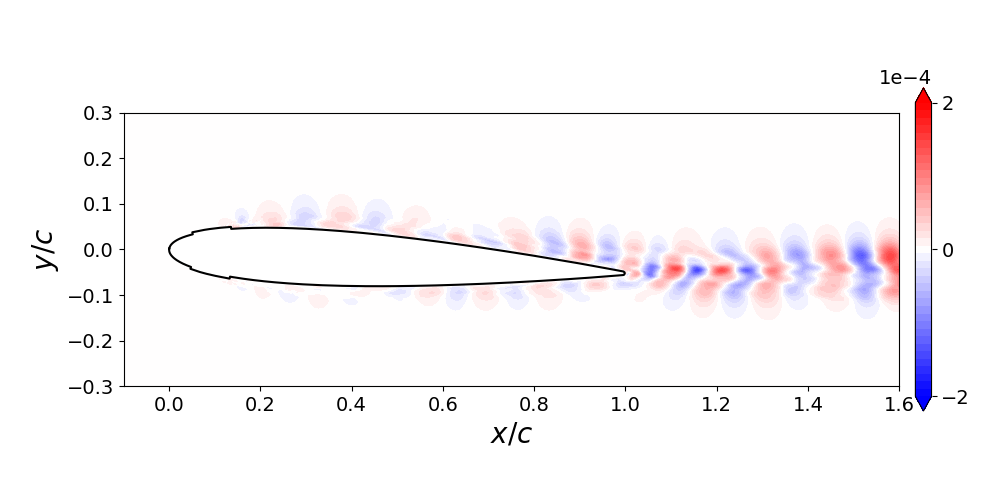}
    \caption{$n_z = 1, ~ \Tilde{v}, $ ASPOD}
    \end{subfigure}
    \begin{subfigure}{0.49\textwidth}
    \centering
    \includegraphics[width = \linewidth]{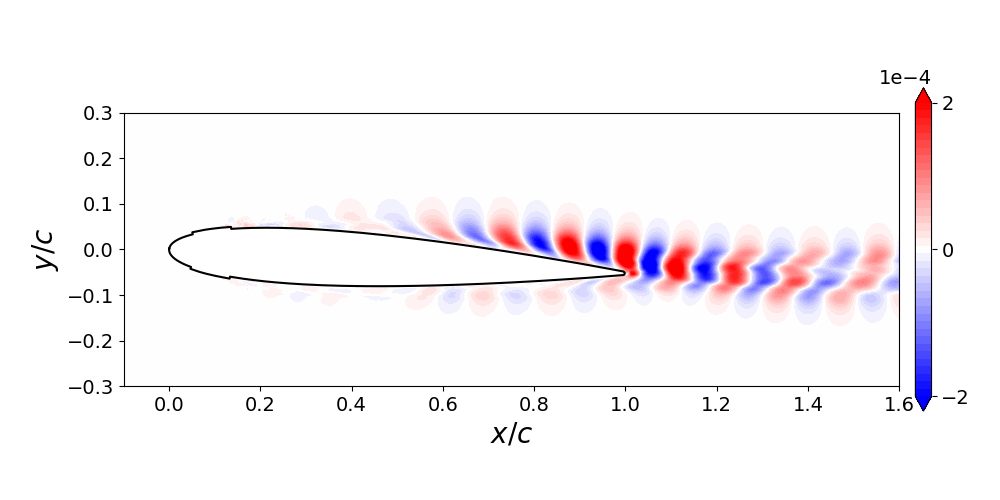}
    \caption{$n_z = 1, ~ \Tilde{v}, $ HSPOD}
    \end{subfigure}
    \caption{The \textit{leading} SPOD mode shape for the first two leading spanwise wavenumbers ($n_z = 0, 1$) at the frequency $He = 11.78$. Both acoustic SPOD and hydrodynamic SPOD are presented, indicated by ASPOD and HSPOD, respectively.}
    \label{fig:compare_Aspod_Hspod_mode_shape}
\end{figure}

This section reports the comparison of the mode shapes from A-SPOD and H-SPOD, focusing on the velocity components ($\Tilde{u}, \Tilde{v}$). The leading SPOD mode shapes at the frequencies $He = 11.78$ and the two leading wavenumbers $n_z = 0, 1$ are shown in figure \ref{fig:compare_Aspod_Hspod_mode_shape}. 

For the case $n_z = 0$, the mode shapes from A-SPOD and H-SPOD look similar, showing wavepacket structures that are propagating from the location of the tripping element to the wake region. The concentration of wavepackets around the trailing-edge can be seen particularly in the $\Tilde{v}$ component. A distinction can be observed in the case of $n_z = 1$. For the frequency $He = 11.78$, the acoustics are evanescent according to the scattering condition. Thus, the reconstruction of the near-field hydrodynamics from the A-SPOD shows a very different behaviour compared to the H-SPOD. The H-SPOD mode shapes still show the wavepacket structures, with envelopes mainly concentrated around the trailing edge. On the other hand, the structure of A-SPOD mode is less organised. 
In this case, A-SPOD captures the sound generated from the trip and vortices in the wake region, which is much weaker than the trailing-edge noise. 
This observation is consistent as discussed in \S \ref{ASPOD}.

The velocity reconstructions from the A-SPOD and H-SPOD modes demonstrate favourable consistency for propagative wavenumbers and frequencies. This agreement indicates that the wavepackets in the turbulent boundary layer can be identified as the driver of the scattering trailing-edge noise. 

\bibliographystyle{jfm}
\bibliography{PartI}

\end{document}